\documentclass[manuscript,acmlarge]{acmart}
\usepackage{booktabs}
\usepackage{caption}
\usepackage{subcaption}
\usepackage{pifont}% http://ctan.org/pkg/pifont
\usepackage{tikz} %package for checkmark
\usepackage{enumitem} 
\usepackage{graphicx}
\usepackage[normalem]{ulem}
\usepackage{makecell}
\usepackage{url} 
\usepackage{wrapfig}

\newcommand{\cmark}{\ding{51}}%
\newcommand{\xmark}{\ding{55}}%

\AtBeginDocument{%
  \providecommand\BibTeX{{%
    \normalfont B\kern-0.5em{\scshape i\kern-0.25em b}\kern-0.8em\TeX}}}

\setcopyright{acmcopyright}
\copyrightyear{2018}
\acmYear{2018}
\acmDOI{XXXXXXX.XXXXXXX}

\acmConference[Conference acronym 'XX]{Make sure to enter the correct
  conference title from your rights confirmation emai}{June 03--05,
  2018}{Woodstock, NY}
\acmPrice{15.00}
\acmISBN{978-1-4503-XXXX-X/18/06}

\begin{document}

\title{SpiroMask: Measuring Lung Function Using Consumer-Grade Masks}

\author{Rishiraj Adhikary}
\affiliation{%
  \institution{IIT Gandhinagar}
  \country{India}
}
\author{Dhruvi Lodhavia}
\affiliation{%
  \institution{IIT Gandhinagar}
  \country{India}
}
\author{Chris Francis}
\affiliation{%
  \institution{IIT Gandhinagar}
  \country{India}
}
\author{Rohit Patil}
\affiliation{%
  \institution{IIT Gandhinagar}
  \country{India}
}
\author{Tanmay Srivastava}
\authornote{The work was done while these authors were a research fellow at IIT Gandhinagar.}
\affiliation{%
  \institution{Stony Brook University}
  \country{USA}
}
\author{Prerna Khanna}
\authornotemark[1]
\affiliation{%
  \institution{Stony Brook University}
  \country{USA}
}
\author{Nipun Batra}
\affiliation{%
  \institution{IIT Gandhinagar}
  \country{India}
}
\author{Joe Breda}
\affiliation{%
  \institution{University of Washington}
  \country{USA}
}
\author{Jacob Peplinski}
\affiliation{%
  \institution{University of Washington}
  \country{USA}
}
\author{Shwetak Patel}
\affiliation{%
  \institution{University of Washington}
  \country{USA}
}

\renewcommand{\shortauthors}{Adhikary et al.}

\begin{abstract}
According to the World Health Organisation (WHO), 235 million people suffer from respiratory illnesses which causes four million deaths annually. Regular lung health monitoring can lead to prognoses about deteriorating lung health conditions. This paper presents our system \emph{SpiroMask} that retrofits a microphone in consumer-grade masks (N95 and cloth masks) for continuous lung health monitoring. We evaluate our approach on 48 participants (including 14 with lung health issues) and find that we can estimate parameters such as lung volume and respiration rate within the approved error range by the American Thoracic Society (ATS). Further, we show that our approach is robust to sensor placement inside the mask.

\end{abstract}

%\zp{suffer from deaths seems inaccurate}




\received{30 October 2021}
\received[revised]{8 May 2022}
\received[accepted]{18 October 2022}

\maketitle

\section{Introduction}
\begin{figure}[ht]
    \centering
    \includegraphics[scale=0.25]{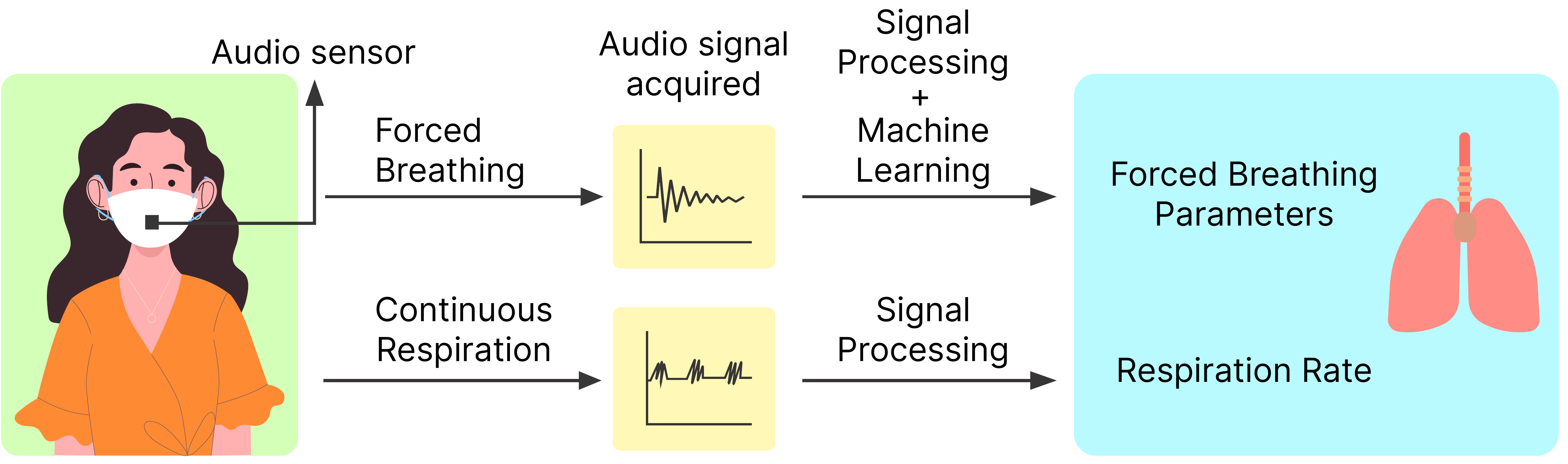}
    \caption{SpiroMask: Our system to estimate lung health parameters by processing the audio data collected from a microphone fitted inside consumer-grade masks. 
    }
     \label{fig:hero-image}
\end{figure}

\noindent According to the World Health Organisation (WHO), 235 million people suffer from respiratory illnesses.  Chronic Obstructive Pulmonary Disease (COPD) is associated with an estimated 3 million deaths per year, and asthma had an estimated 200 thousand deaths per year~\cite{schluger2014lung}. Common infectious respiratory diseases, such as influenza, cause 600 thousand deaths worldwide. 90\% of deaths due to pulmonary disease occur in low-income and middle-income countries\footnote{\url{https://www.who.int/health-topics/chronic-respiratory-diseases}}. COPD development generally starts early in life due to a complex interplay of disadvantageous factors, many of which occur in low and middle income countries~\cite{brakema2019copd}. Studies have shown that lung diagnosis is typically missed or delayed until poor lung health conditions advance. Early diagnosis of lung ailments can positively influence disease course, slowing progression, relieving symptoms and reducing the incidence of exacerbation~\cite{soriano2009screening, van2003early}. But unavailability of lung function equipment hinders proper diagnosis~\cite{tan2008copd,indiaspend}.\\

\noindent Regular lung health monitoring can lead to prognoses about deteriorating lung health conditions~\cite{aliverti2017wearable, tomasic2018continuous}. Typically, two kinds of breathing data are used to derive lung health bio-markers:

\begin{itemize}
    \item  \textbf{Tidal breathing or normal breathing:} Tidal breathing refers to inhalation and exhalation during restful breathing. It is in involuntary action. Lung disease can change the normal characteristics of tidal breathing~\cite{palmer2004tidal}.
    
    \item \textbf{Forced breathing:} It is performed by taking a deep inhalation resulting in full expansion of the chest followed by a forceful exhalation. A person will require to voluntarily perform forceful breathing maneuver.
\end{itemize}

\noindent Forced Expiratory Flow (FEF) measurement deduced from forced breathing is the most widely used method to assess the severity of asthma. Monitoring the forcefully exhaled airflow can help diagnose the onset of asthma, COPD and other conditions that affect breathing~\cite{van1996tidal, larson2011accurate, song2020spirosonic}. But, FEF measurements require a controlled environment. Moreover, a majority of young and old patients with airway obstruction are not able to perform adequate forced breathing maneuvers~\cite{van1996tidal}. Prior research has shown that tidal breathing patterns can also be used to detect and quantify airway obstruction~\cite{walker1990clinical}. 
Respiration rate can be deduced from tidal breathing. It is defined as the number of breaths taken by a person in a minute. Respiratory rate deduced from tidal breathing is an important marker of cardiac arrest, dyspnea~\cite{fieselmann1993respiratory, hodgetts2002identification, wang2021smartphone,kristiansen2021machine}, accessing sleep quality and monitoring stress~\cite{dai2021respwatch}. Lung inflammation caused by COPD deterioration or lung infection leads to a higher respiration rate~\cite{rahman2020breatheasy,larson2011accurate}. \\

\noindent In clinical settings and hospitals, exhaled airflow is measured using a spirometry test (Figure~\ref{fig:all-in-one} a).  During a spirometry test, a patient performs forceful breathing through a flow-monitoring device (a tube or mouthpiece), which measures instantaneous flow and cumulative exhaled volume. However, spirometry tests performed at hospitals are not transient, and the recent global pandemic has led to the suspension of certain non-urgent healthcare services such as routine diagnostic testing~\cite{kouri2020chest, johnson2020impact}. Although home spirometry tests are available~\cite{otulana1990use}, even the cheapest hand-held digital spirometer cost about USD 300.  Previous work has accurately estimated forced breathing parameters using a smartphone~\cite{larson2012spirosmart}. In smartphone spirometry, a person must do a maneuver of forceful breathing towards the cellphone. But, the approach to microphone-based smartphone spirometry is susceptible to users' way of holding the phone, users' lip posture, environmental variability, and phone's make and model~\cite{mayankgoel-thesis}.\\  

\noindent Previous studies~\cite{rahman2020breatheasy} have also used accelerometers in smartphones for continuous respiration monitoring. However, the approach requires a controlled environment where the participant must put the smartphone in a particular location on the chest. Recently, studies~\cite{zhou2020accurate} have shown that wearable spirometry can be conducted using a pressure sensor inside a specialised mask meant for athletes. Progress on mask spirometry is limited because, i) the evaluations were done on healthy adults alone, ii) there was no continuous monitoring of respiration rate, iii) athlete training masks  are relatively costly (40-50 USD) compared to consumer-grade masks (5-10 USD For N95 mask), iv) athlete training masks are meant to restrict the oxygen received by a person to create a high-altitude environment\footnote{\url{https://www.healthline.com/health/training-mask-benefits\#benefits}}, as such they cannot be a replacement for generally used cloth or N95 mask. In Section~\ref{feedback}, we qualitatively show that the general population would not prefer a specialised mask as a daily wearable.\\

\noindent This paper shows that \emph{consumer-grade masks (N95 and cloth masks) can be used for Spirometry and continuous respiration rate monitoring} by processing the signal from the microphone retrofitted inside the mask (Figure~\ref{fig:hero-image}). The main intuition behind using audio is to leverage the relationship between variation in air flow rate and the intensity change in nasal sound~\cite{yap2002acoustic,morrison1969nasal,kumar2021estimating,nam2015estimation}. Our work addresses the limitations of prior research~\cite{larson2011accurate, adhikary2020naqaab, zhou2020accurate}. In particular, our approach to SpiroMask i) provides a relatively controlled environment for Spirometry as well as tidal breathing monitoring, ii) can classify between tidal breathing, noise and speech, iii) is accurate for participants with lung ailments as well as for healthy individuals.
\\

\noindent Our approach uses the audio signal to derive vital lung parameters and continually monitor respiration rate. We used machine learning with sequential forward selection techniques to learn a set of audio features that accurately estimate lung health. We have two separate pipelines for estimating parameters from forced breathing and tidal breathing. To estimate respiration rate (from tidal breathing), we used a neural network that distinguishes tidal breathing from speech and noise. We estimated the average peak to peak time from the tidal breathing signal to derive the respiration rate. The parameters estimated from forced breathing are described in Section~\ref{sec:related-work}.\\

\noindent Our study was approved by the Institutional Review Board (IRB). We recruited 48 participants, including 15 female participants. A total of 14 participants had restrictive and obstructive lung ailment. The number of healthy and unhealthy participants in our trial is comparable to study population of past studies~\cite{goel2016spirocall, larson2011accurate, whitlock2020spiro, zhou2020accurate, yue2018extracting, zubaydi2017mobspiro, rahman2014bodybeat}. For each participant the study lasted between 45 minutes to about an hour. The mean percentage errors on forceful breathing parameters for cloth masks were between 5.2\% to 6.7\%. For the N95 mask, they were between 5.8\% to 6.3\%. We achieved an accuracy of 94.7\% on classifying tidal breathing from noise and speech. The Mean Absolute Error (MAE) on the estimation of respiration rate was 0.68 for the cloth mask and 0.49 for the N95 mask. Our results on forceful breathing are within the acceptable error range as endorsed by the American Thoracic Society or ATS.\\

\noindent We have performed sensitivity analysis on the position of the sensor inside the mask. Our approach is robust to sensor placement for forced breathing. However, certain positions (directly below the nose) inside the mask are ideal for estimating respiration rate from tidal breathing monitoring.  \\

\noindent To summarise, the \textbf{main contributions} of this paper are:\vspace{-5pt}
\begin{itemize}
    \item \emph{SpiroMask:} a novel mask-based system for estimating forced and tidal breathing to assess lung health parameters using a microphone
    \begin{itemize}
        \item that works on consumer-grade masks
        \item is accurate within the ATS guidelines
        \item works well for both healthy and unhealthy subjects
        \item is robust to the microphone placement
    \end{itemize}
    \item \emph{Public dataset:} We publicly release our dataset at Github\footnote{\url{https://github.com/rishi-a/SpiroMask}} We believe ours is the first such large publicly available dataset that measures both the tidal and forced breathing parameters and ground truth for 48 participants (including 14 with lung ailments). We believe that our dataset can help advance research in the community.
    \item \emph{Reproducibility:} We believe that our work is fully reproducible. We use the same repository as above for the code. All the generated tables and graphs have corresponding scripts to reproduce all the results. We believe our efforts towards reproducibility will lower the effort towards replication and building on top of our work.
    \end{itemize}

\section{Background and Related Work}
\label{sec:related-work}
This section underlines the basics of spirometry and lung function indices, followed by an overview of prior work on audio-based and pressure-based sensing of lung indices.
\subsection{Spirometry}
Spirometry is a widely used pulmonary function test. It measures how fast and how much air the patient can breathe out and is the most widely employed objective measure of lung function~\cite{townsend2011spirometry}. Spirometry tests are usually performed in clinics or hospitals.  Currently, the most commonly used devices for respiratory evaluation are hand-held spirometers. A spirometry test consists of the following sequence of events~\cite{graham2019standardization}.

\begin{itemize}
    \item A soft clip is placed on the patient's nose to allow normal breathing through her mouth.
    \item The patient wraps her lips tightly around the spirometer mouthpiece shown in Figure~\ref{fig:all-in-one} (a), ensuring that all the exhaled air goes through into the spirometer for accurate measurement.
    \item The patient then takes the deepest possible breath, filling her lungs to the maximum. 
    \item The patient exhales hard and fast and continues exhaling into the spirometer until no more air comes out.  
\end{itemize}

\noindent Spirometry results are effort dependent. Patients need to be coached by a trained professional to perform a forceful exhalation for a successful spirometry test. A digital spirometry test produces the flow versus volume plot of the lung\footnote{\url{https://pftforum.com/review/tutorials/spirometry-tutorials/assessing-flow-volume-loops/}} from which vital lung parameters are extracted.
A forceful breathing maneuver requires a controlled environment with proper guidance from doctors.\\

\noindent A standard spirometer measures the volume and flow of air that can be inhaled and exhaled. Spirometry generates pneumotachographs, which plot the volume and flow of air coming in and out of the lungs. The most common parameters measured by spirometry are forced vital capacity (FVC), forced expiratory volume at one second (FEV1), and peak expiratory flow (PEF). FVC is the total air volume exhaled denoted by C in Table~\ref{tab:parameters}, FEV1 is the exhaled air volume in the first second of exhalation, denoted by B. PEF is the maximum airflow velocity in exhalation, denoted by A.\\ 

\begin{table}[ht]
\begin{tabular}{@{}lll@{}}
\toprule
 & \multicolumn{2}{c}{
 
\includegraphics[scale=0.3]{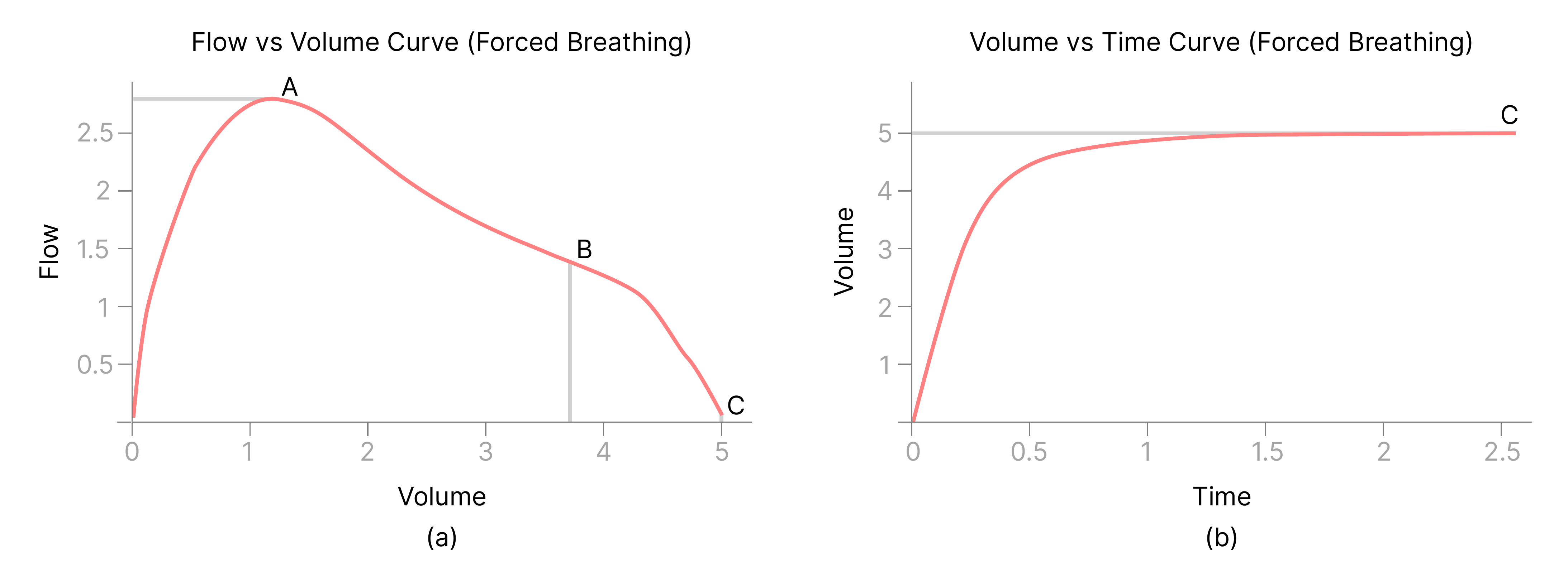}

} \\ \midrule
\textbf{ID} & \textbf{Abbreviation} & \textbf{Definition} \\ \midrule
A & PEF & \begin{tabular}[c]{@{}l@{}}Peak expiratory flow (PEF) is the maximum airflow \\  velocity in exhalation. (L/s)\end{tabular} \\ \midrule
B & FEV1 & \begin{tabular}[c]{@{}l@{}}Forced expiratory volume in 1 second (FEV1) is the \\ exhaled air volume in the first second of exhalation (L)\end{tabular} \\ \midrule
C & FVC & Forced vital capacity (FVC) is the total air volume exhaled (L) \\ \midrule
D & FEV1/FVC & \begin{tabular}[c]{@{}l@{}}FEV1/FVC is the ratio of FEV1 and FVC.\\ This ratio should be > 80\% among healthy~\cite{carey2010current}\end{tabular} \\ \midrule
E & RR & Respiratory Rate (samples/min) \\
\bottomrule
\end{tabular}
\caption{Forced breathing and tidal breathing lung function parameters.}
\label{tab:parameters}
\end{table}

\subsection{Smartphone Spirometry}
\noindent Researchers have explored the potential to turn existing mobile devices and smartphones into portable electronic spirometers using their inbuilt microphones supplemented with machine learning techniques~\cite{sim2019mobile,rahman2019towards,larson2012spirosmart,viswanath2018spiroconfidence,thap2016high,zubaydi2017mobspiro, song2020spirosonic}. These studies use the features from the Hilbert transformation~\cite{johansson1999hilbert}, linear predictive coding~\cite{o1988linear} and the spectrogram of an audio signal to train a linear regression model for each of the vital lung parameters, i.e.  FVC, FEV1 and PEF. Compared to the clinical spirometer, the reported mean error is 5.1\%, 5.2\% and 6.3\% for FEV1, FVC and PEF, respectively. In our work, we show that Hilbert transformation with a finite impulse response filter gives a comparable or better estimate of lung function parameters, and we can avoid non-trivial modelling like linear predictive coding. Also, given that the audio sensor is fixed in the mask, we are not required to compensate for pressure losses sustained over the variable distance from the mouth to the microphone, and reverberation/reflections caused in and around the subject's body.\\
%\nb{the following para is too detailed. compress it by 50 percent or so. don't give details, broadly mention what they have done. no need for accuracy numbers also.}
\noindent Researchers have also proposed a variable frequency complex demodulation method (VFCDM) technique to extract the FEV1/FVC ratio~\cite{thap2016high} from audio. A built-in smartphone microphone was also used in~\cite{zubaydi2017mobspiro} to estimate FEV1 and FVC. However, they do not estimate PEF or tidal breathing parameters like respiration rate. Commodity smartphone has also been used to measures the humans’ chest wall motion via acoustic sensing. Lung function indices are deduced from the measured  motion~\cite{song2020spirosonic}. Existing literature on extracting PFT parameters requires a person to perform the forced breathing maneuver in a controlled noiseless environment. Researchers have investigated the scope of extracting PFT events from speech.  Previous literature~\cite{rahman2019towards} studied subjects having varying pulmonary conditions to collect mobile sensor data, including audio, to extract pulmonary biomarkers such as breathing, coughs, spirometry, and breathlessness.
\begin{table}[ht]
\begin{tabular}{@{}lllrrlll@{}}
\toprule
\multicolumn{1}{c}{} & \multicolumn{2}{c}{Type of mask}                                                                         & \multicolumn{2}{c}{Participants}                            & \multicolumn{3}{c}{Lung health parameters}                                                                                                            \\ \midrule
\multicolumn{1}{c}{} & \multicolumn{1}{c}{\begin{tabular}[c]{@{}c@{}}Cloth/\\ Surgical\end{tabular}} & \multicolumn{1}{c}{N95} & \multicolumn{1}{c}{Healthy} & \multicolumn{1}{c}{Unhealthy} & \multicolumn{1}{c}{\begin{tabular}[c]{@{}c@{}}Continuous \\ monitoring\\ of respiration rate\end{tabular}} & \multicolumn{1}{c}{\begin{tabular}[c]{@{}c@{}}Forced \\ breathing\\ parameters\end{tabular}} & \multicolumn{1}{c}{\begin{tabular}[c]{@{}c@{}}Discussion on \\ sensor \\ positioning\\ inside the mask\end{tabular}} \\ \midrule
Zhou et al.~\cite{zhou2020accurate}         & \xmark                                                                             & \xmark                      & 20                          & 0                             & \xmark                                                                                                         & \begin{tabular}[c]{@{}l@{}}PEF, FEV1, \\ FVC\end{tabular}                                    & \xmark                                                                                                                \\ \hline
Adhikary et al.~\cite{adhikary2020naqaab}     & \cmark                                                                            & \xmark                      & 8                           & 0                             & \xmark                                                                                                         & FVC                                                                                          & \xmark                                                                                                                \\ \hline
Curtiss,\\ Rothrock et al.~\cite{curtiss2021facebit}     & \cmark                                                                            & \cmark                      & 9                           & 0                             & \cmark                                                                                                         & \xmark                                                                                          & \xmark                                                                                                                \\ \hline
SpiroMask \textbf{(Our work)}             & \cmark                                                                            & \cmark                     & \textbf{34}                          & \textbf{14}                            & \cmark                                                                                                        & \begin{tabular}[c]{@{}l@{}}\textbf{PEF}, \textbf{FEV1}, \\ \textbf{FVC}\end{tabular}                                    & \cmark                                                                                                               \\ \bottomrule
\end{tabular}
\caption{Previous studies on masks based lung function measurement: i) do not work on consumer grade masks; ii) are not robust to sensor placement; iii) can estimate respiration rate only in controlled settings; and iv) did not consider participants with lung ailments.}
\label{tab:novel-table}
\end{table}
Smartphone spirometry has the following limitations:
\begin{itemize}
    \item It requires accounting for variability in the distance between microphone and mouth. In Section~\ref{sec:comparison-with-smartphone}, we have described how this variability affects the flow-volume curve.
    \item Not all smartphones are created equal. The flow detected by the microphone in smartphones~\cite{larson2011accurate} relies on the mechanical transduction of sound, which is affected by the position of the microphone and the physical casing surrounding it~\cite{mariakakis2019challenges}.
    \item It is not suited for sensing tidal breathing as a person will not keep the smartphone near their nose for a long duration.
    \item It is relatively less accurate for participants with lung ailments~\cite{larson2011accurate, song2020spirosonic}.
    \item It requires a user to actively interact with the smartphone.
\end{itemize}

\subsection{Wearable  Spirometry}
Recent efforts have been made to integrate face masks with audio and pressure sensors to extract vital lung parameters~\cite{zhou2020accurate, adhikary2020naqaab}. Researchers have used a MEMS-based barometric pressure sensor inside an athlete training mask\footnote{\url{https://www.trainingmask.com/training-masks/training-mask-3-0/} Last accessed: 15th August 2021}. Such masks are specially designed for athletes, restricting breathing, making an athlete feel like they are at a high altitude.
Accurate wearable spirometry has been performed in athlete masks with error margins of 2.9\% and 3.3\% for FVC and FEV1 respectively~\cite{zhou2020accurate}. Researchers have also experimented by integrating an audio sensor inside a surgical mask to estimate FVC from the energy of the audio signal but they have not quantified the error~\cite{adhikary2020naqaab}. However, progress in mask spirometry is limited in the following ways: 
\begin{itemize}
    \item Continuous monitoring of tidal parameters was absent.
    \item The proposed approaches are not suitable for consumer-grade N95 and cloth masks.
    \item Previous work did not consider participants with lung ailments in their user-study and thus the efficacy is not known.
    \item There was no discussion on the robustness of sensor positioning inside the mask. 
\end{itemize}
Table~\ref{tab:novel-table} compares our work with previous literature, and shows the novelty of our work for: i) estimating lung parameters for consumer-grade cloth and N95 masks; ii) single system for measuring both forced and tidal breathing parameters; iii) a larger study with unhealthy and healthy subjects; and a discussion on robustness of sensor placement.

\subsection{Respiration Rate Detection}
Most of the prior work estimates respiration rate using Inertial Measurement Unit (IMU). Röddiger et al.~\cite{roddiger2019towards} estimated respiration rate from accelerometer and gyroscope placed inside in-ear headphones. Hernandez et al.~\cite{hernandez2014bioglass} installed IMU in Google glass to estimate pulse rate and respiration rate from ballistocardiogram. Sun et al.~\cite{sun2017sleepmonitor} used accelerometer data from smartwatches to deduce respiration rate when a person is asleep. Liaqat et al.~\cite{liaqat2019wearbreathing} used machine learning algorithms to filter out IMU data that can be used to estimate respiration rate in the wild using smartwatch. Aly et al.~\cite{aly2016zephyr} used the approach of placing a smartphone over the chest to estimate respiration rate using accelerometer and gyroscope. Islam et al.~\cite{islam2021breathtrack} used sensor fusion data from both smartwatch and smartphone. All of these works except the work done by Sun et al.~\cite{sun2017sleepmonitor} focused on estimating respiration rate with different posture of a user. For a fair comparison, we compared the error of only `Sitting' posture with our work and found that our results are comparable or better to previous works. We have summarized many previous works in Table~\ref{tab:tidalresults} below.\\

\noindent In the past, microphones have also been used to estimate respiration rate. Kumar et al.~\cite{kumar2021estimating} used short audio segments from in-ear headphones to estimate respiration rate when a person is subject to strenuous exercises. Nam et al.~\cite{nam2015estimation} used a microphone placed below the nose to record nasal sound. We cannot directly compare the error between SpiroMask and previous work that used a microphone due to different metrics used but we believe that SprioMask performance is acceptable given that our evaluation was on a relatively larger set of participants.

\begin{table}[ht]
\begin{tabular}{@{}llll@{}}
\toprule
% \multicolumn{1}{c}{\textbf{Grouping Variable}}   & \multicolumn{2}{c}{\textbf{p-value}} \\ \midrule
\multicolumn{1}{c}{} & \multicolumn{1}{c}{\textbf{Metric | Error}} & \multicolumn{1}{c}{\textbf{Hardware}} & \multicolumn{1}{c}{\textbf{Participants}}\\ \midrule
Röddiger et al. \cite{roddiger2019towards} & MAE = 3.10  & Bluetooth earphone (via Accelerometer) & 12 Participants \\ 
& MAE = 2.74 & (via Gyroscope) & (2 Females)\\
\hline
Hernandez et al. \cite{hernandez2014bioglass} & MAE = 1.15 & Head mounted device (via Gyroscope) & 12 Participants\\ 
& MAE = 1.1 & (via Accelerometer) & (6 Females)\\
& MAE = 1.1 & (via Camera) & \\
\hline
Kumar et al. \cite{kumar2021estimating} & MSE = 0.2 & Headphones (via Microphone) & 21 Participants\\ 
\hline
Sun et al. \cite{sun2016quadratic} & MAE = 0.75 &  Smartwatch (via Accelerometer only) & 16 Participants\\ 
\hline
Hernandez et al. \cite{hernandez2015biowatch} & MAE = 0.22 & Smartwatch (via Gyroscope) & 12 Participants \\
& MAE = 0.41 & (via Accelerometer) &\\
& MAE = 0.22 & (both) & \\
\hline
Liaqat et al. \cite{liaqat2019wearbreathing} & MAE = 1.9 & Smartwatch (via Accelerometer ) & 14 Participants\\ 
\hline
Islam et al. \cite{islam2021breathtrack} & MAPE = 8.26\% & Smartphone ((via Accelerometer) & 131 Participants\\ 
\hline
Aly et al. \cite{aly2016zephyr} & & Smartphone (via Accelerometer) & 7 participants\\ 
&  MdAE* = 0.04 & (with metronome) & (4 females)  \\ 
& MdAE = 0.53 & (without metronome) & \\
\hline
\textbf{SpiroMask (our work)} & \textbf{MAE = 0.6} & \textbf{Consumer grade mask (via microphone)} & \textbf{37 Participants}
\\
& &  & \textbf{(13 Female)} \\
 & &  & \makecell{(14 with \\lung ailments)}\\
\bottomrule
\end{tabular}
\caption{SpiroMask's performance on  respiration rate detection is better or comparable to previous work. (MAE is Mean Absolute Error and MdAE is Median Absolute Error)}
\label{tab:tidalresults}
\end{table}

\subsection{Audio and Pressure sensor}
\begin{figure}[ht]
    \centering
    \includegraphics[scale=.75]{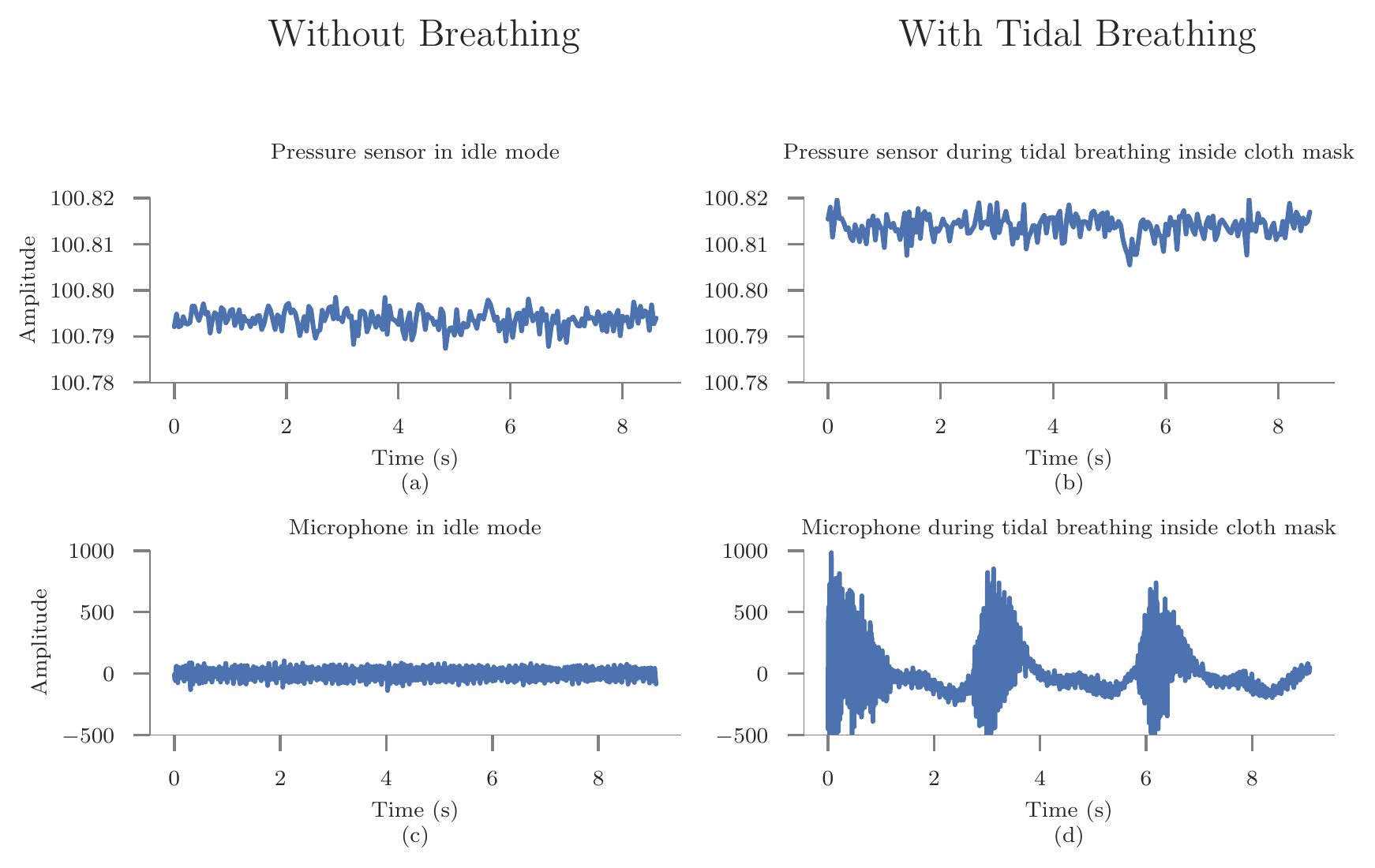}
    \caption{(a) and (b) shows that tidal breathing is indistinguishable from a pressure sensor placed inside the cloth mask whereas (c) and (d) show that microphone is more suitable inside cloth mask to monitor tidal breathing. Thus, previous work~\cite{zhou2020accurate} in mask based spirometry does not work on consumer-grade masks.}
      \label{fig:pressureANDmicrophone}
\end{figure}

Prior literature has successfully used both an in-situ microphone and a MEMS barometer inside a very tightly sealed athlete mask to sense breathing and perform portable spirometry~\cite{zhou2020accurate, adhikary2020naqaab}. However, our experiments show that a pressure sensor cannot be used to sense breathing inside a consumer-grade cloth mask. Figure~\ref{fig:pressureANDmicrophone} (a) and (b) shows that the pressure sensor outputs a similar signal (Pearson correlation coefficient, $r=88\%$) when the cloth mask is worn and not worn. The amplitude in Figure~\ref{fig:pressureANDmicrophone} (b) is higher than in  Figure~\ref{fig:pressureANDmicrophone} (a) because the pressure inside a mask is higher than the atmospheric pressure. However, there is little to no change in the signal to distinguish between inhalation and exhalation. Previous work~\cite{zhou2020accurate} leveraged the change in pressure to detect forceful breathing inside a special kind of mask. However, the differential pressure is insignificant in N95 and cloth masks making the breathing signal unrecognisable. It implies that pressure sensors are ineffective in distinguishing tidal breathing in cloth masks unless the sensor is sealed from atmospheric pressure.  Figure~\ref{fig:pressureANDmicrophone} (c) and (d) shows that a microphone does a better job (Pearson correlation coefficient, $r=1\%$) in sensing tidal breathing inside a cloth mask. These results suggest that commodity-grade pressure sensors placed on standard masks are likely to result in poorer lung health estimation than audio sensing. Thus, we do not baseline against this previous work~\cite{zhou2020accurate} in mask based spirometry as the approach does not work on consumer-grade masks.
\\

\section{Data Collection Procedure}
\label{sec:study-design}
We now describe our data collection procedure.
\subsection{IRB approval and screening criteria}
Our user study on SpiroMask was approved by the Institutional Review Board (IRB). All participants between 18 to 70 years of age could become a participant in the study. People who are severely ill and have been advised bed rest by doctors were not allowed to participate in the study. We recruited 52 participants for the study, out of which 16 participants reported having lung ailments. All participants were remunerated as per institute guidelines.

\subsection{Entry survey}
Every participant filled out an entry survey before the user study. We asked them to declare their age, height, weight, any recent illness, history of contracting COVID19 and if they had a meal before coming for the user study. We also asked them if they had a clinically validated history of lung ailments and if they had done a spirometry test in the past.

\subsection{Sensing hardware}
\label{sec:sensing-hardware}
\subsubsection{Microphone}
\noindent We recorded the audio of forceful expiration and tidal breathing using an Arduino nano sense microcontroller\footnote{\url{https://store.arduino.cc/usa/nano-33-ble-sense}}
due to the availability of embedded sensors like the MEMS microphone. Its compact size makes it possible to affix the microcontroller on a face mask. The microphone has a sampling rate of 16 kHz. We placed our microcontroller inside a 3D printed enclosure (Figure~\ref{fig:all-in-one} (f) to protect it from static discharge and wear and tear. The 3D enclosure is affixed to the mask using velcro and double-sided tape (Figure~\ref{fig:all-in-one} (e). We advised the participants to affix the microcontroller on their mask to ensure their comfort and safety while wearing it. We confirmed that the mask was appropriately worn before starting the SpiroMask experiment. Our future system would be smaller and hand-sewable on the fabric.

\subsubsection{Spirometer}
Our setup also consisted of the Helios 401\footnote{\url{https://labsakha.com/wp-content/uploads/2019/03/RMS-Helios-401-Features-Specifications.pdf}} medical-grade (ISO 14971:2019~\cite{teferra2017iso}) hand-held spirometer. We used the
spirometer to collect the ground truth lung capacity of every participant. 

\subsubsection{Smartphone}
We used a Samsung Galaxy M20 smartphone to collect accelerometer data for the expansion and contraction of the chest. The accelerometer had a sampling rate of 100 Hz. 

\subsection{Data communication and storage}
The sensor inside the mask was connected to a desktop computer via USB cable. To ensure no data loss and interruption (due to draining battery) during the user study, we refrained from using any wireless mode of audio data transfer. Transmitting high quality audio with limited power is a challenge because our sensing hardware samples audio at 16 kHz which amounts to 768 Kbps of data~\cite{ei-forum-1}. But the practical throughput of BLE is around 125 Kbps or less~\cite{nordic-blog-1}. For every participant, the investigator checked if the microphone and the smartphone are responding to remote commands. We used the MATLAB mobile application\footnote{\url{https://www.mathworks.com/products/matlab-mobile.html}} to collect accelerometer data from the smartphone.  An investigator could issue remote commands from their laptop to retrieve audio and IMU data from the microphone and smartphone. The sensitive audio data was stored securely in the cloud. 

\subsection{COVID19 norms} We followed all local COVID19 guidelines during the entire process. The tests were done in a well-ventilated room with a single participant at a time. The investigator was double vaccinated and wore an N95 mask during the entire duration of the experiment. On the participant's arrival, we asked the participant to sanitise their hands. In the spirometry test, each participant used a new mouthpiece. We did not reuse any N95 or cloth masks. The investigator gave a fresh piece of both the masks.

\begin{table}[ht]
\begin{tabular}{@{}lll@{}}
\toprule
                                        & Forced Breathing                                                                      & Tidal Breathing                                                                      \\ \midrule
Total Participants (n)                  & 48                                                                                    & 37                                                                                   \\
Participants with lung ailments (n, \%) & 14 (29.1\%)                                                                           & 14 (37.8\%)                                                                          \\
Females (n, \%)                         & 15 (31.3\%)                                                                           & 13 (35.1\%)                                                                          \\
Age (yrs) (mean, range)                 & 28.05 (21-68)                                                                         & 26.19 (20-32)                                                                        \\
Height (cm) (mean, range)               & 167.39 (142.4-182.9)                                                                  & 167.0 (152.4-182.88)                                                                 \\ \midrule
Participants with lung ailments         & \multicolumn{2}{l}{\begin{tabular}[c]{@{}l@{}}Asthma: 3\\ Other Restrictive lung disorder: 4\\ Obstructive lung ailment: 2\\ Post COVID lung infection: 4\\ Wheeze: 1\end{tabular}} \\ \bottomrule
\end{tabular}
\caption{Demographic information for the participants}
\label{tab:participant-info}
\end{table}

\subsection{Data collection for forced breathing}
\begin{figure}[ht]
    \centering
    \includegraphics[scale=0.34]{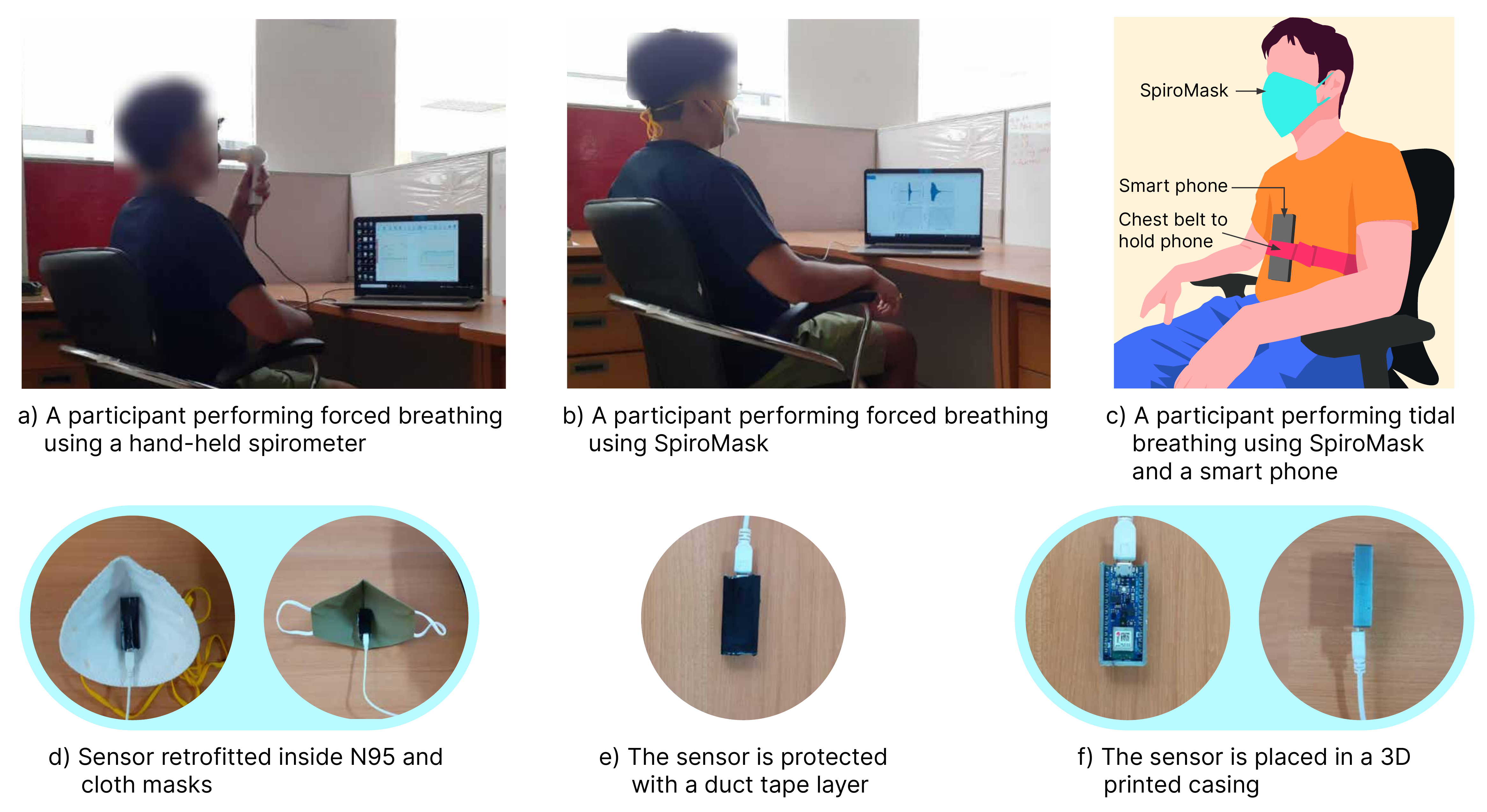}
    \caption{a) and b) shows the spirometry test followed by a SpiroMask test. c) shows tidal breathing user-study where a smartphone is placed on the sternum with the help of a belt. d) shows how the retrofit device was fixed into both the type of mask. e) shows the protective velcro tape layer above the sensor to protect it from mucus, and f) shows the microcontroller placed inside the 3D printed casing.}
     \label{fig:all-in-one}
\end{figure}

\noindent We started the user study with data collection for forced breathing, as shown in Figure~\ref{fig:all-in-one}. We conducted the study over two phases. Phase 1 was in January - February 2021, and Phase 2 began in July 2021\footnote{Our country was very significantly impacted by COVID19 between March and June 2021 and thus we suspended data collection during that time}. Table~\ref{tab:participant-info} summarises the participant demography.\\

\begin{figure}[ht]
    \centering
    \includegraphics[scale=0.30]{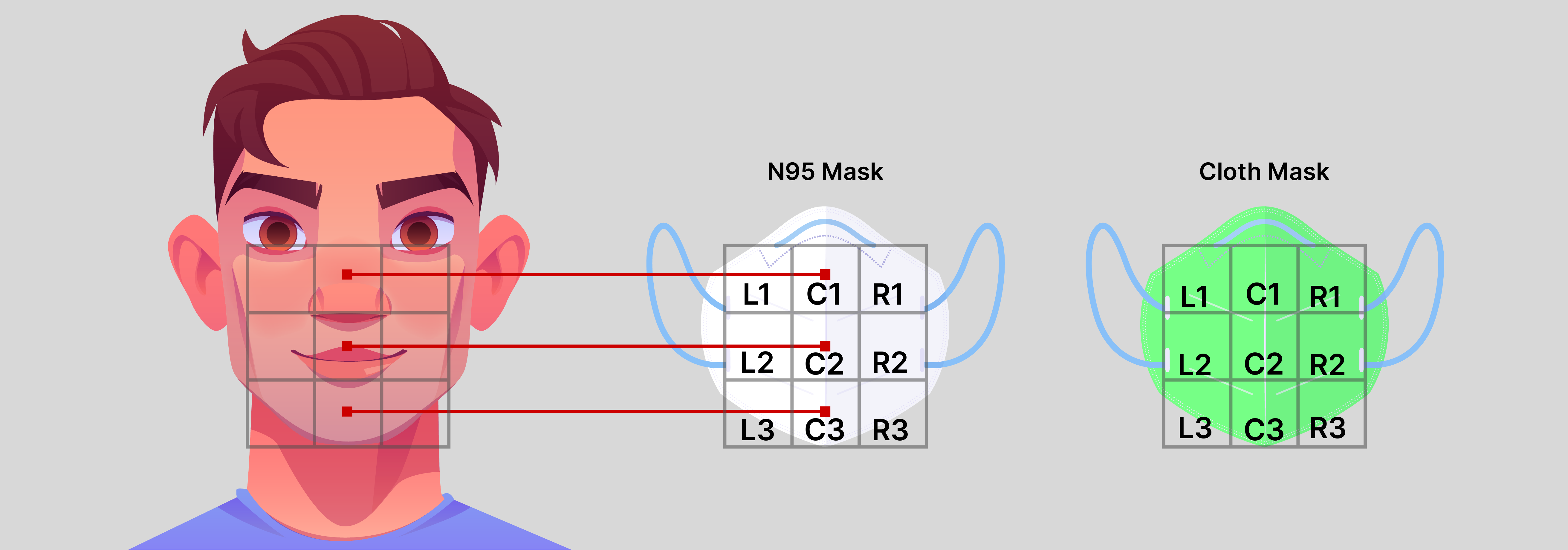}
    \caption{We collected data from different position of the sensor inside the mask. The positions are shown relative to the chin, lip and nose.}
      \label{fig:face-position}
\end{figure}

\noindent The investigators explained the entire user study to the participants.  We demonstrated the spirometry test to the participants. For most of the participants, we could obtain the spirometry ground truth in the first attempt itself. The proprietary spirometer software flagged incorrect attempts for some participants. For every wrong maneuver, we repeated the spirometry test at least once, after which we obtained the ground truth.  SpiroMask test followed the spirometry test. Forceful exhalation using the mouth is made possible in a hand-held spirometer due to the availability of a dedicated mouthpiece. Based on some pilots experiments, we realised that it is easier to do a forceful exhalation using the nose with mouth closed when a face mask is worn. We asked the participants to attach the sensor themselves on the mask. We issued the remote command to start data collection and instructed the participant to begin deep inhalation and forceful exhalation. Each forceful breath lasted for 6 to 8 seconds. For 12 participants, we collected samples of forceful breathing audio for different positions inside the mask (L1, C1, R1, L3, R3 in Figure~\ref{fig:face-position}). Eight (66.6\%) out of these 12 participants preferred to start with the R3 position. Most of the other participants, who did not participate in the sensor positioning study also preferred the R3 position. They all had an unanimous opinion that R3 was the most comfortable position to place the sensor as it did not lead to any discomfort. To ensure that forced breathing performance does not degrade in the course of multiple maneuver, we asked all the 12 participants to rest for at least 5 minutes after each attempt.  The ATS guideline~\cite{graham2019standardization} permits upto 8  maneuver in adults if the correct flow-volume curve is not achieved. The study duration for these subset of 12 participants was relatively longer compared to other participants. We then repeated the SpiroMask test using the cloth mask.\\ 

\noindent Despite several attempts, four participants (not included in the pool of 48 participants) were unable to perform the spirometry test. One participant had buccinator muscle pain making it impossible for her to hold the mouthpiece of the spirometer. Three other participants could not perform a proper forceful breathing maneuver.  They complained about lung discomfort when asked to attempt a forceful exhalation. The investigator decided not to continue with the spirometry test for these three individuals anticipating worsening medical conditions. It should be noted that while these three people expressed relative comfort using our system SpiroMask, we still had to discard their samples owing to the lack of verified ground truth.\\ 

\noindent We finally had data for 48 participants, out of which 14 had lung ailments. The number of healthy and unhealthy participants in our study is comparable or more than in several similar studies~\cite{goel2016spirocall, larson2011accurate, whitlock2020spiro, zhou2020accurate, yue2018extracting, zubaydi2017mobspiro, rahman2014bodybeat}.

\subsection{Data collection for tidal breathing}
\label{sec:datacollection_tidal}

\noindent Data collection on tidal breathing started in the second phase (July 2021) of our user study. Table~\ref{tab:participant-info} summarises the participant demography. First, we asked the participants to place the smartphone in the sternum with the help of a chest belt, as shown in Figure~\ref{fig:all-in-one}.  Previous work~\cite{chu2019respiration} has shown that putting an accelerometer in the sternum can help us retrieve the respiration rate by leveraging the movement of the rib cage and the movement of the abdomen. Accelerometer data are susceptible to motion artifacts, therefore we had positioned the participants on stationary chairs without any movable parts in it and advised them to remain as still as possible. To filter out noises due to involuntary movements, we used convolution filter. Convolution has the effect of a low pass filter~\cite{downey2016think}. We experimentally found out that using a window size of 70 samples denoises the signal. The accelerometer's sampling rate was 100 Hz. Previous work has also used a stretch sensor combined with a motion sensor in a chest belt to measure breathing parameters, even when the user is ambulatory~\cite{whitlock2020spiro}. Proprietary respiration belts were used in some studies~\cite{yue2018extracting, shih2019breeze} to collect the ground truth on respiration rate. Such ground truth systems are very expensive and unavailable in our country.\\

\noindent Similar to the SpiroMask test for forced breathing, we asked the participant to wear the mask and start tidal breathing. We issued remote commands from a laptop to start data collection from the smartphone and the microphone simultaneously. We also instructed the participants to begin counting their exhalations after issuing the remote command. Each sample of tidal breathing lasted for 20 seconds. We used a development platform~\cite{edgeimpulse} in which the time length of audio recording from the microcontroller was restricted to 20 seconds. The length of the audio recording could be extended had we used the Software Developemnt Toolkit (SDK) of the microprocessor. We did not choose to do so due to engineering challenges. Prior work~\cite{roddiger2019towards} has shown that 20 seconds of IMU data is enough to estimate respiration rate. In Appendix~\ref{appendix-1}, we have demonstrated why 20 seconds of audio is enough to estimate respiration rate. For 18 out of 37 participants\footnote{We should recall that 48 users participated in the Forced breathing test and 37 users participated in the tidal breathing test as summarised in Table~\ref{tab:participant-info}}, we took at least two samples at each position (L1, C1, R1, L3, R3 in Figure~\ref{fig:face-position}) of the mask. These positions covered the entire space of both the masks where the sensor could be placed. We repeated the SpiroMask test for tidal breathing on a cloth mask. Some participants were not comfortable in strapping the smartphone to the chest with a belt. For all such participants, we relied on their self-count and a metronome test.\\

\begin{figure}[ht]
    \centering
    \includegraphics[scale=0.75]{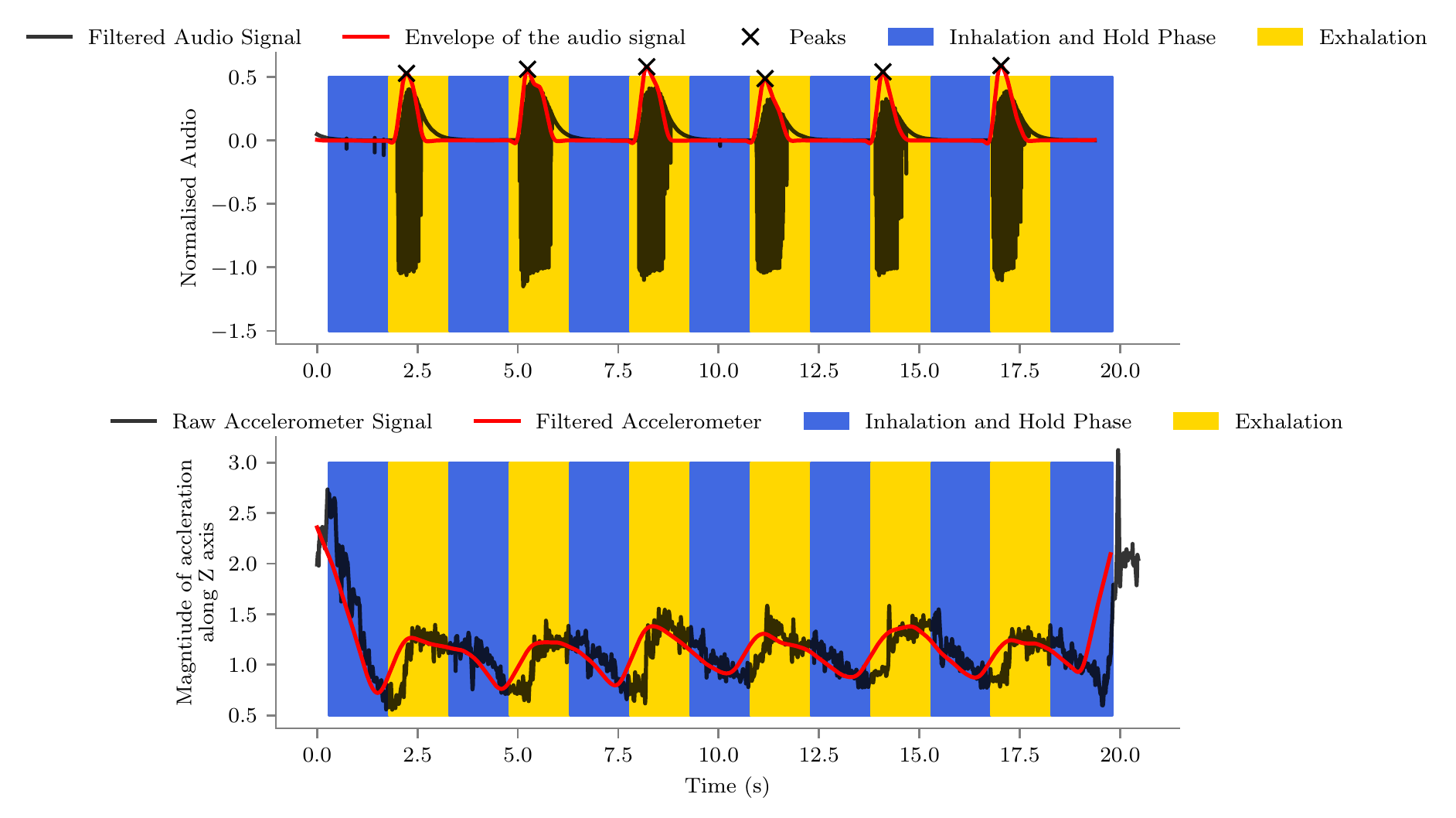}
    \caption{Six breathing cycles were detected from a 20 second window of tidal breathing. The accelerometer data validates that there were six cycles of chest expansion and contraction.}
    \label{fig:audio-and-accelerometer}
\end{figure}

\noindent We also performed a metronome test~\cite{semjen1998getting} for a subset of participants to validate the use of accelerometers to detect actual respiration rate.
Previous studies have explicitly used a metronome as a ground truth to monitor respiration rate~\cite{chu2019respiration}. The metronome test is similar to the procedure of tidal breathing. But, in addition, we asked the participant to inhale and exhale as per the clicks of a 40 Beat Per Minute (BPM) metronome (40 beats corresponds to 20 exhalations and 20 inhalations). The metronome clicks were played on quietly so that they are not captured in the audio of tidal breathing. We choose 40 BPM because the participants were more comfortable at a lower breathing pace. Figure~\ref{fig:audio-and-accelerometer} shows a 20-second window of audio and IMU data for a participant performing breathing as per the metronome. We filtered the IMU data using a moving average filter with a window size of 20 data points. The average peak to peak time is 3.2 second which gives us 6.25 breathing cycles ($\frac{20s}{3.25s}$) which is close to the theoretical number of $\dfrac{20\textrm{beats}}{60\textrm{s}}*20s = 6.66 \textrm{beats}$. Figure~\ref{fig:audio-and-accelerometer} validates the use of accelerometer as ground truth for respiration rate. \\ 

\subsection{Exit survey}
The participants finally filled out an exit survey where they gave their feedback and opinion on the comfort of masks and spirometry tests. Particularly, we asked the following optional questions:
\begin{enumerate}
    \item Can you compare the spirometer and SpiroMask test—which did you prefer and why?
    \item Which mask (out of N95 and cloth masks) feels more comfortable to you and why?
    \item Can you compare the SpiroMask and phone with chest belt for respiration rate monitoring—which did you prefer and why?
\end{enumerate}
The entire user study took approximately an hour per participant.

\section{Approach}
\label{sec:approach}
The goal of SpiroMask is to estimate vital lung parameters and respiration rate using audio of breathing maneuver. Our work is inspired by the advancement of smartphone~\cite{larson2012spirosmart} and wearable spirometry~\cite{zhou2020accurate}. The main intuition behind using audio is to leverage the relationship between variation in air flow rate and the intensity change in tracheal sound~\cite{yap2002acoustic}. We now describe the pipeline for estimating the forced breathing and tidal breathing parameters.
\subsection{Forced Breathing}
\label{sec:approach-forced-breathing}
\begin{figure}[ht]
    \centering
    \includegraphics[scale=.33]{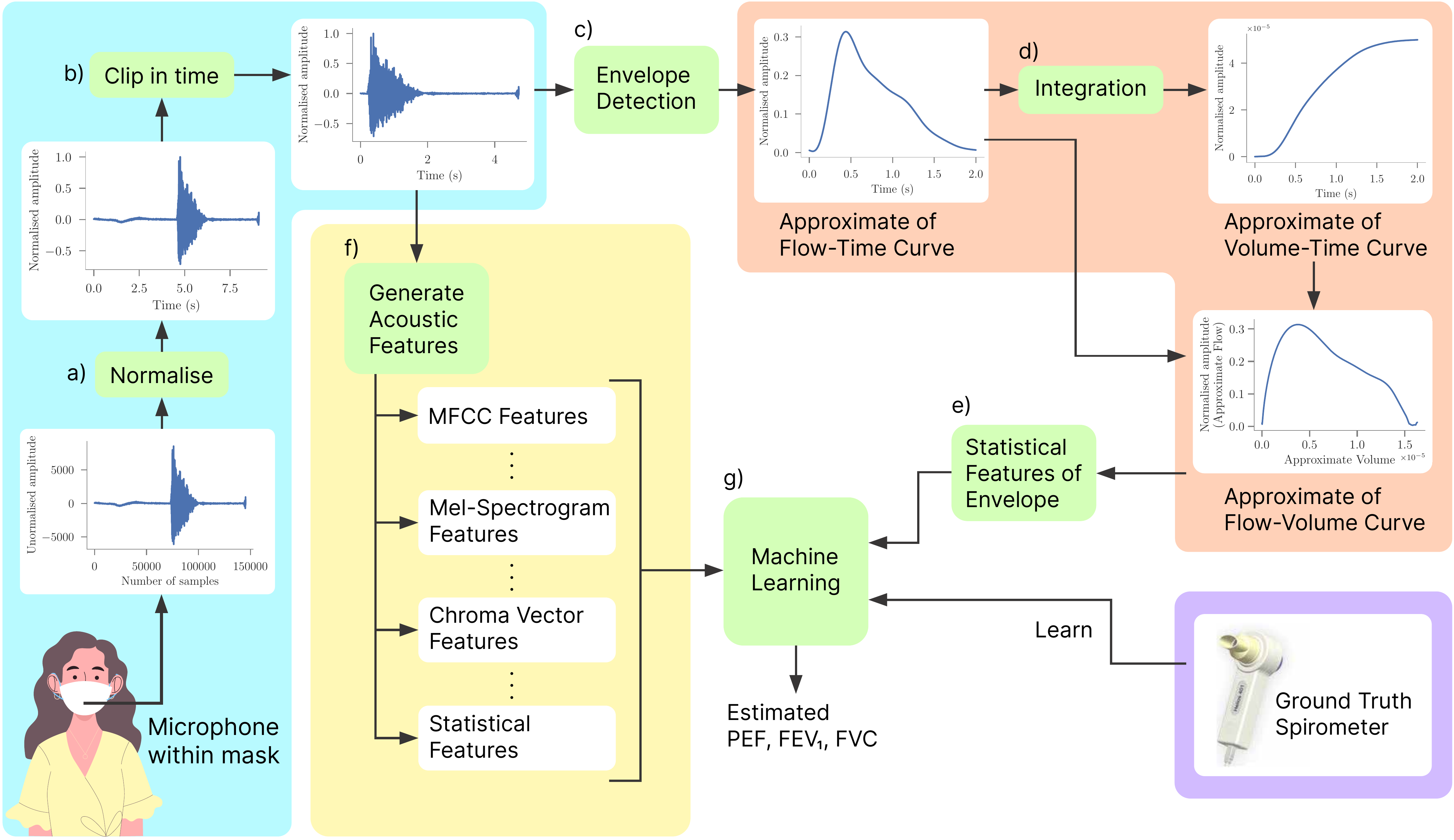}
    \caption{Pipeline showing the process to extract forced breathing parameters.}
      \label{fig:forced-pipeline}
\end{figure}

A spirometry test comprises of forceful exhalation through a flow monitoring device that measures instantaneous flow and cumulative exhaled volume (Table~\ref{tab:parameters}). Similar to a spirometry test, a user is required to wear our smart mask, breathe in their full lung volume, and forcefully exhale.  The entire pipeline of extracting forced breathing parameters is shown in Figure~\ref{fig:forced-pipeline}. We now explain the steps below. 

\begin{enumerate}[label=(\roman*)]
  \item \textbf{Recording Audio:} The microphone inside the mask records the exhalation and sends the audio data to a computer.
  \item \textbf{Normalising amplitude:} The audio data is normalised between -1 and 1~\cite{taylor2018estimation}.
  \item \textbf{Clipping audio:} Keeping in terms with the ATS guidelines~\cite{culver2017recommendations}, we extract the part of the signal which represents a second before the start of forceful exhalation till the end of it. The start of forceful exhalation is detected using a threshold amplitude.
  \item \textbf{Envelope Detection:} 
  The audio signal's envelope can be assumed to be a reasonable approximation of the flow rate because it is a measure of the overall signal power (or amplitude) at low frequency~\cite{larson2012spirosmart}. We obtain an estimation of the acoustic envelope of the forceful exhalation using Hilbert Transform (HT)~\cite{johansson1999hilbert}. This method has been employed in previous acoustical flow estimation studies~\cite{larson2012spirosmart, taylor2018estimation}. To validate HT's estimated envelope for a signal with multiple harmonics, we generated a synthetic amplitude modulated signal using a message signal (the envelope) and a carrier wave~\cite{seydnejad1997real, sun2016quadratic, potamianos1994comparison}. Figure~\ref{fig:am-carrier} (in Appendix) shows that the envelope estimated is a poor fit to the true envelope. We improve the estimated envelope using an finite impulse response (FIR) filter with Kaiser window~\cite{potamianos1994comparison}. Since an audio signal comprises multiple harmonics, we will need to process the HT envelope using an FIR filter with 
  Kaiser window. The estimated lung parameters' error margin would depend on stopband attenuation and transition width of the filter.
  \item \textbf{Approximate of `Volume-Time' (VT) curve:} The cumulative sum of flow rate over time gives us the approximate measure of the volume of the lung.
  \item \textbf{Approximate of `Flow-Volume' (FV) curve:} At this stage we have the approximate `flow versus time' curve and the approximate `volume versus time' curve. These two results can be combined to derive the `flow versus volume' curve.

\item \textbf{Feature Generation:} 
We created two sets of features. The first set of features were from the audio waveform and the second set of features was from the FV curve. These features were input into a regression model to predict PEF, FEV1 and FVC.\\

  As shown in Figure~\ref{fig:forced-pipeline} (f), we generated acoustic features with a window size of 30ms and a step size of 15ms for each audio sample. The window and the step size were based on prior literature~\cite{larson2011accurate}. The features are Mel Filter Bank Energy (MFE), 10 mean and variance normalised Mel Frequency Cepstral Coefficient (MFCC MVN), the power spectrum, Mel Bands of spectogram. Besides these features, we also generated temporal features from the audio waveform and the FV curve~\cite{barandas2020tsfel}. Temporal features include information about entropy, peak to peak distance, the centroid of the signal etc. Previous literature~\cite{barandas2020tsfel} lists all the temporal features and how they are computed for time series data. All the features are described in Table~\ref{tab:all-features} in the Appendix.\\
  
  MFE features are known to distinguish speakers by modelling the shape of the vocal tract~\cite{tak2017novel, madikeri2011mel}. Previous literature has used MFCC, temporal and statistical features to classify abnormal lung sound and non-speech sounds~\cite{chatterjee2019mlung++, rahman2014bodybeat, schuller2010interspeech}. As stated in previous work~\cite{larson2011accurate}, the features in the frequency domain can be assumed to be amplitudes excited by reflection in the vocal track and the mouth opening and therefore should be proportional to the flow rate.\\ 
  
  \item \textbf{Machine Learning:}
   Given the extracted features and ground truth from spirometry, we train three supervised regression models: linear regression, random forest (RF) and support vector regression (SVR) models to predict PEF, FEV1 and FVC from 260 features including acoustic and temporal features from each audio files. There are two main reasons for choosing these three algorithms. i) Linear regression is easy to implement and thus would be best suited for  edge devices with limited compute capabilities. ii) RF and SVR can learn non-linear decision surfaces. They are also non-parametric algorithms (SVR with Radial Basis Function (RBF) is non-parametric) which means that their complexity grows as more data is made available. \\
  
\noindent Specific features: To predict PEF, we used the cumulative sum of MFE, Log MFE, MFCC MVN, Power Spectrum and Melspectogram as an input to the regression model. To predict FEV1, we clipped the portion of the audio waveform after 1 second of the start of exhalation. These changes in features were inspired by previous literature~\cite{larson2011accurate}. For FVC, we used the features from entire audio waveform.\\
  
\noindent Feature selection: We used sequential forward selection technique (SFS)~\cite{james2013introduction} to reduce the feature space for faster computation and prevent overfitting.
\end{enumerate}

\subsection{Tidal Breathing}
\label{tidal_approach}

\begin{figure}[ht]
        \centering
        \includegraphics[scale = 0.25]{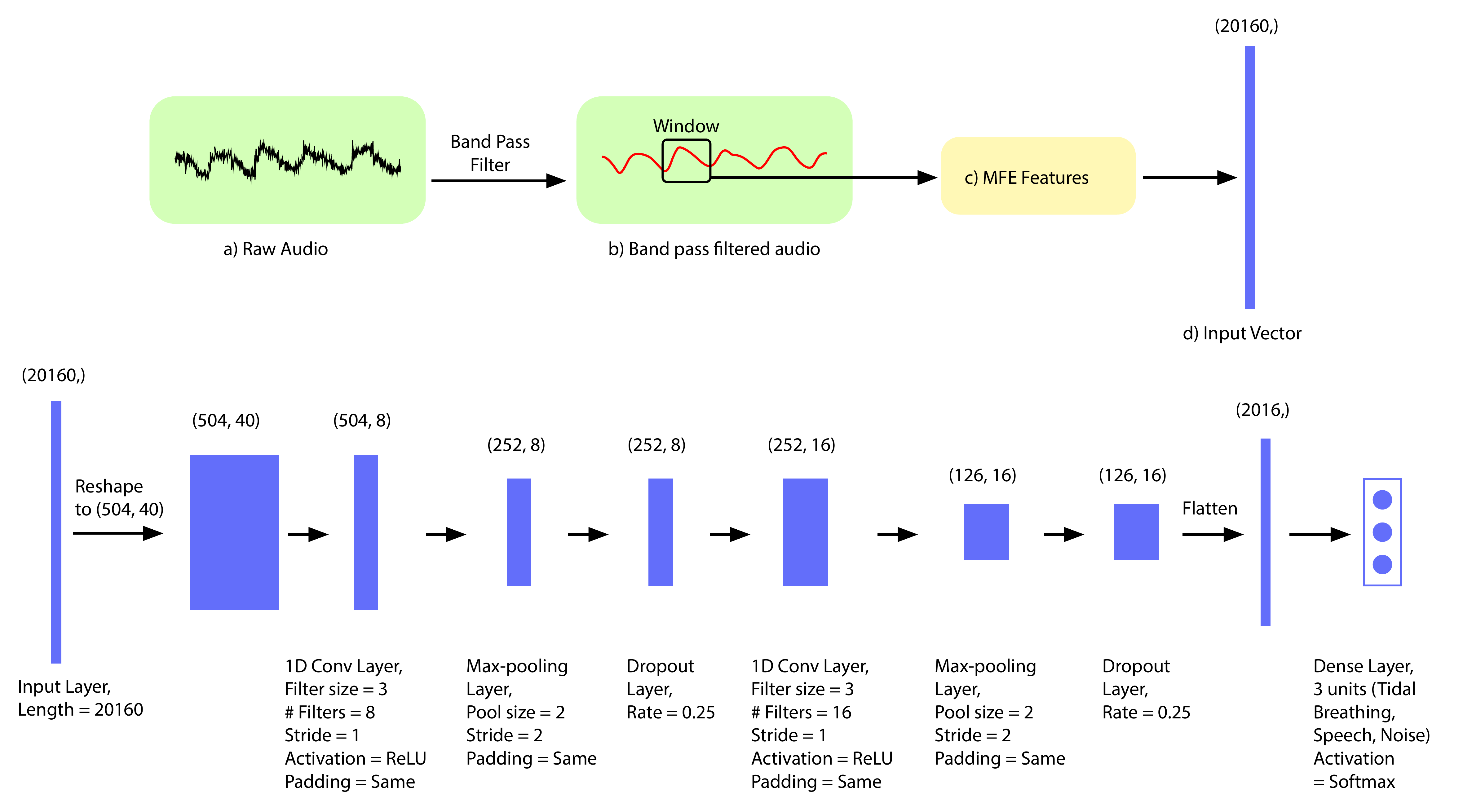}
        \caption{1D Convolutional Neural Network used to classify tidal breathing, speech and noise.}
        \label{fig:nn}
\end{figure}

\noindent  Our objective to to estimate respiration rate from tidal breathing. The first step in this process is to classify the audio samples as speech, tidal breathing or noise. Thereafter, an algorithm calculated the average peak to peak time for every sample classified as tidal breathing to derive the respiration rate.

\noindent \textbf{Classification Task:} 
The entire pipeline of estimating respiration rate is explained below.

\begin{itemize}
    \item \textbf{Feature Generation:} Pulmonary nasal sounds lie in the frequency range of 20Hz to 2500 Hz whereas speech which is located in the range of 300-4000 Hz~\cite{chatterjee2019mlung++, seren2005frequency}. For each audio sample, we used a Butterworth band pass filter with a low cut off frequency of 50Hz and high cut off frequency of 500Hz. We experimentally found out that these choices of frequencies conserved all the tidal breathing information as shown in Figure~\ref{fig:speech-and-breathing}. The audio samples were sliced using a rectangular window as shown in Figure~\ref{fig:nn} (b). We describe the choice of window size and the size of offset between subsequent window in Section~\ref{subsubsec:exp-setting-tidal}. For each window, we generate Mel Filter Bank Energy (MFE) features. MFE features isolate vocal track components which are crucial for estimating lung health~\cite{chauhan2017breathprint, larson2011accurate, rahman2014bodybeat}.

    \begin{figure}[ht]
        \centering
        \includegraphics[scale=.65]{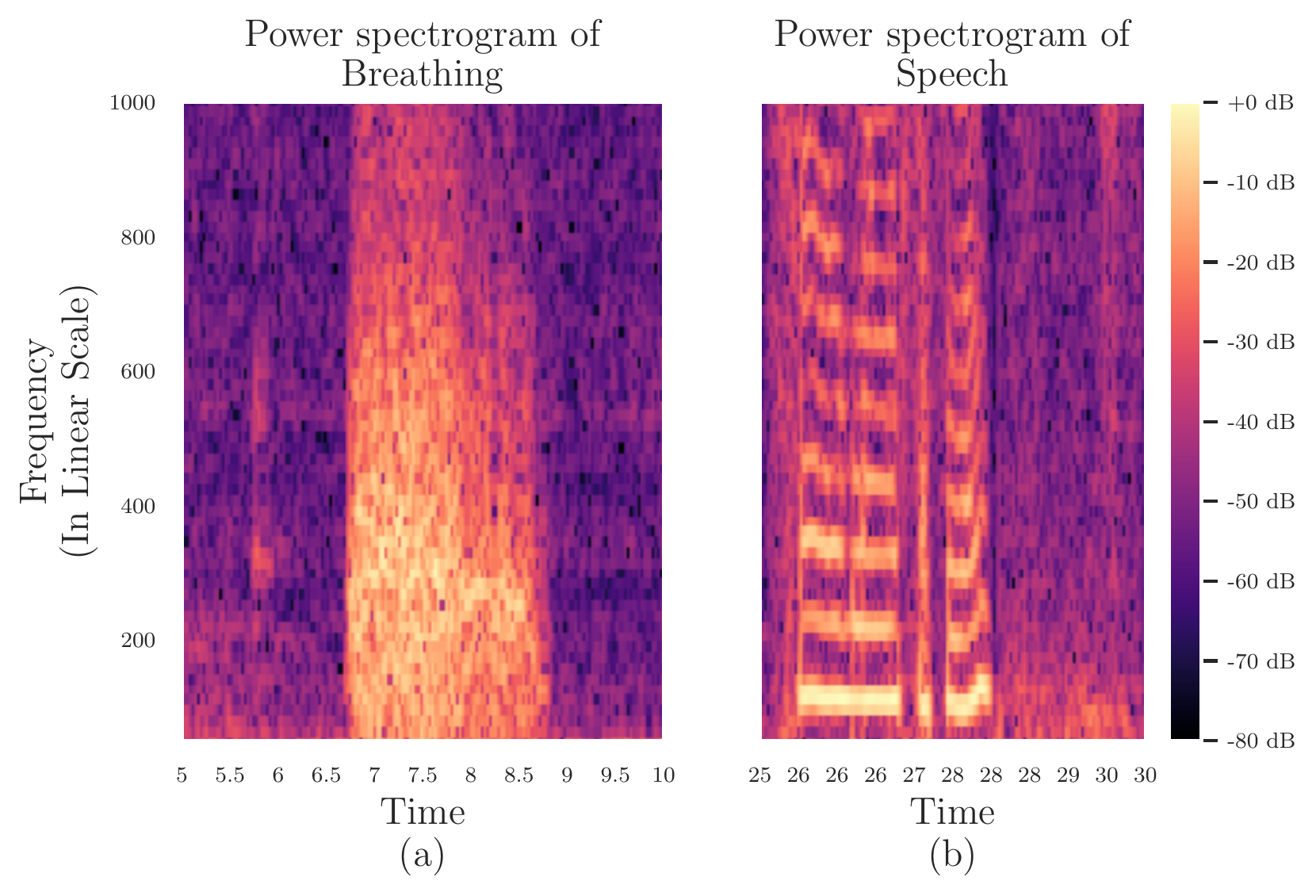}
        \caption{(a) The majority of breathing information is withing 500 Hz. In (b), horizontal lines or harmonics signifies speech information which is different from (a). At a cutoff of 500 Hz, a significant amount of speech information is lost. The lower harmonics aren't enough to decipher speech audio.}
        \label{fig:speech-and-breathing}
    \end{figure}

    \item \textbf{Classification: } We used a 1D  Convolutional Neural Network (1D CNN) to classify speech, tidal breathing and noise~\cite{zeng2018understanding, xu2021listen2cough}. We collected 40 minutes of tidal breathing data during our user study. To ensure a relatively balanced dataset, we also had 36 minutes of speech and 40 minutes of noise data. The details about noise data is described in the next section. 1D CNNs can be efficiently trained with a limited dataset of 1D signals where a temporal dependence exists between the values~\cite{kiranyaz20191}. As shown in Figure~\ref{fig:nn}, we used a 2 layer CNN with dropout. 
    
    \item \textbf{Estimating Respiratory Rate:} Once we have identified and isloated tidal breathing audio, we used a respiratory rate detection algorithm to detect the respiratory rate for samples labelled as tidal breathing by the 1D CNN. We applied the Hilbert Transform envelope detection algorithm to the audio samples as described in Section~\ref{sec:approach-forced-breathing}. We used a peak detection algorithm~\cite{virtanen2020scipy} over the Hilbert envelope. The total length of the tidal signal divided by the average peak to peak time gives us the respiration rate.
\end{itemize}
\section{Evaluation}
\label{sec:eval}
\subsection{Experimental Setting For Forced Breathing}
\label{experimental_setting_forced}
\begin{figure}[ht]
    \centering
    \includegraphics[scale=0.75]{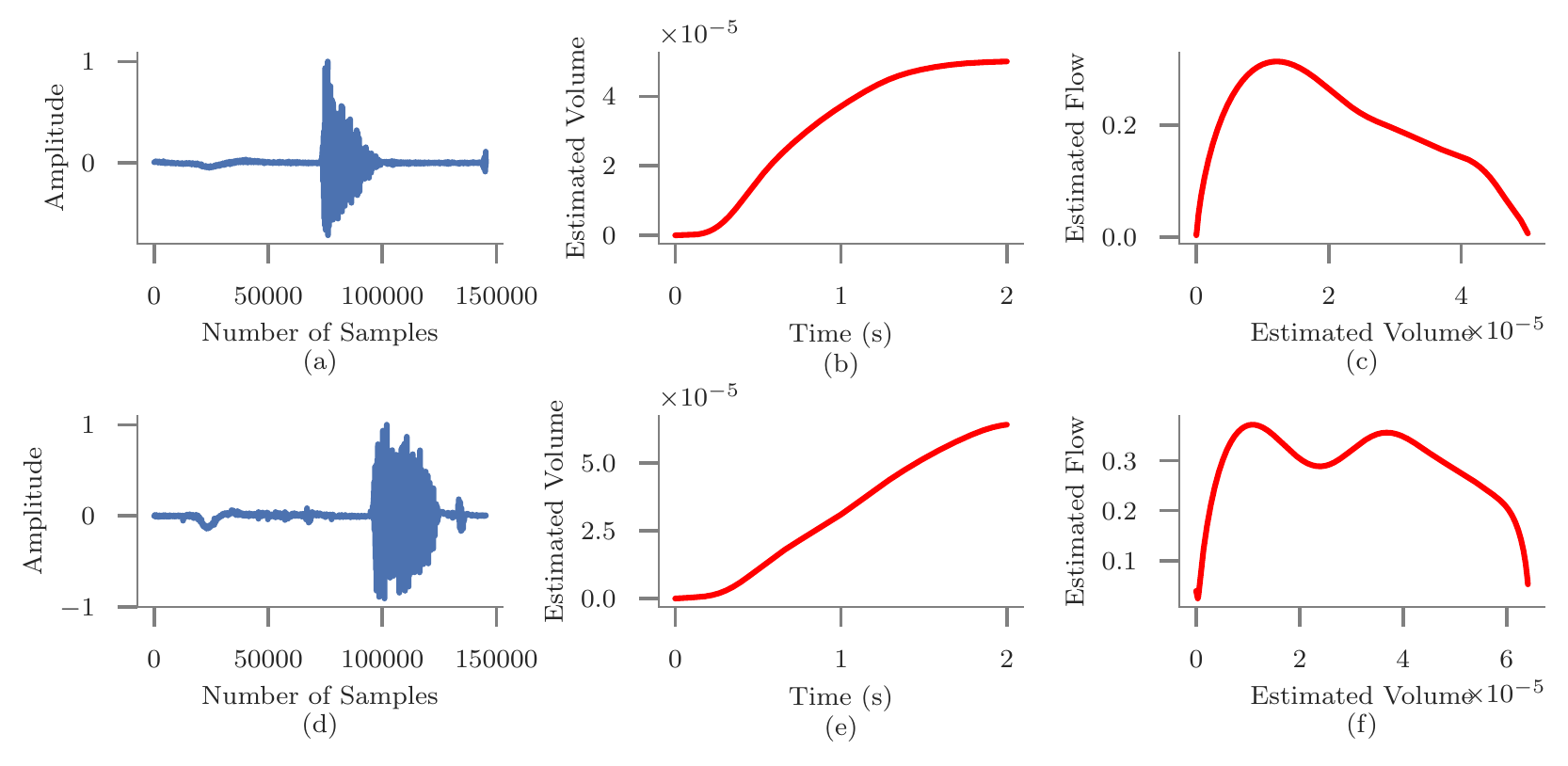}
    \caption{(c) and (f) shows a successful and unsuccessful forced breathing maneuver due to expiratory curvilinearity~\cite{miller2005standardisation}, respectively. (b) and (e) are the corresponding volume-time plot. (a) and (d) are the corresponding audio recording.}
      \label{fig:flowVSVolume}
\end{figure}
\noindent \textbf{Signal Processing:} The amplitude of the forced breathing audio recordings was normalised between -1 and 1, and clipped in time. Then, Hilbert transform was applied to deduce the approximate flow versus time curve by applying the pipeline explained in Section~\ref{sec:approach}. The shape of the `flow-volume' curve depends on the design of a minimum-order FIR filter. We need a minimum order filter to save on processing time for each sample and to ensure numerical stability~\cite{winder2002analog,shenoi2006introduction}. The transition width ($w$) and the stopband ripple ($s$) of a FIR filter decide the order of the filter. We used $w = \frac{2}{n}$ as the transition width, (where $n$ is the sampling rate which is 12 kHz) and $s = -10 dB$ as the stopband ripple to obtain the correct shape of flow-volume curve. Finally, one of the investigators visually analysed the flow-volume curve of every participant for its correct shape~\cite{knudson1976maximal,luo2017automatic}.  In our future work, a machine learning model would classify correct and incorrect flow-volume curve based on the approach in the previous work ~\cite{luo2017automatic}. Figure~\ref{fig:flowVSVolume}(c) and Figure ~\ref{fig:flowVSVolume}(f) shows a correct and incorrect shape of the flow-volume curve, respectively. While Figure~\ref{fig:flowVSVolume}(c) shows end expiratory curvilinearity~\cite{miller2005standardisation}, the same curvilinearity is missing in Figure ~\ref{fig:flowVSVolume}(f). Beside curvilinearity we also relied on previous literature~\cite{luo2017automatic} to distinguish between correct and incorrect shape of flow-volume curve. The proper shape implies a successful forceful breathing maneuver.\\

\noindent \textbf{Cross validation: }
In both RF and SVR, we use a nested leave-one-out (subject) cross-validation strategy (LOOCV). The outer loop is used for predicting the lung parameters for a test participant, where all but that participant is used in the train set. The inner loop is used to fine-tune the hyper-parameters. Leave One Out Cross Validation (LOOCV) is preferred because it has far less bias compared to the validation set approach where we randomly split the dataset into train/test/validation or use K-Fold cross validation. LOOCV tends not to overestimate the test error rate as much as the validation approach does~\cite{james2013introduction}.\\  

\noindent \textbf{Metric:} We reported the mean percentage error across all participants for FEV1, FVC and PEF. The percentage error is given by $|{\dfrac{v_a-v_e}{v_a}}|*100$ where $v_a$ is the ground truth value and $v_e$ is the estimated value. Percentage error is chosen as a metric because the guidelines on standardized spirometry~\cite{miller2005standardisation} as well as subsequent research~\cite{rubini2010daily,larson2012spirosmart} report the error between two spirometry efforts in percentages. Using the same error metric across Spirometry research helps us in a fair comparison of the performances.\\

\noindent \textbf{Tuning Hyperparameters:} The number of trees in RF span from 5 to 500 in steps of 10. The nodes in RF were expanded until leaves contain less than two samples. For SVR, the hyperparameter space consisted of the regularisation parameter and the type of kernel among linear, radial basis function and polynomial. We used grid search strategy to find the optimal hyperparameters. We discuss the results of forced breathing in Section~\ref{sec:result-forced}.\\ 
%\nb{you suddenly mention SFS in the results for forced breathing..but do not mention it either in the approach nor in the experimental settings.}

\subsection{Results For Forced Breathing}
\label{sec:result-forced}
\begin{figure}[ht]
    \centering
    \includegraphics[scale=0.75]{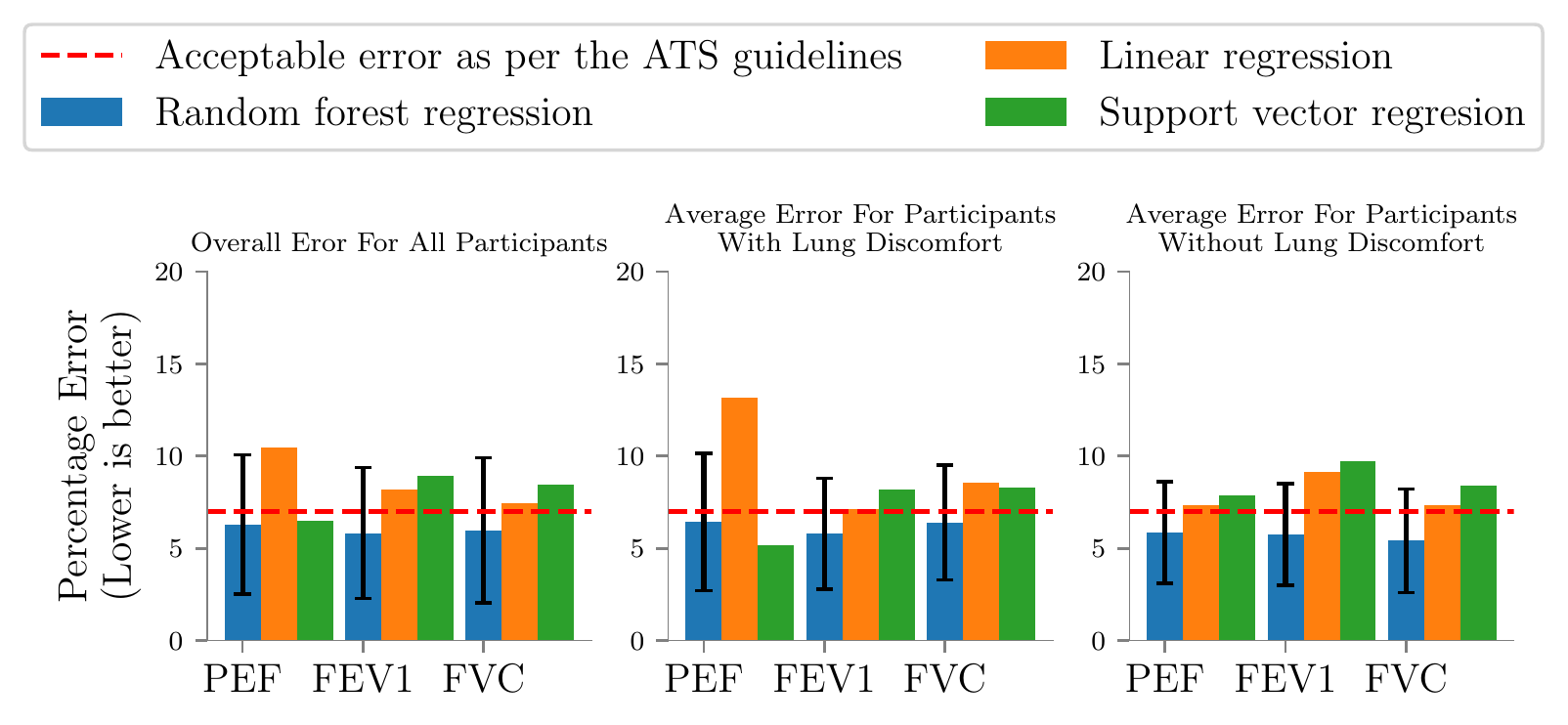}
    \caption{For \textbf{N95 mask}, the prediction error for participants with healthy and unhealthy lung condition is acceptable as per the ATS guidelines~\cite{miller2005standardisation} for the Random Forest regression model.}
      \label{fig:FVCErrorN95}
\end{figure}

\begin{figure}[ht]
    \centering
    \includegraphics[scale=1]{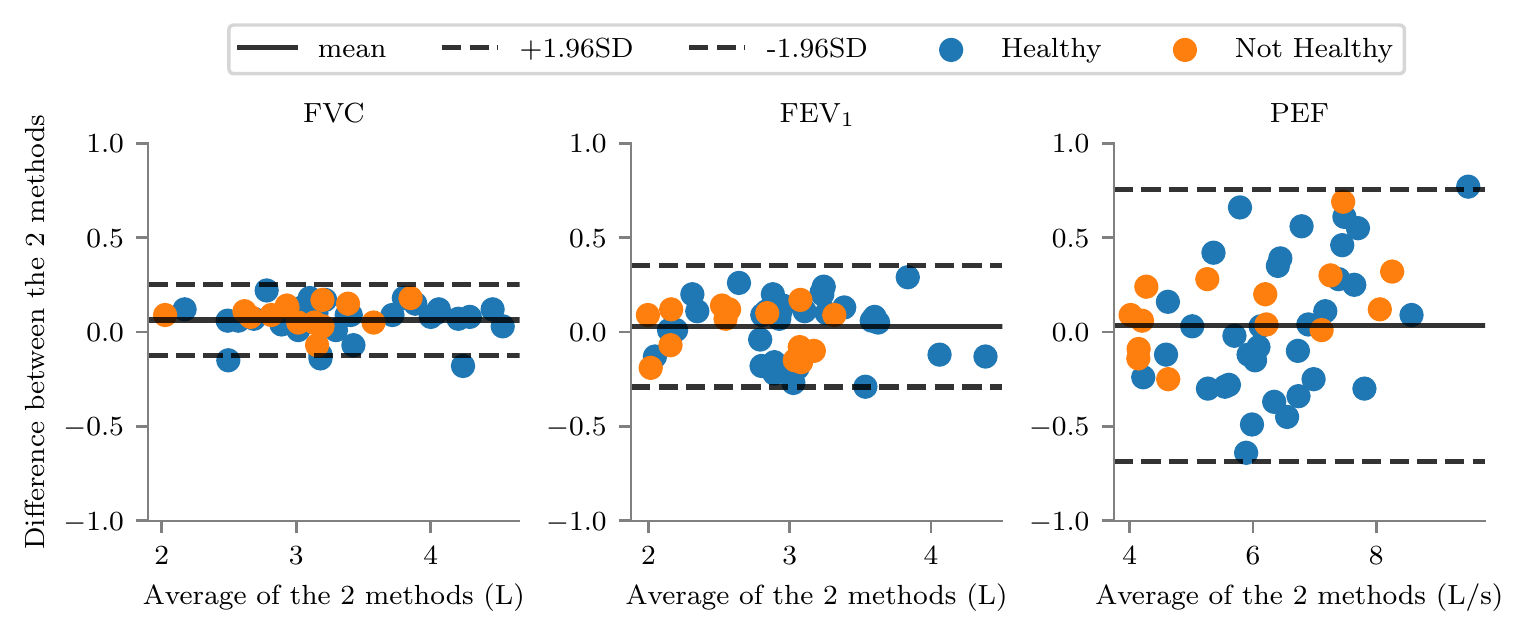}
    \caption{\textbf{Bland-Altman Plot For N95 Mask:} For FVC and FEV1, there is a agreement between Spirometer and SpiroMask for participants with healthy and unhealthy lung. The same agreement exists for PEF but with relatively higher variability.}
      \label{fig:bland-altman-n95}
\end{figure}

Our overall result for N95 mask is shown in Figure~\ref{fig:FVCErrorN95}. The RF regression performs the best across participants with a healthy and unhealthy lung condition.  Across all the participants, we have an Mean Percentage Error (MPE) of  6.30\% on PEF, 5.82\% on FEV1 and 5.98\% on FVC for the N95 mask. To explain the direction of bias, Figure~\ref{fig:bland-altman-n95} shows the Bland-Altman~\cite{bland1986statistical} plot of each lung function measure. The vertical axis shows the difference between SpiroMask and Spirometer. The horizontal axis shows the mean value of the two methods. Measures taken for healthy participants are shown in blue dots and unhealthy participants in orange dots. Lines indicating the $\pm2\sigma$ are shown as dashed lines. From the plot, it can be seen that SpiroMask generalises well across healthy and unhealthy participants. On average, the ground truth Spirometer has slightly higher value for all the three parameters. Although, the PEF has higher variability compared to FEV1 and FVC, but agreement holds true for PEF as well.\\

\begin{figure}[ht]
    \centering
    \includegraphics[scale=0.75]{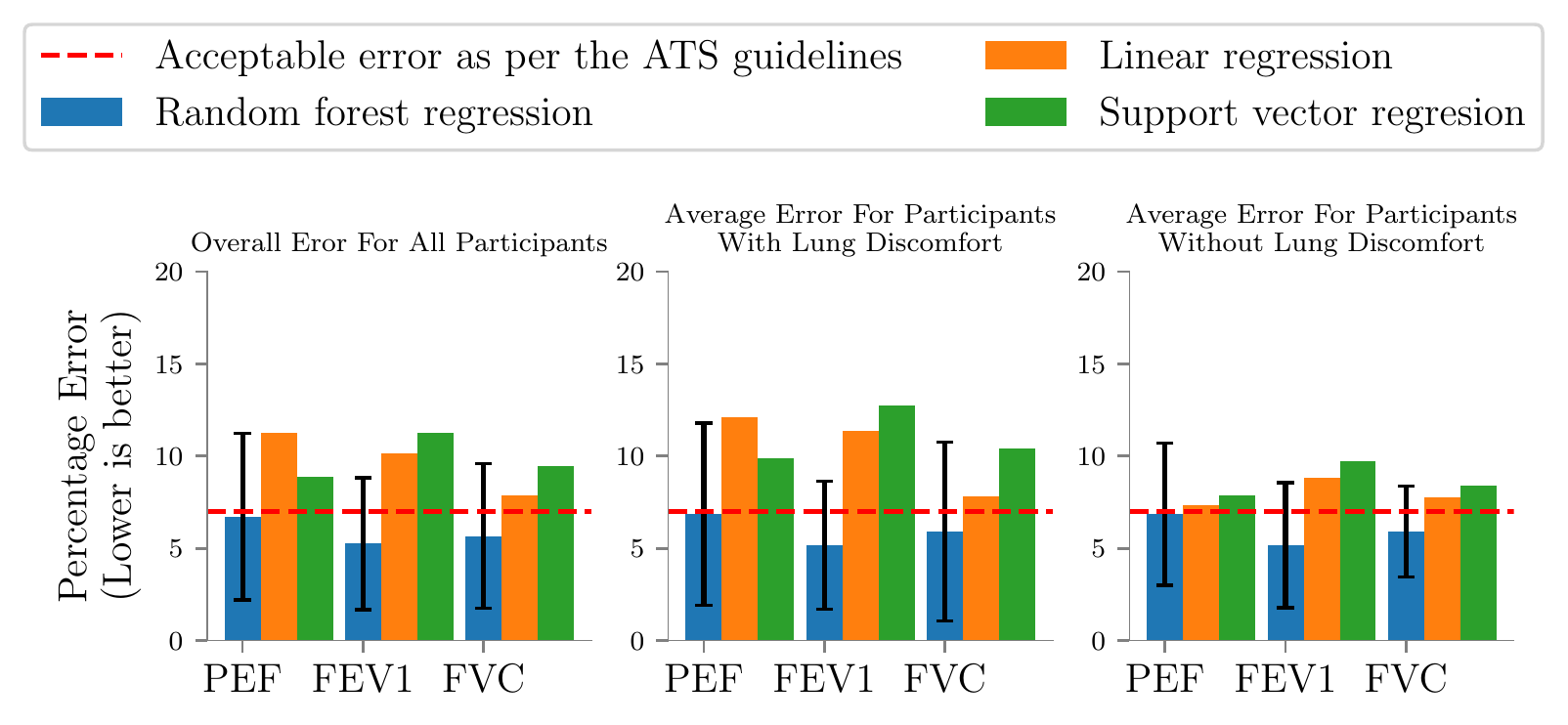}
    \caption{For \textbf{Cloth Mask}, the prediction errors for cloth mask is comparable to N95 mask. The standard deviation of FVC is higher compared to N95 mask. The prediction error across all participants is within the ATS guidelines~\cite{miller2005standardisation} for the Random Forest regression model}
      \label{fig:FVCErrorCloth}
\end{figure}

\begin{figure}[ht]
    \centering
    \includegraphics[scale=1]{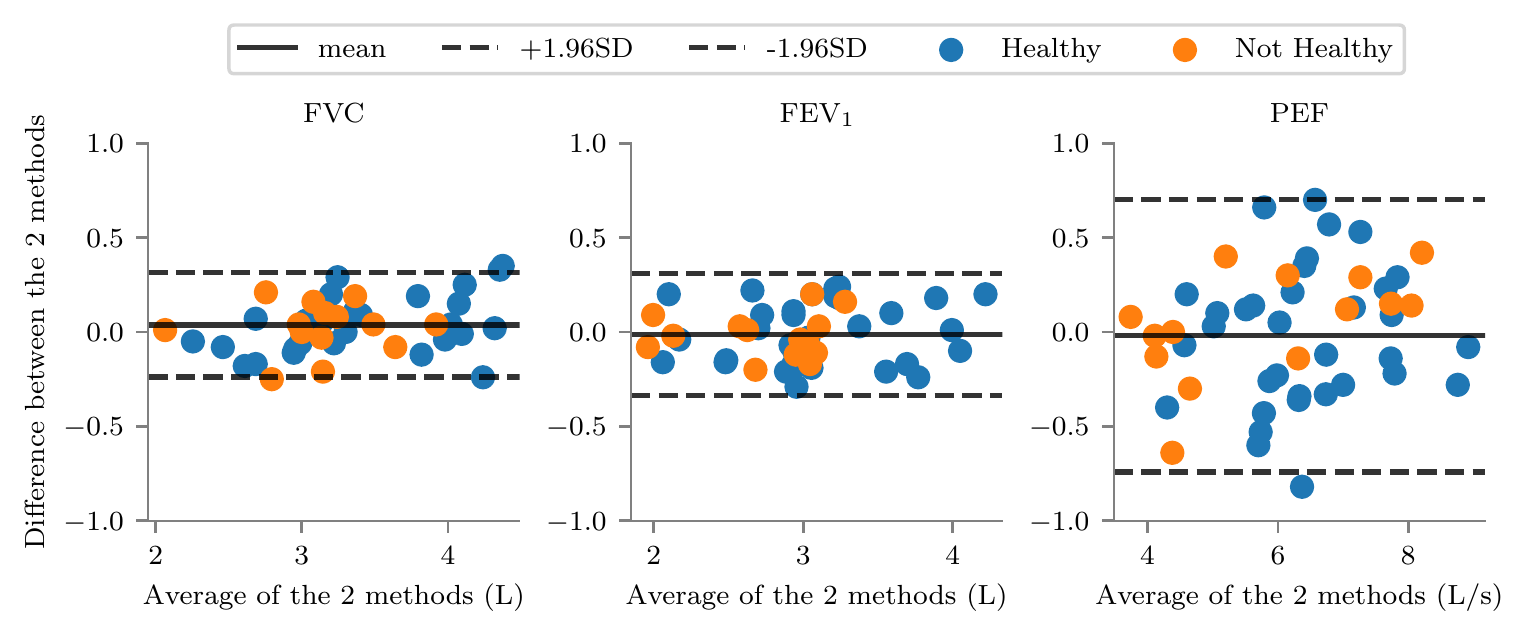}
    \caption{\textbf{Bland-Altman Plot For Cloth Mask:} For FVC and FEV1, there is a agreement between Spirometer and SpiroMask for participants with healthy and unhealthy lung. The same agreement exists for PEF but with relatively higher variability. On average, only FVC parameter from the ground truth Spirometer is higher compared to SpiroMask as indicated by the black horizontal line.}
      \label{fig:bland-altman-cloth}
\end{figure}

\noindent For the cloth mask (Figure~\ref{fig:FVCErrorCloth}), we have MPE of 6.71\% on PEF, 5.25\% on FEV1 and 5.67\% on FVC. The Bland-Altman plot for cloth mask (Figure~\ref{fig:bland-altman-cloth}) shows a similar agreement between spirometer and SpiroMask. Our results for all the lung parameters fall within the clinically relevant range. As mentioned in previous work~\cite{larson2011accurate}, a clinically relevant range is used because a participant cannot simultaneously use a spirometer and SpiroMask, so actual ground truth is unattainable. The limit of variability for a measure of lung function value should be within 7\% over short duration as mandated by the ATS guidelines~\cite{miller2005standardisation}. \textbf{Our result on both the masks are well within the ATS guidelines for both healthy and unhealthy participants.} Details of confounding factors and trends are described in Section~\ref{sec:confounding}.\\

\noindent Our result on PEF is better on the N95 mask. FEV1 and FVC are marginally better on cloth masks. In the future, we plan to study the comparative performance of the two masks using statistical tests. \\

\noindent The category of features selected using Sequential Forward Selection for PEF, FEV1 and FVC are listed in the Appendix~\ref{appendix-1}.

\subsection{Experimental Setting For Tidal Breathing}
\label{subsubsec:exp-setting-tidal}
\subsubsection{Classification Task}
We had 120 samples of tidal breathing across 37 participants. Speech data was available from 108 participants (36 Minutes of data). There were 120 samples of noise data (40 Minutes). The noise class consists of noise collected from different places like crowded cafeteria (30 samples), artificial traffic noise played via YouTube in a Bluetooth speaker in different volume levels (40 samples), white noise (30 samples generated fromAudacity\footnote{\url{https://www.audacityteam.org/}}) and 20 samples collected from a garden (mostly involving children playing and bird chirping). We have the following set of hyperparameters: the window size, offset size and Fast Fourier Transform (FFT) length (The choice of the hyperparameter space is described in Appendix). The window size cannot be bigger than the shortest audio sample, and the offset size can at maximum be as large as the window size.  We used 6-fold cross-validation where the inner loop was used for tuning hyperparameters.

\subsubsection{Estimating Respiration Rate}
For all samples that were classified as tidal breathing, we applied Hilbert Transformation over the sample, followed by peak detection. Peak detection was also applied on the acceleromter signal. An algorithm compared the average peak to peak time between the ground truth accelerometer signal and the audio signal. The total length of the signal divided by the average peak to peak time gave us the respiration rate. The respiration rate was compared using Mean Absolute Error (MAE) for each participant. Our choice of error is in line with most of the previous literature listed in Table~\ref{tab:tidalresults} in Section~\ref{sec:related-work}.

\subsection{Results For Tidal Breathing}
\subsubsection{Classification task}
\label{results_classification}
\begin{figure}[ht]
    \centering
    \includegraphics[scale=1]{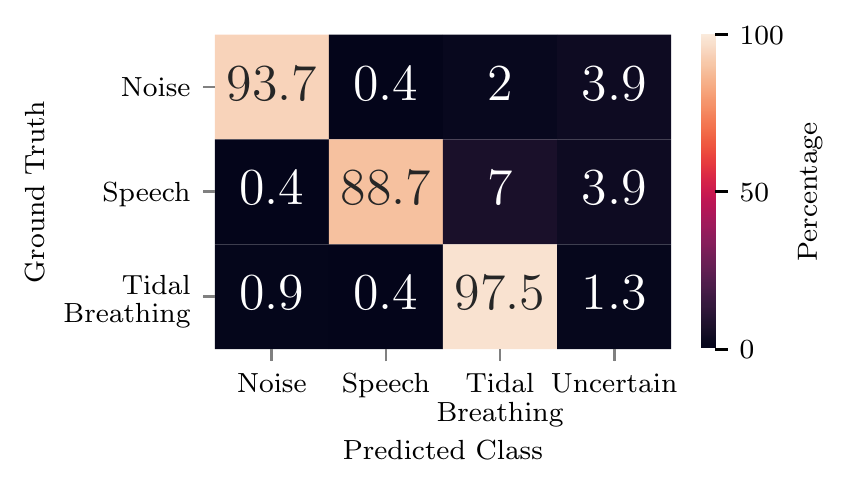}
    \caption{Tidal breathing was correctly classified for 97.5\% of the samples. Only 7\%  of speech samples and 2\% of noise samples were misclassified as tidal samples. Few samples which could not be classified into any classes are labelled as uncertain.}
    \label{fig:confusion_matrix}
\end{figure}

Our result in Figure~\ref{fig:confusion_matrix} shows that we achieve an accuracy of 94.7\%.  The confusion matrix shows that the CNN misclassified 2\% of noise samples and 7\% speech samples as tidal breathing. As described in Section~\ref{tidal_approach}, we segmented each audio sample into smaller windows, and the classifier predicts the class of every window. The CNN assigns one class (Tidal Breathing, Noise or Speech) to an audio sample if 90\% of the segmented windows of that sample belong to that class. We have set a high threshold to ensure a fewer number of false positives. Figure~\ref{fig:confusion_matrix} also shows the percentage of samples which the CNN could not classify. A higher threshold leads to better accuracy, but at the cost of some samples labelled as uncertain. A lower threshold would lead to higher misclassification. For example, the peak detection algorithm would return an inaccurate respiration rate if the CNN labels a speech sample as a tidal sample. A higher threshold lessens misclassification by segregating audio samples that cannot be classified into one class.

\subsubsection{Estimating respiration rate}
We achieved a Mean Absolute Error (MAE) of 0.49 on the N95 mask. The MAE on the cloth mask was 0.68. Our results are comparable to previous work on estimating respiration rate using smartphone~\cite{rahman2020breatheasy} and WiFi signals~\cite{yue2018extracting}. We estimated the respiration rate from the average peak to peak time. It must be noted that the Hilbert Transform envelope for samples incorrectly classified as tidal breathing samples had an amplitude of zero resulting in no peaks, and we thus reject such cycles.

\section{Sensitivity analysis}
\label{sec:sensitivity}
\subsection{Sensor Positioning:}
In this section, we analyse the robustness of our methods with respect to sensor placement inside the mask. The motivation to perform this experiment is that a person might retrofit the sensor inside any position inside the mask, and SpiroMask will be expected to monitor forced and tidal breathing without any drop in accuracy of lung parameters. \\

\begin{figure}[ht]
    \centering
    \includegraphics[scale=0.30]{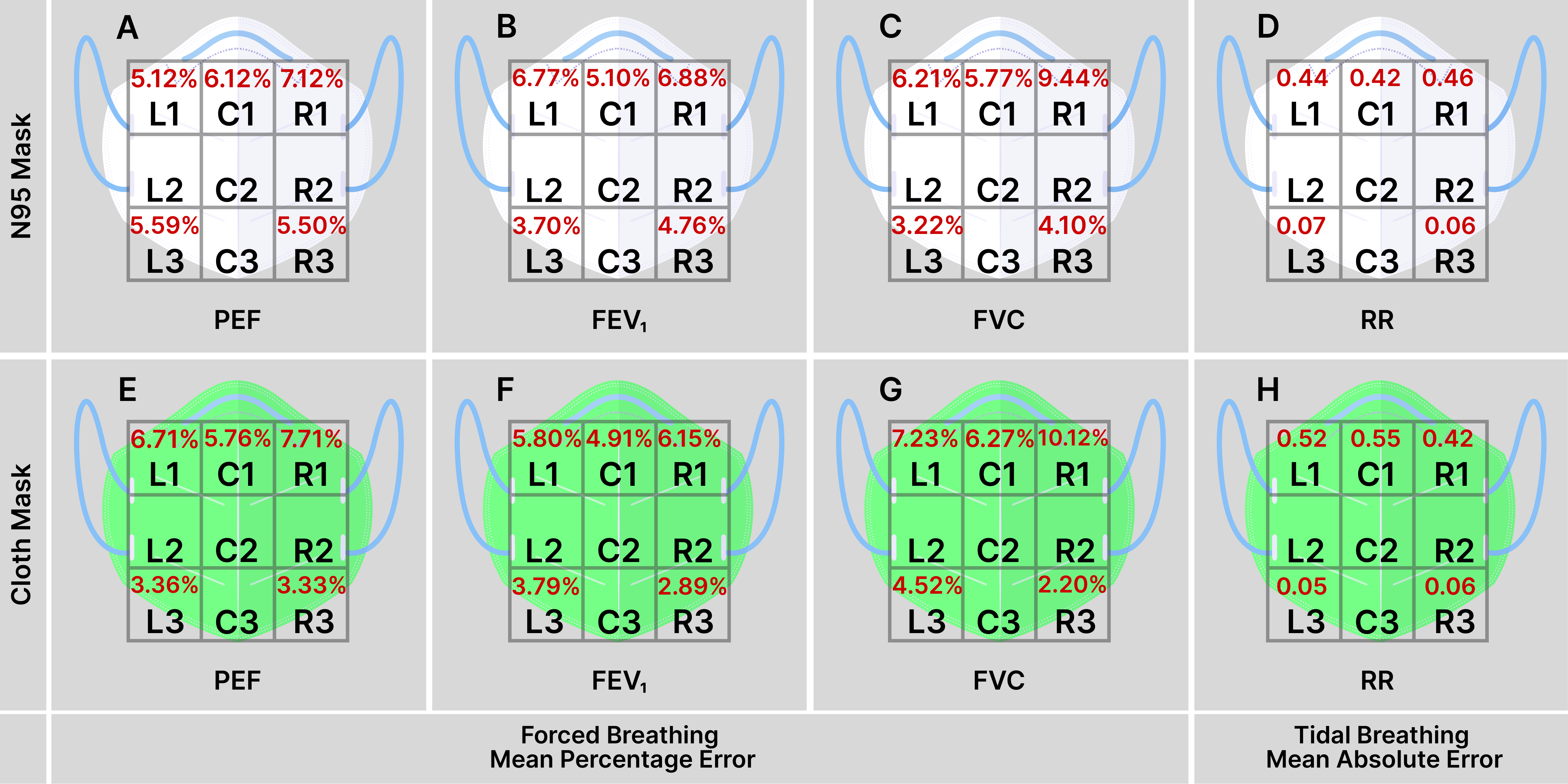}
    \caption{Forced breathing is robust to sensor placement, with relatively better performance at position L3 and R3. Tidal breathing can be accurately monitored at position L3 and R3}
      \label{fig:sensitivity}
\end{figure}

\noindent We described in Section~\ref{sec:study-design} that we collected forced breathing and tidal breathing samples for some participant by placing the sensor in five different positions inside the mask (Figure~\ref{fig:face-position}). We had sensor position data for 12 participants (including 8 participants with unhealthy lungs). For tidal breathing, we had sensor positioning data for 18 participants. We used the model trained on audio samples collected by placing the sensor at position L1 and predicted on all other samples that belong to locations C1, R1, L3, R3. We used the same cross-validation strategy as described in Section~\ref{sec:eval}. Figure~\ref{fig:sensitivity} shows the main results on 12 participants (Forced Breathing) and 18 participants (Tidal Breathing). For forced breathing, we observe that SpiroMask is robust to sensor placement where the MPE for all positions is below 7\% (the ATS acceptable error). However, for tidal breathing, location L3 and R3 have the lowest MAE. The low MAE on L3 and R3 can be attributed to the microphone being below the apex of the nose. None of the audio samples from L3 or R3 were misclassified during the classification task explained in Section~\ref{sec:eval}. A detailed break-up of MPE and MAE for forced and tidal breathing among healthy and unhealthy participants is given in Figure~\ref{fig:position-appendix_1} and Figure~\ref{fig:position-appendix} in the Appendix. Note that the error metrics were chosen based on prior work on Smartphone spirometry and respiration rate estimation as explained in Secion~\ref{experimental_setting_forced} and Section~\ref{subsubsec:exp-setting-tidal}.

\subsection{Sampling Rate:}
For continuous respiration rate monitoring, SpiroMask does not analyse speech signals and is concerned only with tidal breathing signals. The information content in tidal breathing lies between 50 Hz to 500 Hz.

\begin{figure}[ht]
    \centering
    \includegraphics[scale=1]{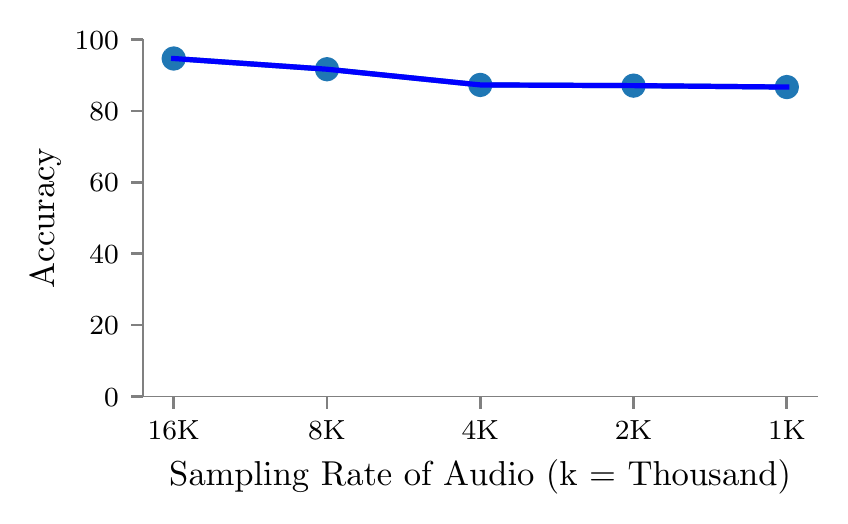}
    \caption{Speech becomes indistinguishable below 8 kHz~\cite{french1947factors}. The classification accuracy drops from 94.7\% to 91.7\% when sampling rate is decreased from 16kHz to 8kHZ}
      \label{fig:sampling-accuracy}
\end{figure}

\begin{figure}[ht]
    \centering
    \includegraphics[scale=1]{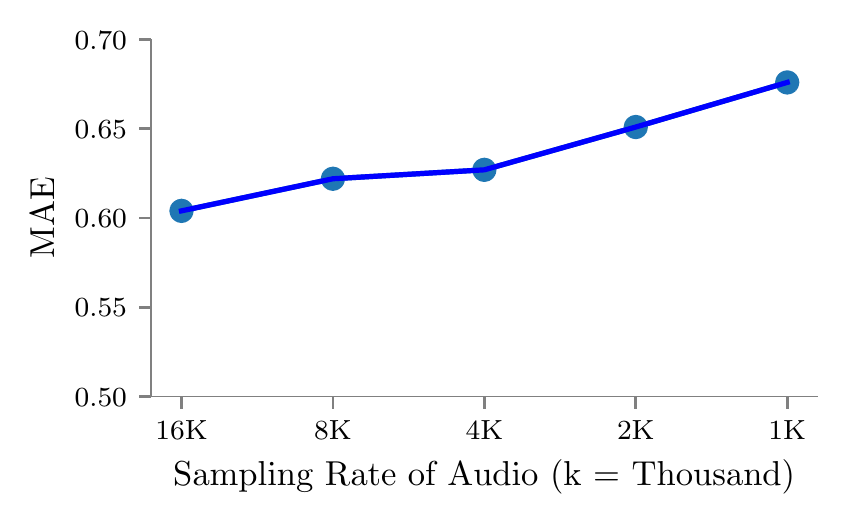}
    \caption{Mean Absolute Error of respiration rate detection on tidal samples increases on decreasing the sampling rate. MAE is 0.6 at 16kHz, it increases to 0.68 at 1kHz}
      \label{fig:MAE-sampling}
\end{figure}

Figure~\ref{fig:sampling-accuracy} shows the classification accuracy for different sampling rates. The classification accuracy remains above 80\% even when the sampling rate is reduced to 1 KHz. The decrease in accuracy is attributed to misclassification of noise and speech samples as `uncertain' samples at a lower sampling rate. Figure~\ref{fig:MAE-sampling} shows that the mean absolute error of average peak to peak time (consequently the respiration rate) calculated on tidal samples increases on decreasing the sampling rate. Reducing the sampling rate to 1 KHz also aids in aiding privacy concerns as speeches are not intelligible when audio we sample audio at 1 KHz. To validate speech intelligibility, we asked three participants to hear two speech recordings. Each of these recordings were sampled at 2 KHz and 1 KHz. Each sample size is 16-bits long. None of the participants could decipher the speeches recorded at 1 KHz. One of the three participants could decipher one of the speech recording sampled at 2 KHz. Speech spoken relatively slowly are easier to decipher at lower sampling rate.  Therefore for preserving privacy, we can use lower sampling rates at the cost of a slight decrease in the accuracy of the respiration rate. 

\section{Energy Consumption of Classifier}
\label{sec:energy}
To study the battery life and the usability of our system we arrived at an estimated power consumption of SpiroMask's tidal breathing monitoring system. We evaluated the power consumption of the classifier running on the microcontroller. We used an oscilloscope to calculate the voltage drop across a $68 \Omega$ shunt resistor. We arrived at the value of shunt resistor on basis of the voltage resolution of the oscilloscope so that a small current draw (in order of $mA$) leads to a high voltage drop (in order of 100$mV$) across the shunt. Consequently, we could arrive at the current consumption. The microcontroller first samples the audio, after which the model inference predicts whether the sampled audio represents `Noise', `Tidal Breathing' or `Speech'. The current consumption at different stages of the microcontroller can be seen in Table~\ref{tab:current-consumption}. With a duty cycle of one respiratory rate every minute, the average current consumption is 1.64 mA. To arrive at a battery life estimate, we assume 11 hrs/day of active respiration rate measurement with a 240 mAh coin cell battery. In this scenario, SpiroMask's respiration rate system is expected to last for approximately thirteen days on battery power alone.

\begin{table}[ht]
\begin{tabular}{llll}
\hline
\multicolumn{3}{l}{\textbf{\begin{tabular}[c]{@{}l@{}}Current \\ Consumption (mA)\end{tabular}}} & \textbf{\begin{tabular}[c]{@{}l@{}}Battery Life \\ With 11h/day of Operation\end{tabular}} \\ \hline
No Operation                     & Sampling                     & Classifier                     &                                                                                            \\
0.96                             & 1.6                          & 4.7                            & 13 Days                                                                                    \\ \hline
\end{tabular}
\caption{Current consumption of the CNN classifier. With a 240 mAh battery, the classifier can run for 13 days.}
\label{tab:current-consumption}
\end{table}
\section{Confounding Factors and Trends}
\label{sec:confounding}
%  \nb{I think you need to clarify this is only for forced breathing. have we done any such similar analysis for respiration rate? }
We performed 8-way multivariate analysis of variance (MANOVA) \cite{warne2014primer} 
%\nb{citation?} 
tests to determine if other variables significantly contributed to the difference (in terms of percent error) between SpiroMask and the spirometer in estimating the forced breathing parameters FEV1, FVC and PEF. MANOVA is more suitable than univariate analysis of variance (ANOVA) tests for forced breathing since we have 3 dependent variables. We performed separate MANOVA tests for N95 and cloth masks. We used height, weight, gender, age, whether the subject has performed spirometry tests before, whether the subject reported any lung ailments, whether the subject has the habit of smoking, and whether the subject had a meal before appearing for the test as the 8 grouping variables. The test for cloth mask shows that none of the grouping variables have a significant effect on the percent error between SpiroMask and spirometer measures (all p-values were greater than 0.05, the significance level of the test). 
The test for N95 mask suggests that height could be a significant variable  (p-value $\approx$ 0.04). Since this is only suggested by the test for N95 mask and not by the test for cloth mask, we need to investigate further to determine how significant the variable height is. We have provided scatter plots showing the variation of percent error for the forced breathing parameters with respect to height (for N95 mask) in Section \ref{sec:manova} of the Appendix. The presence of trend between FVC and FEV1 with height suggests incorporating such information whenever available will further improve our models. We leave the detailed ablation study with participant features like health and weight for the future.\\
 
\noindent We also performed 7-way analysis of variance (ANOVA) tests to determine if other variables significantly contributed to the difference (in terms of percent error) between SpiroMask and a smartphone strapped to the chest in estimating the tidal breathing parameter (respiration rate). We performed separate ANOVA tests for N95 and cloth masks. We used height, weight, gender, age, whether the subject reported any lung ailments, whether the subject has the habit of smoking, and whether the subject had a meal before appearing for the test as the 7 grouping variables. The tests for both N95 and cloth masks show that none of the grouping variables have a  significant effect on the percent error between SpiroMask and smartphone measures (all p-values were greater than 0.05, the significance level of the test). Refer to Section \ref{sec:manova} in the Appendix for more detailed results of the MANOVA and ANOVA tests. 
\section{Participant feedback}
\label{feedback}
We collected feedback through an exit survey with three optional questions: whether they preferred a spirometer or SpiroMask (for measuring forced breathing parameters), whether they preferred SpiroMask or a phone strapped to the chest with a belt (for measuring tidal breathing parameters), and whether they preferred a cloth mask or an N95 mask. 

\subsection{SpiroMask and spirometer}
Around 58\% of the responses preferred SpiroMask over a spirometer, while around 21\% of the responses were equally comfortable with both methods. The main reasons cited by the responses preferring SpiroMask over a spirometer included: \textit{``mask was more comfortable''}, \textit{``forced exhalation was easier to perform in a mask''}, and \textit{``spirometer was hard to hold and operate, it is bulky''}. Some of the older aged subjects commented that \textit{``it becomes hard to hold the spirometer in the mouth for a long time"} (subject no. 16), and \textit{``mask is preferable as spirometer required breathing using mouthpiece which was a bit difficult''} (subject no. 7). Interestingly, 75\% preferred a cloth mask over an N95 mask, mainly based on comfort.  \\

\subsection{SpiroMask and chest belt}

As mentioned in Section~\ref{sec:approach}, we used the accelerometer sensor of a smartphone to collect ground truth on respiration rate. But, we could attain the ground truth via accelerometer only for a few participants. For the rest, we relied on self-counting as the ground truth. Among those who wore the smartphone strapped belt, 91\% preferred SpiroMask over the phone to monitor tidal breathing during the user study. A participant said, `` the chest belt seemed to restrict breathing,'' and others voiced similar opinions. When asked by the investigator if they would like to wear a chest belt or SpiroMask as a part of their daily life, the opinion unanimously favoured SpiroMask.

\section{Limitations and Future Work}

We now discuss some limitations of our work and proposed future work to address them
\begin{itemize}
    \item Human motions hinder breathing measurement. Prior research has shown that sensors worn in the chest can be used to measure breathing parameters during activities such as walking~\cite{whitlock2020spiro}. But, we found that wearing a chest belt is not always comfortable and is susceptible to jittering activity. Currently, SpiroMask can classify noise, breathing audio and speech but does not attempt to extract breathing signal from noisy audio recordings. We investigated audio signals which were collected from the microphone placed inside a user donned mask in simultaneous presence of background noise and user activity. In our follow up work~\cite{adhikary2022Towards}, we demonstrated  the challenges in extracting useful breathing information from such signals.
    
    \item Our confounding variable analysis suggests that incorporating personal parameters such as height can help improve the estimation for forced breathing parameters. In the future, we plan to conduct an ablation study to study the potential improvements in modelling when personal health parameters are available. Previous smartphone spirometry work has shown improvement via personalisation~\cite{larson2012spirosmart}, though their notion of personalisation was different.
    \item A detailed large scale user study on the usability of SpiroMask is currently out of the scope of our current research. In our current work, we evaluated the percentage of participants who would prefer SpiroMask over a traditional spirometer or a chest belt.
    \item Currently, SpiroMask could not detect inhalations due to low amplitude. This is a known problem in smartphone spirometry~\cite{larson2012spirosmart}. However, unlike the smartphone, since we use a custom microphone, we plan to experiment with multi-microphone setup, one with high gain for inhalation and the other for exhalation. 
    \item In our current work, we do not estimate the tidal volume. This is primarily due to lack of clinical-grade ground truth device for tidal volume estimation. We believe that our existing pipeline for forced volume should be easily adaptable to estimate tidal volume.
    \item In the current work, we predict specific points on the flow-volume curve (such as FEV1, FVC). In the future, we plan to predict the whole flow-volume curve instead of only these points. This problem can be naturally mapped to a sequence-to-sequence learning problem~\cite{sutskever2014sequence}. We plan to leverage recent advances in neural networks based sequence to sequence methods for this task~\cite{zhang2018sequence, nguyen2020improving}.
    \item Our current prototype requires external power. Recent advances in the community have leveraged triboelectric nanogenerator to develop self powered acoustic sensor~\cite{arora2018saturn}. In the future, we would like to explore these recent advances towards making the system self-powered.
    \item Our work shows that lung health can be diagnosed by monitoring forceful breathing or tidal breathing via a microphone placed inside a consumer-grade N95 or cloth mask. We optimised the Convolutions Neural Network (CNN) classifier described in Section~\ref{tidal_approach} to work in real-time in the Arduino Nano 33 BLE Sense\footnote{\url{https://store-usa.arduino.cc/products/arduino-nano-33-ble-sense}} microcontroller. We used offline algorithms to deduce forced and tidal breathing parameters. We envision that a SpiroMask will feed the lung health parameters on a wearer's smartphone over Bluetooth Low Energy (BLE).
\end{itemize}

\section{Conclusion}
In this paper, we presented a system for performing spirometry and continuous respiration rate monitoring using consumer-grade N95 and cloth masks. Forced and tidal breathing are used for deriving lung health bio-markers like estimating the respiration rate and volume of exhaled air. We showed that a retrofit sensor placed inside an N95 or a two-layered cloth mask can estimate forced and tidal breathing. Our evaluation of over 48 participants of forced breathing implies that the accuracy of wearable spirometry is well within the clinically accepted range for participants with and without lung ailments. Moreover, our work is comparable to existing research on portable spirometry and requires less complex modelling, making it possible to deploy it on microcontrollers running machine learning models. Our subjective evaluation in our study population shows acceptability and ease of use compared to traditional spirometry.

\section*{Acknowledgments}
The authors thank the reviewers for their constructive and actionable feedback. The authors would also thank Prof. Mayank Goel for his insights on some of the experiments. Rishiraj is supported by the Prime Minister's Research Fellowship (PMRF) awarded by the Government of India and the Fulbright fellowship awarded jointly by the Indian and the US governments.

\bibliographystyle{ACM-Reference-Format}
\bibliography{base}


\begin{thebibliography}{92}

%%% the Web address.

\ifx \showCODEN    \undefined \def \showCODEN     #1{\unskip}     \fi
\ifx \showDOI      \undefined \def \showDOI       #1{#1}\fi
\ifx \showISBNx    \undefined \def \showISBNx     #1{\unskip}     \fi
\ifx \showISBNxiii \undefined \def \showISBNxiii  #1{\unskip}     \fi
\ifx \showISSN     \undefined \def \showISSN      #1{\unskip}     \fi
\ifx \showLCCN     \undefined \def \showLCCN      #1{\unskip}     \fi
\ifx \shownote     \undefined \def \shownote      #1{#1}          \fi
\ifx \showarticletitle \undefined \def \showarticletitle #1{#1}   \fi
\ifx \showURL      \undefined \def \showURL       {\relax}        \fi
% invisible to TeX
\providecommand\bibfield[2]{#2}
\providecommand\bibinfo[2]{#2}
\providecommand\natexlab[1]{#1}
\providecommand\showeprint[2][]{arXiv:#2}

\bibitem[Adhikary et~al\mbox{.}(2020)]%
        {adhikary2020naqaab}
\bibfield{author}{\bibinfo{person}{Rishiraj Adhikary}, \bibinfo{person}{Tanmay
  Srivastava}, \bibinfo{person}{Prerna Khanna}, \bibinfo{person}{Aabhas~Asit
  Senapati}, {and} \bibinfo{person}{Nipun Batra}.}
  \bibinfo{year}{2020}\natexlab{}.
\newblock \showarticletitle{Naqaab: towards health sensing and persuasion via
  masks}. In \bibinfo{booktitle}{\emph{Adjunct Proceedings of the 2020 ACM
  International Joint Conference on Pervasive and Ubiquitous Computing and
  Proceedings of the 2020 ACM International Symposium on Wearable Computers}}.
  \bibinfo{pages}{5--8}.
\newblock


\bibitem[Adhikary et~al\mbox{.}(2022)]%
        {adhikary2022Towards}
\bibfield{author}{\bibinfo{person}{Rishiraj Adhikary}, \bibinfo{person}{Aryan
  Varshney}, {and} \bibinfo{person}{Nipun Batra}.}
  \bibinfo{year}{2022}\natexlab{}.
\newblock \showarticletitle{Towards Continuous Respiration Rate Detection While
  Walking}. In \bibinfo{booktitle}{\emph{Adjunct Proceedings of the 2022 ACM
  International Joint Conference on Pervasive and Ubiquitous Computing and
  Proceedings of the 2022 ACM International Symposium on Wearable Computers}}.
\newblock


\bibitem[Aliverti(2017)]%
        {aliverti2017wearable}
\bibfield{author}{\bibinfo{person}{Andrea Aliverti}.}
  \bibinfo{year}{2017}\natexlab{}.
\newblock \showarticletitle{Wearable technology: role in respiratory health and
  disease}.
\newblock \bibinfo{journal}{\emph{Breathe}} \bibinfo{volume}{13},
  \bibinfo{number}{2} (\bibinfo{year}{2017}), \bibinfo{pages}{e27--e36}.
\newblock


\bibitem[Aly and Youssef(2016)]%
        {aly2016zephyr}
\bibfield{author}{\bibinfo{person}{Heba Aly} {and} \bibinfo{person}{Moustafa
  Youssef}.} \bibinfo{year}{2016}\natexlab{}.
\newblock \showarticletitle{Zephyr: Ubiquitous accurate multi-sensor
  fusion-based respiratory rate estimation using smartphones}. In
  \bibinfo{booktitle}{\emph{IEEE INFOCOM 2016-The 35th Annual IEEE
  International Conference on Computer Communications}}. IEEE,
  \bibinfo{pages}{1--9}.
\newblock


\bibitem[Arora et~al\mbox{.}(2018)]%
        {arora2018saturn}
\bibfield{author}{\bibinfo{person}{Nivedita Arora}, \bibinfo{person}{Steven~L
  Zhang}, \bibinfo{person}{Fereshteh Shahmiri}, \bibinfo{person}{Diego Osorio},
  \bibinfo{person}{Yi-Cheng Wang}, \bibinfo{person}{Mohit Gupta},
  \bibinfo{person}{Zhengjun Wang}, \bibinfo{person}{Thad Starner},
  \bibinfo{person}{Zhong~Lin Wang}, {and} \bibinfo{person}{Gregory~D Abowd}.}
  \bibinfo{year}{2018}\natexlab{}.
\newblock \showarticletitle{SATURN: A thin and flexible self-powered microphone
  leveraging triboelectric nanogenerator}.
\newblock \bibinfo{journal}{\emph{Proceedings of the ACM on Interactive,
  Mobile, Wearable and Ubiquitous Technologies}} \bibinfo{volume}{2},
  \bibinfo{number}{2} (\bibinfo{year}{2018}), \bibinfo{pages}{1--28}.
\newblock


\bibitem[Barandas et~al\mbox{.}(2020)]%
        {barandas2020tsfel}
\bibfield{author}{\bibinfo{person}{Mar{\'\i}lia Barandas},
  \bibinfo{person}{Duarte Folgado}, \bibinfo{person}{Let{\'\i}cia Fernandes},
  \bibinfo{person}{Sara Santos}, \bibinfo{person}{Mariana Abreu},
  \bibinfo{person}{Patr{\'\i}cia Bota}, \bibinfo{person}{Hui Liu},
  \bibinfo{person}{Tanja Schultz}, {and} \bibinfo{person}{Hugo Gamboa}.}
  \bibinfo{year}{2020}\natexlab{}.
\newblock \showarticletitle{TSFEL: Time series feature extraction library}.
\newblock \bibinfo{journal}{\emph{SoftwareX}}  \bibinfo{volume}{11}
  (\bibinfo{year}{2020}), \bibinfo{pages}{100456}.
\newblock


\bibitem[Bland and Altman(1986)]%
        {bland1986statistical}
\bibfield{author}{\bibinfo{person}{J~Martin Bland} {and}
  \bibinfo{person}{DouglasG Altman}.} \bibinfo{year}{1986}\natexlab{}.
\newblock \showarticletitle{Statistical methods for assessing agreement between
  two methods of clinical measurement}.
\newblock \bibinfo{journal}{\emph{The lancet}} \bibinfo{volume}{327},
  \bibinfo{number}{8476} (\bibinfo{year}{1986}), \bibinfo{pages}{307--310}.
\newblock


\bibitem[Brakema et~al\mbox{.}(2019)]%
        {brakema2019copd}
\bibfield{author}{\bibinfo{person}{EA Brakema}, \bibinfo{person}{FA
  Van~Gemert}, \bibinfo{person}{RMJJ Van~der Kleij}, \bibinfo{person}{S Salvi},
  \bibinfo{person}{Milo Puhan}, {and} \bibinfo{person}{NH Chavannes}.}
  \bibinfo{year}{2019}\natexlab{}.
\newblock \showarticletitle{COPD’s early origins in low-and-middle income
  countries: what are the implications of a false start?}
\newblock \bibinfo{journal}{\emph{NPJ primary care respiratory medicine}}
  \bibinfo{volume}{29}, \bibinfo{number}{1} (\bibinfo{year}{2019}),
  \bibinfo{pages}{1--3}.
\newblock


\bibitem[Carey and Foundation(2010)]%
        {carey2010current}
\bibfield{author}{\bibinfo{person}{W.D. Carey} {and}
  \bibinfo{person}{Cleveland~Clinic Foundation}.}
  \bibinfo{year}{2010}\natexlab{}.
\newblock \bibinfo{booktitle}{\emph{Current Clinical Medicine}}.
\newblock \bibinfo{publisher}{Saunders Elsevier}.
\newblock
\showISBNx{9781416066439}
\showLCCN{2010019237}
\urldef\tempurl%
\url{https://books.google.co.in/books?id=qo-rpwAACAAJ}
\showURL{%
\tempurl}


\bibitem[Chatterjee et~al\mbox{.}(2019)]%
        {chatterjee2019mlung++}
\bibfield{author}{\bibinfo{person}{Soujanya Chatterjee},
  \bibinfo{person}{Md~Mahbubur Rahman}, \bibinfo{person}{Ebrahim Nemati},
  \bibinfo{person}{Viswam Nathan}, \bibinfo{person}{Korosh Vatanparvar}, {and}
  \bibinfo{person}{Jilong Kuang}.} \bibinfo{year}{2019}\natexlab{}.
\newblock \showarticletitle{mLung++ automated characterization of abnormal lung
  sounds in pulmonary patients using multimodal mobile sensors}. In
  \bibinfo{booktitle}{\emph{Adjunct Proceedings of the 2019 ACM International
  Joint Conference on Pervasive and Ubiquitous Computing and Proceedings of the
  2019 ACM International Symposium on Wearable Computers}}.
  \bibinfo{pages}{474--481}.
\newblock


\bibitem[Chauhan et~al\mbox{.}(2017)]%
        {chauhan2017breathprint}
\bibfield{author}{\bibinfo{person}{Jagmohan Chauhan}, \bibinfo{person}{Yining
  Hu}, \bibinfo{person}{Suranga Seneviratne}, \bibinfo{person}{Archan Misra},
  \bibinfo{person}{Aruna Seneviratne}, {and} \bibinfo{person}{Youngki Lee}.}
  \bibinfo{year}{2017}\natexlab{}.
\newblock \showarticletitle{BreathPrint: Breathing acoustics-based user
  authentication}. In \bibinfo{booktitle}{\emph{Proceedings of the 15th Annual
  International Conference on Mobile Systems, Applications, and Services}}.
  \bibinfo{pages}{278--291}.
\newblock


\bibitem[Chu et~al\mbox{.}(2019)]%
        {chu2019respiration}
\bibfield{author}{\bibinfo{person}{Michael Chu}, \bibinfo{person}{Thao Nguyen},
  \bibinfo{person}{Vaibhav Pandey}, \bibinfo{person}{Yongxiao Zhou},
  \bibinfo{person}{Hoang~N Pham}, \bibinfo{person}{Ronen Bar-Yoseph},
  \bibinfo{person}{Shlomit Radom-Aizik}, \bibinfo{person}{Ramesh Jain},
  \bibinfo{person}{Dan~M Cooper}, {and} \bibinfo{person}{Michelle Khine}.}
  \bibinfo{year}{2019}\natexlab{}.
\newblock \showarticletitle{Respiration rate and volume measurements using
  wearable strain sensors}.
\newblock \bibinfo{journal}{\emph{NPJ digital medicine}} \bibinfo{volume}{2},
  \bibinfo{number}{1} (\bibinfo{year}{2019}), \bibinfo{pages}{1--9}.
\newblock


\bibitem[Culver et~al\mbox{.}(2017)]%
        {culver2017recommendations}
\bibfield{author}{\bibinfo{person}{Bruce~H Culver}, \bibinfo{person}{Brian~L
  Graham}, \bibinfo{person}{Allan~L Coates}, \bibinfo{person}{Jack Wanger},
  \bibinfo{person}{Cristine~E Berry}, \bibinfo{person}{Patricia~K Clarke},
  \bibinfo{person}{Teal~S Hallstrand}, \bibinfo{person}{John~L Hankinson},
  \bibinfo{person}{David~A Kaminsky}, \bibinfo{person}{Neil~R MacIntyre},
  {et~al\mbox{.}}} \bibinfo{year}{2017}\natexlab{}.
\newblock \showarticletitle{Recommendations for a standardized pulmonary
  function report. An official American Thoracic Society technical statement}.
\newblock \bibinfo{journal}{\emph{American journal of respiratory and critical
  care medicine}} \bibinfo{volume}{196}, \bibinfo{number}{11}
  (\bibinfo{year}{2017}), \bibinfo{pages}{1463--1472}.
\newblock


\bibitem[Curtiss et~al\mbox{.}(2021)]%
        {curtiss2021facebit}
\bibfield{author}{\bibinfo{person}{Alexander Curtiss}, \bibinfo{person}{Blaine
  Rothrock}, \bibinfo{person}{Abu Bakar}, \bibinfo{person}{Nivedita Arora},
  \bibinfo{person}{Jason Huang}, \bibinfo{person}{Zachary Englhardt},
  \bibinfo{person}{Aaron-Patrick Empedrado}, \bibinfo{person}{Chixiang Wang},
  \bibinfo{person}{Saad Ahmed}, \bibinfo{person}{Yang Zhang}, {et~al\mbox{.}}}
  \bibinfo{year}{2021}\natexlab{}.
\newblock \showarticletitle{FaceBit: Smart Face Masks Platform}.
\newblock \bibinfo{journal}{\emph{Proceedings of the ACM on Interactive,
  Mobile, Wearable and Ubiquitous Technologies}} \bibinfo{volume}{5},
  \bibinfo{number}{4} (\bibinfo{year}{2021}), \bibinfo{pages}{1--44}.
\newblock


\bibitem[Dai et~al\mbox{.}(2021)]%
        {dai2021respwatch}
\bibfield{author}{\bibinfo{person}{Ruixuan Dai}, \bibinfo{person}{Chenyang Lu},
  \bibinfo{person}{Michael Avidan}, {and} \bibinfo{person}{Thomas
  Kannampallil}.} \bibinfo{year}{2021}\natexlab{}.
\newblock \showarticletitle{RespWatch: Robust Measurement of Respiratory Rate
  on Smartwatches with Photoplethysmography}. In
  \bibinfo{booktitle}{\emph{Proceedings of the International Conference on
  Internet-of-Things Design and Implementation}}. \bibinfo{pages}{208--220}.
\newblock


\bibitem[Downey(2016)]%
        {downey2016think}
\bibfield{author}{\bibinfo{person}{Allen Downey}.}
  \bibinfo{year}{2016}\natexlab{}.
\newblock \bibinfo{booktitle}{\emph{Think DSP: digital signal processing in
  Python}}.
\newblock \bibinfo{publisher}{" O'Reilly Media, Inc."}.
\newblock


\bibitem[Fieselmann et~al\mbox{.}(1993)]%
        {fieselmann1993respiratory}
\bibfield{author}{\bibinfo{person}{John~F Fieselmann},
  \bibinfo{person}{Michael~S Hendryx}, \bibinfo{person}{Charles~M Helms}, {and}
  \bibinfo{person}{Douglas~S Wakefield}.} \bibinfo{year}{1993}\natexlab{}.
\newblock \showarticletitle{Respiratory rate predicts cardiopulmonary arrest
  for internal medicine inpatients}.
\newblock \bibinfo{journal}{\emph{Journal of general internal medicine}}
  \bibinfo{volume}{8}, \bibinfo{number}{7} (\bibinfo{year}{1993}),
  \bibinfo{pages}{354--360}.
\newblock


\bibitem[French and Steinberg(1947)]%
        {french1947factors}
\bibfield{author}{\bibinfo{person}{Norman~R French} {and}
  \bibinfo{person}{John~C Steinberg}.} \bibinfo{year}{1947}\natexlab{}.
\newblock \showarticletitle{Factors governing the intelligibility of speech
  sounds}.
\newblock \bibinfo{journal}{\emph{The journal of the Acoustical society of
  America}} \bibinfo{volume}{19}, \bibinfo{number}{1} (\bibinfo{year}{1947}),
  \bibinfo{pages}{90--119}.
\newblock


\bibitem[Goel(2016)]%
        {mayankgoel-thesis}
\bibfield{author}{\bibinfo{person}{Mayank Goel}.}
  \bibinfo{year}{2016}\natexlab{}.
\newblock \bibinfo{title}{{Extending the capabilities of smartphone sensors for
  applications in interaction and health}}.
\newblock
  \bibinfo{howpublished}{\url{https://digital.lib.washington.edu/researchworks/handle/1773/37080}}.
\newblock
\newblock
\shownote{[Online; accessed 10-April-2022]}.


\bibitem[Goel et~al\mbox{.}(2016)]%
        {goel2016spirocall}
\bibfield{author}{\bibinfo{person}{Mayank Goel}, \bibinfo{person}{Elliot Saba},
  \bibinfo{person}{Maia Stiber}, \bibinfo{person}{Eric Whitmire},
  \bibinfo{person}{Josh Fromm}, \bibinfo{person}{Eric~C Larson},
  \bibinfo{person}{Gaetano Borriello}, {and} \bibinfo{person}{Shwetak~N
  Patel}.} \bibinfo{year}{2016}\natexlab{}.
\newblock \showarticletitle{Spirocall: Measuring lung function over a phone
  call}. In \bibinfo{booktitle}{\emph{Proceedings of the 2016 CHI conference on
  human factors in computing systems}}. \bibinfo{pages}{5675--5685}.
\newblock


\bibitem[Graham et~al\mbox{.}(2019)]%
        {graham2019standardization}
\bibfield{author}{\bibinfo{person}{Brian~L Graham}, \bibinfo{person}{Irene
  Steenbruggen}, \bibinfo{person}{Martin~R Miller}, \bibinfo{person}{Igor~Z
  Barjaktarevic}, \bibinfo{person}{Brendan~G Cooper}, \bibinfo{person}{Graham~L
  Hall}, \bibinfo{person}{Teal~S Hallstrand}, \bibinfo{person}{David~A
  Kaminsky}, \bibinfo{person}{Kevin McCarthy}, \bibinfo{person}{Meredith~C
  McCormack}, {et~al\mbox{.}}} \bibinfo{year}{2019}\natexlab{}.
\newblock \showarticletitle{Standardization of spirometry 2019 update. An
  official American thoracic Society and European respiratory Society technical
  statement}.
\newblock \bibinfo{journal}{\emph{American journal of respiratory and critical
  care medicine}} \bibinfo{volume}{200}, \bibinfo{number}{8}
  (\bibinfo{year}{2019}), \bibinfo{pages}{e70--e88}.
\newblock


\bibitem[Hernandez et~al\mbox{.}(2014)]%
        {hernandez2014bioglass}
\bibfield{author}{\bibinfo{person}{Javier Hernandez}, \bibinfo{person}{Yin Li},
  \bibinfo{person}{James~M Rehg}, {and} \bibinfo{person}{Rosalind~W Picard}.}
  \bibinfo{year}{2014}\natexlab{}.
\newblock \showarticletitle{Bioglass: Physiological parameter estimation using
  a head-mounted wearable device}. In \bibinfo{booktitle}{\emph{2014 4th
  International Conference on Wireless Mobile Communication and
  Healthcare-Transforming Healthcare Through Innovations in Mobile and Wireless
  Technologies (MOBIHEALTH)}}. IEEE, \bibinfo{pages}{55--58}.
\newblock


\bibitem[Hernandez et~al\mbox{.}(2015)]%
        {hernandez2015biowatch}
\bibfield{author}{\bibinfo{person}{Javier Hernandez}, \bibinfo{person}{Daniel
  McDuff}, {and} \bibinfo{person}{Rosalind~W Picard}.}
  \bibinfo{year}{2015}\natexlab{}.
\newblock \showarticletitle{Biowatch: estimation of heart and breathing rates
  from wrist motions}. In \bibinfo{booktitle}{\emph{2015 9th International
  Conference on Pervasive Computing Technologies for Healthcare
  (PervasiveHealth)}}. IEEE, \bibinfo{pages}{169--176}.
\newblock


\bibitem[Hodgetts et~al\mbox{.}(2002)]%
        {hodgetts2002identification}
\bibfield{author}{\bibinfo{person}{Timothy~J Hodgetts}, \bibinfo{person}{Gary
  Kenward}, \bibinfo{person}{Ioannis~G Vlachonikolis}, \bibinfo{person}{Susan
  Payne}, {and} \bibinfo{person}{Nicolas Castle}.}
  \bibinfo{year}{2002}\natexlab{}.
\newblock \showarticletitle{The identification of risk factors for cardiac
  arrest and formulation of activation criteria to alert a medical emergency
  team}.
\newblock \bibinfo{journal}{\emph{Resuscitation}} \bibinfo{volume}{54},
  \bibinfo{number}{2} (\bibinfo{year}{2002}), \bibinfo{pages}{125--131}.
\newblock


\bibitem[Islam et~al\mbox{.}(2021)]%
        {islam2021breathtrack}
\bibfield{author}{\bibinfo{person}{Bashima Islam}, \bibinfo{person}{Md~Mahbubur
  Rahman}, \bibinfo{person}{Tousif Ahmed}, \bibinfo{person}{Mohsin~Yusuf
  Ahmed}, \bibinfo{person}{Md~Mehedi Hasan}, \bibinfo{person}{Viswam Nathan},
  \bibinfo{person}{Korosh Vatanparvar}, \bibinfo{person}{Ebrahim Nemati},
  \bibinfo{person}{Jilong Kuang}, {and} \bibinfo{person}{Jun~Alex Gao}.}
  \bibinfo{year}{2021}\natexlab{}.
\newblock \showarticletitle{BreathTrack: Detecting Regular Breathing Phases
  from Unannotated Acoustic Data Captured by a Smartphone}.
\newblock \bibinfo{journal}{\emph{Proceedings of the ACM on Interactive,
  Mobile, Wearable and Ubiquitous Technologies}} \bibinfo{volume}{5},
  \bibinfo{number}{3} (\bibinfo{year}{2021}), \bibinfo{pages}{1--22}.
\newblock


\bibitem[James et~al\mbox{.}(2013)]%
        {james2013introduction}
\bibfield{author}{\bibinfo{person}{Gareth James}, \bibinfo{person}{Daniela
  Witten}, \bibinfo{person}{Trevor Hastie}, {and} \bibinfo{person}{Robert
  Tibshirani}.} \bibinfo{year}{2013}\natexlab{}.
\newblock \bibinfo{booktitle}{\emph{An introduction to statistical learning}}.
  Vol.~\bibinfo{volume}{112}.
\newblock \bibinfo{publisher}{Springer}.
\newblock


\bibitem[Jan~Jongboom(2019)]%
        {edgeimpulse}
\bibfield{author}{\bibinfo{person}{Zach~Shelby Jan~Jongboom}.}
  \bibinfo{year}{2019}\natexlab{}.
\newblock \bibinfo{title}{{Edge Impulse}}.
\newblock \bibinfo{howpublished}{\url{https://www.edgeimpulse.com/}}.
\newblock
\newblock
\shownote{[Online; accessed 13-April-2022]}.


\bibitem[Johansson(1999)]%
        {johansson1999hilbert}
\bibfield{author}{\bibinfo{person}{Mathias Johansson}.}
  \bibinfo{year}{1999}\natexlab{}.
\newblock \showarticletitle{The hilbert transform}.
\newblock \bibinfo{journal}{\emph{Mathematics Master’s Thesis. V{\"a}xj{\"o}
  University, Suecia. Disponible en internet: http://w3. msi. vxu.
  se/exarb/mj\_ex. pdf, consultado el}}  \bibinfo{volume}{19}
  (\bibinfo{year}{1999}).
\newblock


\bibitem[Johnson et~al\mbox{.}(2020)]%
        {johnson2020impact}
\bibfield{author}{\bibinfo{person}{Karin~G Johnson}, \bibinfo{person}{Shannon~S
  Sullivan}, \bibinfo{person}{Afua Nti}, \bibinfo{person}{Vida Rastegar}, {and}
  \bibinfo{person}{Indira Gurubhagavatula}.} \bibinfo{year}{2020}\natexlab{}.
\newblock \showarticletitle{The impact of the COVID-19 pandemic on sleep
  medicine practices}.
\newblock \bibinfo{journal}{\emph{Journal of Clinical Sleep Medicine}}
  (\bibinfo{year}{2020}), \bibinfo{pages}{jcsm--8830}.
\newblock


\bibitem[Jongboom(2022)]%
        {ei-forum-1}
\bibfield{author}{\bibinfo{person}{Jan Jongboom}.}
  \bibinfo{year}{2022}\natexlab{}.
\newblock \bibinfo{title}{{Edge Impulse Forum}}.
\newblock
  \bibinfo{howpublished}{\url{https://forum.edgeimpulse.com/t/what-pre-processing-step-is-applied-to-raw-audio-samples-by-edge-impulse/2706/4?u=rishi}}.
\newblock
\newblock
\shownote{[Online; accessed 7-May-2022]}.


\bibitem[Kiranyaz et~al\mbox{.}(2019)]%
        {kiranyaz20191}
\bibfield{author}{\bibinfo{person}{Serkan Kiranyaz}, \bibinfo{person}{Turker
  Ince}, \bibinfo{person}{Osama Abdeljaber}, \bibinfo{person}{Onur Avci}, {and}
  \bibinfo{person}{Moncef Gabbouj}.} \bibinfo{year}{2019}\natexlab{}.
\newblock \showarticletitle{1-d convolutional neural networks for signal
  processing applications}. In \bibinfo{booktitle}{\emph{ICASSP 2019-2019 IEEE
  International Conference on Acoustics, Speech and Signal Processing
  (ICASSP)}}. IEEE, \bibinfo{pages}{8360--8364}.
\newblock


\bibitem[Knudson et~al\mbox{.}(1976)]%
        {knudson1976maximal}
\bibfield{author}{\bibinfo{person}{Ronald~J Knudson}, \bibinfo{person}{Ronald~C
  Slatin}, \bibinfo{person}{Michael~D Lebowitz}, {and}
  \bibinfo{person}{Benjamin Burrows}.} \bibinfo{year}{1976}\natexlab{}.
\newblock \showarticletitle{The maximal expiratory flow-volume curve: normal
  standards, variability, and effects of age}.
\newblock \bibinfo{journal}{\emph{American Review of Respiratory Disease}}
  \bibinfo{volume}{113}, \bibinfo{number}{5} (\bibinfo{year}{1976}),
  \bibinfo{pages}{587--600}.
\newblock


\bibitem[Kouri et~al\mbox{.}(2020)]%
        {kouri2020chest}
\bibfield{author}{\bibinfo{person}{Andrew Kouri}, \bibinfo{person}{Samir
  Gupta}, \bibinfo{person}{Azadeh Yadollahi}, \bibinfo{person}{Clodagh~M Ryan},
  \bibinfo{person}{Andrea~S Gershon}, \bibinfo{person}{Teresa To},
  \bibinfo{person}{Susan~M Tarlo}, \bibinfo{person}{Roger~S Goldstein},
  \bibinfo{person}{Kenneth~R Chapman}, {and} \bibinfo{person}{Chung-Wai Chow}.}
  \bibinfo{year}{2020}\natexlab{}.
\newblock \showarticletitle{CHEST Reviews: Addressing reduced laboratory-based
  pulmonary function testing during a pandemic}.
\newblock \bibinfo{journal}{\emph{Chest}} (\bibinfo{year}{2020}).
\newblock


\bibitem[Kristiansen et~al\mbox{.}(2021)]%
        {kristiansen2021machine}
\bibfield{author}{\bibinfo{person}{Stein Kristiansen},
  \bibinfo{person}{Konstantinos Nikolaidis}, \bibinfo{person}{Thomas
  Plagemann}, \bibinfo{person}{Vera Goebel}, \bibinfo{person}{Gunn~Marit
  Traaen}, \bibinfo{person}{Britt {\O}verland}, \bibinfo{person}{Lars
  Aaker{\o}y}, \bibinfo{person}{Tove-Elizabeth Hunt},
  \bibinfo{person}{Jan~P{\aa}l Loennechen}, \bibinfo{person}{Sigurd~Loe
  Steinshamn}, {et~al\mbox{.}}} \bibinfo{year}{2021}\natexlab{}.
\newblock \showarticletitle{Machine Learning for Sleep Apnea Detection with
  Unattended Sleep Monitoring at Home}.
\newblock \bibinfo{journal}{\emph{ACM Transactions on Computing for
  Healthcare}} \bibinfo{volume}{2}, \bibinfo{number}{2} (\bibinfo{year}{2021}),
  \bibinfo{pages}{1--25}.
\newblock


\bibitem[Kumar et~al\mbox{.}(2021)]%
        {kumar2021estimating}
\bibfield{author}{\bibinfo{person}{Agni Kumar}, \bibinfo{person}{Vikramjit
  Mitra}, \bibinfo{person}{Carolyn Oliver}, \bibinfo{person}{Adeeti Ullal},
  \bibinfo{person}{Matt Biddulph}, {and} \bibinfo{person}{Irida Mance}.}
  \bibinfo{year}{2021}\natexlab{}.
\newblock \showarticletitle{Estimating Respiratory Rate From Breath Audio
  Obtained Through Wearable Microphones}. In \bibinfo{booktitle}{\emph{2021
  43rd Annual International Conference of the IEEE Engineering in Medicine \&
  Biology Society (EMBC)}}. IEEE, \bibinfo{pages}{7310--7315}.
\newblock


\bibitem[Larson et~al\mbox{.}(2012)]%
        {larson2012spirosmart}
\bibfield{author}{\bibinfo{person}{Eric~C Larson}, \bibinfo{person}{Mayank
  Goel}, \bibinfo{person}{Gaetano Boriello}, \bibinfo{person}{Sonya Heltshe},
  \bibinfo{person}{Margaret Rosenfeld}, {and} \bibinfo{person}{Shwetak~N
  Patel}.} \bibinfo{year}{2012}\natexlab{}.
\newblock \showarticletitle{SpiroSmart: using a microphone to measure lung
  function on a mobile phone}. In \bibinfo{booktitle}{\emph{Proceedings of the
  2012 ACM Conference on ubiquitous computing}}. \bibinfo{pages}{280--289}.
\newblock


\bibitem[Larson et~al\mbox{.}(2011)]%
        {larson2011accurate}
\bibfield{author}{\bibinfo{person}{Eric~C Larson}, \bibinfo{person}{TienJui
  Lee}, \bibinfo{person}{Sean Liu}, \bibinfo{person}{Margaret Rosenfeld}, {and}
  \bibinfo{person}{Shwetak~N Patel}.} \bibinfo{year}{2011}\natexlab{}.
\newblock \showarticletitle{Accurate and privacy preserving cough sensing using
  a low-cost microphone}. In \bibinfo{booktitle}{\emph{Proceedings of the 13th
  international conference on Ubiquitous computing}}.
  \bibinfo{pages}{375--384}.
\newblock


\bibitem[Liaqat et~al\mbox{.}(2019)]%
        {liaqat2019wearbreathing}
\bibfield{author}{\bibinfo{person}{Daniyal Liaqat}, \bibinfo{person}{Mohamed
  Abdalla}, \bibinfo{person}{Pegah Abed-Esfahani}, \bibinfo{person}{Moshe
  Gabel}, \bibinfo{person}{Tatiana Son}, \bibinfo{person}{Robert Wu},
  \bibinfo{person}{Andrea Gershon}, \bibinfo{person}{Frank Rudzicz}, {and}
  \bibinfo{person}{Eyal~De Lara}.} \bibinfo{year}{2019}\natexlab{}.
\newblock \showarticletitle{WearBreathing: Real world respiratory rate
  monitoring using smartwatches}.
\newblock \bibinfo{journal}{\emph{Proceedings of the ACM on Interactive,
  Mobile, Wearable and Ubiquitous Technologies}} \bibinfo{volume}{3},
  \bibinfo{number}{2} (\bibinfo{year}{2019}), \bibinfo{pages}{1--22}.
\newblock


\bibitem[Luo et~al\mbox{.}(2017)]%
        {luo2017automatic}
\bibfield{author}{\bibinfo{person}{Andrew~Z Luo}, \bibinfo{person}{Eric
  Whitmire}, \bibinfo{person}{James~W Stout}, \bibinfo{person}{Drew Martenson},
  {and} \bibinfo{person}{Shwetak Patel}.} \bibinfo{year}{2017}\natexlab{}.
\newblock \showarticletitle{Automatic characterization of user errors in
  spirometry}. In \bibinfo{booktitle}{\emph{2017 39th Annual International
  Conference of the IEEE Engineering in Medicine and Biology Society (EMBC)}}.
  IEEE, \bibinfo{pages}{4239--4242}.
\newblock


\bibitem[Madikeri and Murthy(2011)]%
        {madikeri2011mel}
\bibfield{author}{\bibinfo{person}{Srikanth~R Madikeri} {and}
  \bibinfo{person}{Hema~A Murthy}.} \bibinfo{year}{2011}\natexlab{}.
\newblock \showarticletitle{Mel filter bank energy-based slope feature and its
  application to speaker recognition}. In \bibinfo{booktitle}{\emph{2011
  National Conference on Communications (NCC)}}. IEEE, \bibinfo{pages}{1--4}.
\newblock


\bibitem[Mariakakis et~al\mbox{.}(2019)]%
        {mariakakis2019challenges}
\bibfield{author}{\bibinfo{person}{Alex Mariakakis}, \bibinfo{person}{Edward
  Wang}, \bibinfo{person}{Shwetak Patel}, {and} \bibinfo{person}{Mayank Goel}.}
  \bibinfo{year}{2019}\natexlab{}.
\newblock \showarticletitle{Challenges in realizing smartphone-based health
  sensing}.
\newblock \bibinfo{journal}{\emph{IEEE Pervasive Computing}}
  \bibinfo{volume}{18}, \bibinfo{number}{2} (\bibinfo{year}{2019}),
  \bibinfo{pages}{76--84}.
\newblock


\bibitem[Miller et~al\mbox{.}(2005)]%
        {miller2005standardisation}
\bibfield{author}{\bibinfo{person}{Martin~R Miller}, \bibinfo{person}{JATS
  Hankinson}, \bibinfo{person}{V Brusasco}, \bibinfo{person}{F Burgos},
  \bibinfo{person}{R Casaburi}, \bibinfo{person}{A Coates}, \bibinfo{person}{R
  Crapo}, \bibinfo{person}{Pvd Enright}, \bibinfo{person}{CPM Van Der~Grinten},
  \bibinfo{person}{P Gustafsson}, {et~al\mbox{.}}}
  \bibinfo{year}{2005}\natexlab{}.
\newblock \showarticletitle{Standardisation of spirometry}.
\newblock \bibinfo{journal}{\emph{European respiratory journal}}
  \bibinfo{volume}{26}, \bibinfo{number}{2} (\bibinfo{year}{2005}),
  \bibinfo{pages}{319--338}.
\newblock


\bibitem[Morrison(1969)]%
        {morrison1969nasal}
\bibfield{author}{\bibinfo{person}{Murray~D Morrison}.}
  \bibinfo{year}{1969}\natexlab{}.
\newblock \showarticletitle{Nasal spirometry: a volumetric study of nasal air
  flow capacity}.
\newblock \bibinfo{journal}{\emph{Archives of Otolaryngology}}
  \bibinfo{volume}{90}, \bibinfo{number}{5} (\bibinfo{year}{1969}),
  \bibinfo{pages}{636--640}.
\newblock


\bibitem[Nam et~al\mbox{.}(2015)]%
        {nam2015estimation}
\bibfield{author}{\bibinfo{person}{Yunyoung Nam}, \bibinfo{person}{Bersain~A
  Reyes}, {and} \bibinfo{person}{Ki~H Chon}.} \bibinfo{year}{2015}\natexlab{}.
\newblock \showarticletitle{Estimation of respiratory rates using the built-in
  microphone of a smartphone or headset}.
\newblock \bibinfo{journal}{\emph{IEEE journal of biomedical and health
  informatics}} \bibinfo{volume}{20}, \bibinfo{number}{6}
  (\bibinfo{year}{2015}), \bibinfo{pages}{1493--1501}.
\newblock


\bibitem[Nguyen et~al\mbox{.}(2020)]%
        {nguyen2020improving}
\bibfield{author}{\bibinfo{person}{Thai-Son Nguyen}, \bibinfo{person}{Sebastian
  Stueker}, \bibinfo{person}{Jan Niehues}, {and} \bibinfo{person}{Alex
  Waibel}.} \bibinfo{year}{2020}\natexlab{}.
\newblock \showarticletitle{Improving sequence-to-sequence speech recognition
  training with on-the-fly data augmentation}. In
  \bibinfo{booktitle}{\emph{ICASSP 2020-2020 IEEE International Conference on
  Acoustics, Speech and Signal Processing (ICASSP)}}. IEEE,
  \bibinfo{pages}{7689--7693}.
\newblock


\bibitem[O'Shaughnessy(1988)]%
        {o1988linear}
\bibfield{author}{\bibinfo{person}{Douglas O'Shaughnessy}.}
  \bibinfo{year}{1988}\natexlab{}.
\newblock \showarticletitle{Linear predictive coding}.
\newblock \bibinfo{journal}{\emph{IEEE potentials}} \bibinfo{volume}{7},
  \bibinfo{number}{1} (\bibinfo{year}{1988}), \bibinfo{pages}{29--32}.
\newblock


\bibitem[Otulana et~al\mbox{.}(1990)]%
        {otulana1990use}
\bibfield{author}{\bibinfo{person}{Babatunde~A Otulana}, \bibinfo{person}{Tim
  Higenbottam}, \bibinfo{person}{Lilie Ferrari}, \bibinfo{person}{John Scott},
  \bibinfo{person}{Gilbert Igboaka}, {and} \bibinfo{person}{John Wallwork}.}
  \bibinfo{year}{1990}\natexlab{}.
\newblock \showarticletitle{The use of home spirometry in detecting acute lung
  rejection and infection following heart-lung transplantation}.
\newblock \bibinfo{journal}{\emph{Chest}} \bibinfo{volume}{97},
  \bibinfo{number}{2} (\bibinfo{year}{1990}), \bibinfo{pages}{353--357}.
\newblock


\bibitem[Palmer et~al\mbox{.}(2004)]%
        {palmer2004tidal}
\bibfield{author}{\bibinfo{person}{John Palmer}, \bibinfo{person}{Julian
  Allen}, {and} \bibinfo{person}{Oscar Mayer}.}
  \bibinfo{year}{2004}\natexlab{}.
\newblock \showarticletitle{Tidal breathing analysis}.
\newblock \bibinfo{journal}{\emph{NeoReviews}} \bibinfo{volume}{5},
  \bibinfo{number}{5} (\bibinfo{year}{2004}), \bibinfo{pages}{e186--e193}.
\newblock


\bibitem[Potamianos and Maragos(1994)]%
        {potamianos1994comparison}
\bibfield{author}{\bibinfo{person}{Alexandros Potamianos} {and}
  \bibinfo{person}{Petros Maragos}.} \bibinfo{year}{1994}\natexlab{}.
\newblock \showarticletitle{A comparison of the energy operator and the Hilbert
  transform approach to signal and speech demodulation}.
\newblock \bibinfo{journal}{\emph{Signal processing}} \bibinfo{volume}{37},
  \bibinfo{number}{1} (\bibinfo{year}{1994}), \bibinfo{pages}{95--120}.
\newblock


\bibitem[Rahman et~al\mbox{.}(2020)]%
        {rahman2020breatheasy}
\bibfield{author}{\bibinfo{person}{Md~Mahbubur Rahman},
  \bibinfo{person}{Mohsin~Yusuf Ahmed}, \bibinfo{person}{Tousif Ahmed},
  \bibinfo{person}{Bashima Islam}, \bibinfo{person}{Viswam Nathan},
  \bibinfo{person}{Korosh Vatanparvar}, \bibinfo{person}{Ebrahim Nemati},
  \bibinfo{person}{Daniel McCaffrey}, \bibinfo{person}{Jilong Kuang}, {and}
  \bibinfo{person}{Jun~Alex Gao}.} \bibinfo{year}{2020}\natexlab{}.
\newblock \showarticletitle{BreathEasy: Assessing Respiratory Diseases Using
  Mobile Multimodal Sensors}. In \bibinfo{booktitle}{\emph{Proceedings of the
  2020 International Conference on Multimodal Interaction}}.
  \bibinfo{pages}{41--49}.
\newblock


\bibitem[Rahman et~al\mbox{.}(2019)]%
        {rahman2019towards}
\bibfield{author}{\bibinfo{person}{Md~Mahbubur Rahman}, \bibinfo{person}{Viswam
  Nathan}, \bibinfo{person}{Ebrahim Nemati}, \bibinfo{person}{Korosh
  Vatanparvar}, \bibinfo{person}{Mohsin Ahmed}, {and} \bibinfo{person}{Jilong
  Kuang}.} \bibinfo{year}{2019}\natexlab{}.
\newblock \showarticletitle{Towards reliable data collection and annotation to
  extract pulmonary digital biomarkers using mobile sensors}. In
  \bibinfo{booktitle}{\emph{Proceedings of the 13th EAI International
  Conference on Pervasive Computing Technologies for Healthcare}}.
  \bibinfo{pages}{179--188}.
\newblock


\bibitem[Rahman et~al\mbox{.}(2014)]%
        {rahman2014bodybeat}
\bibfield{author}{\bibinfo{person}{Tauhidur Rahman},
  \bibinfo{person}{Alexander~Travis Adams}, \bibinfo{person}{Mi Zhang},
  \bibinfo{person}{Erin Cherry}, \bibinfo{person}{Bobby Zhou},
  \bibinfo{person}{Huaishu Peng}, {and} \bibinfo{person}{Tanzeem Choudhury}.}
  \bibinfo{year}{2014}\natexlab{}.
\newblock \showarticletitle{BodyBeat: a mobile system for sensing non-speech
  body sounds.}. In \bibinfo{booktitle}{\emph{MobiSys}},
  Vol.~\bibinfo{volume}{14}. Citeseer, \bibinfo{pages}{2--594}.
\newblock


\bibitem[R{\"o}ddiger et~al\mbox{.}(2019)]%
        {roddiger2019towards}
\bibfield{author}{\bibinfo{person}{Tobias R{\"o}ddiger},
  \bibinfo{person}{Daniel Wolffram}, \bibinfo{person}{David Laubenstein},
  \bibinfo{person}{Matthias Budde}, {and} \bibinfo{person}{Michael Beigl}.}
  \bibinfo{year}{2019}\natexlab{}.
\newblock \showarticletitle{Towards respiration rate monitoring using an in-ear
  headphone inertial measurement unit}. In
  \bibinfo{booktitle}{\emph{Proceedings of the 1st International Workshop on
  Earable Computing}}. \bibinfo{pages}{48--53}.
\newblock


\bibitem[Rubini et~al\mbox{.}(2010)]%
        {rubini2010daily}
\bibfield{author}{\bibinfo{person}{Alessandro Rubini}, \bibinfo{person}{Andrea
  Parmagnani}, \bibinfo{person}{Marco Redaelli}, \bibinfo{person}{Michela
  Bond{\`\i}}, \bibinfo{person}{Daniele Del~Monte}, {and}
  \bibinfo{person}{Vincenzo Catena}.} \bibinfo{year}{2010}\natexlab{}.
\newblock \showarticletitle{Daily variations of spirometric indexes and maximum
  expiratory pressure in young healthy adults}.
\newblock \bibinfo{journal}{\emph{Biological Rhythm Research}}
  \bibinfo{volume}{41}, \bibinfo{number}{2} (\bibinfo{year}{2010}),
  \bibinfo{pages}{105--112}.
\newblock


\bibitem[Schluger and Koppaka(2014)]%
        {schluger2014lung}
\bibfield{author}{\bibinfo{person}{Neil~W Schluger} {and} \bibinfo{person}{Ram
  Koppaka}.} \bibinfo{year}{2014}\natexlab{}.
\newblock \showarticletitle{Lung disease in a global context. A call for public
  health action}.
\newblock \bibinfo{journal}{\emph{Annals of the American Thoracic Society}}
  \bibinfo{volume}{11}, \bibinfo{number}{3} (\bibinfo{year}{2014}),
  \bibinfo{pages}{407--416}.
\newblock


\bibitem[Schuller et~al\mbox{.}(2010)]%
        {schuller2010interspeech}
\bibfield{author}{\bibinfo{person}{Bj{\"o}rn Schuller}, \bibinfo{person}{Stefan
  Steidl}, \bibinfo{person}{Anton Batliner}, \bibinfo{person}{Felix Burkhardt},
  \bibinfo{person}{Laurence Devillers}, \bibinfo{person}{Christian M{\"u}ller},
  {and} \bibinfo{person}{Shrikanth~S Narayanan}.}
  \bibinfo{year}{2010}\natexlab{}.
\newblock \showarticletitle{The INTERSPEECH 2010 paralinguistic challenge}. In
  \bibinfo{booktitle}{\emph{Eleventh Annual Conference of the International
  Speech Communication Association}}.
\newblock


\bibitem[Semiconductor(2022)]%
        {nordic-blog-1}
\bibfield{author}{\bibinfo{person}{Nordic Semiconductor}.}
  \bibinfo{year}{2022}\natexlab{}.
\newblock \bibinfo{title}{{Bluetooth Low Energy data throughput}}.
\newblock
  \bibinfo{howpublished}{\url{https://infocenter.nordicsemi.com/index.jsp?topic=\%2Fsds_s140\%2FSDS\%2Fs1xx\%2Fble_data_throughput\%2Fble_data_throughput.html}}.
\newblock
\newblock
\shownote{[Online; accessed 7-May-2022]}.


\bibitem[Semjen et~al\mbox{.}(1998)]%
        {semjen1998getting}
\bibfield{author}{\bibinfo{person}{Andras Semjen}, \bibinfo{person}{Dirk
  Vorberg}, {and} \bibinfo{person}{Hans-Henning Schulze}.}
  \bibinfo{year}{1998}\natexlab{}.
\newblock \showarticletitle{Getting synchronized with the metronome:
  Comparisons between phase and period correction}.
\newblock \bibinfo{journal}{\emph{Psychological Research}}
  \bibinfo{volume}{61}, \bibinfo{number}{1} (\bibinfo{year}{1998}),
  \bibinfo{pages}{44--55}.
\newblock


\bibitem[Seren(2005)]%
        {seren2005frequency}
\bibfield{author}{\bibinfo{person}{Erdal Seren}.}
  \bibinfo{year}{2005}\natexlab{}.
\newblock \showarticletitle{Frequency spectra of normal expiratory nasal
  sound}.
\newblock \bibinfo{journal}{\emph{American journal of rhinology}}
  \bibinfo{volume}{19}, \bibinfo{number}{3} (\bibinfo{year}{2005}),
  \bibinfo{pages}{257--261}.
\newblock


\bibitem[Seydnejad and Kitney(1997)]%
        {seydnejad1997real}
\bibfield{author}{\bibinfo{person}{Saeid~Reza Seydnejad} {and}
  \bibinfo{person}{Richard~I Kitney}.} \bibinfo{year}{1997}\natexlab{}.
\newblock \showarticletitle{Real-time heart rate variability extraction using
  the Kaiser window}.
\newblock \bibinfo{journal}{\emph{IEEE Transactions on Biomedical Engineering}}
  \bibinfo{volume}{44}, \bibinfo{number}{10} (\bibinfo{year}{1997}),
  \bibinfo{pages}{990--1005}.
\newblock


\bibitem[Shenoi(2006)]%
        {shenoi2006introduction}
\bibfield{author}{\bibinfo{person}{Belle~A Shenoi}.}
  \bibinfo{year}{2006}\natexlab{}.
\newblock \bibinfo{booktitle}{\emph{Introduction to digital signal: Processing
  and filter design}}.
\newblock \bibinfo{publisher}{Wiley Online Library}.
\newblock


\bibitem[Shih et~al\mbox{.}(2019)]%
        {shih2019breeze}
\bibfield{author}{\bibinfo{person}{Chen-Hsuan Shih}, \bibinfo{person}{Naofumi
  Tomita}, \bibinfo{person}{Yanick~X Lukic},
  \bibinfo{person}{{\'A}lvaro~Hern{\'a}ndez Reguera}, \bibinfo{person}{Elgar
  Fleisch}, {and} \bibinfo{person}{Tobias Kowatsch}.}
  \bibinfo{year}{2019}\natexlab{}.
\newblock \showarticletitle{Breeze: Smartphone-based acoustic real-time
  detection of breathing phases for a gamified biofeedback breathing training}.
\newblock \bibinfo{journal}{\emph{Proceedings of the ACM on Interactive,
  Mobile, Wearable and Ubiquitous Technologies}} \bibinfo{volume}{3},
  \bibinfo{number}{4} (\bibinfo{year}{2019}), \bibinfo{pages}{1--30}.
\newblock


\bibitem[Sim(2019)]%
        {sim2019mobile}
\bibfield{author}{\bibinfo{person}{Ida Sim}.} \bibinfo{year}{2019}\natexlab{}.
\newblock \showarticletitle{Mobile devices and health}.
\newblock \bibinfo{journal}{\emph{New England Journal of Medicine}}
  \bibinfo{volume}{381}, \bibinfo{number}{10} (\bibinfo{year}{2019}),
  \bibinfo{pages}{956--968}.
\newblock


\bibitem[Song et~al\mbox{.}(2020)]%
        {song2020spirosonic}
\bibfield{author}{\bibinfo{person}{Xingzhe Song}, \bibinfo{person}{Boyuan
  Yang}, \bibinfo{person}{Ge Yang}, \bibinfo{person}{Ruirong Chen},
  \bibinfo{person}{Erick Forno}, \bibinfo{person}{Wei Chen}, {and}
  \bibinfo{person}{Wei Gao}.} \bibinfo{year}{2020}\natexlab{}.
\newblock \showarticletitle{SpiroSonic: monitoring human lung function via
  acoustic sensing on commodity smartphones}. In
  \bibinfo{booktitle}{\emph{Proceedings of the 26th Annual International
  Conference on Mobile Computing and Networking}}. \bibinfo{pages}{1--14}.
\newblock


\bibitem[Soriano et~al\mbox{.}(2009)]%
        {soriano2009screening}
\bibfield{author}{\bibinfo{person}{Joan~B Soriano}, \bibinfo{person}{Jan
  Zielinski}, {and} \bibinfo{person}{David Price}.}
  \bibinfo{year}{2009}\natexlab{}.
\newblock \showarticletitle{Screening for and early detection of chronic
  obstructive pulmonary disease}.
\newblock \bibinfo{journal}{\emph{The Lancet}} \bibinfo{volume}{374},
  \bibinfo{number}{9691} (\bibinfo{year}{2009}), \bibinfo{pages}{721--732}.
\newblock


\bibitem[Sun et~al\mbox{.}(2016)]%
        {sun2016quadratic}
\bibfield{author}{\bibinfo{person}{Changjun Sun}, \bibinfo{person}{Lijing Li},
  {and} \bibinfo{person}{Wen Chen}.} \bibinfo{year}{2016}\natexlab{}.
\newblock \showarticletitle{Quadratic correlation time delay estimation
  algorithm based on kaiser window and hilbert transform}. In
  \bibinfo{booktitle}{\emph{2016 Sixth International Conference on
  Instrumentation \& Measurement, Computer, Communication and Control
  (IMCCC)}}. IEEE, \bibinfo{pages}{927--931}.
\newblock


\bibitem[Sun et~al\mbox{.}(2017)]%
        {sun2017sleepmonitor}
\bibfield{author}{\bibinfo{person}{Xiao Sun}, \bibinfo{person}{Li Qiu},
  \bibinfo{person}{Yibo Wu}, \bibinfo{person}{Yeming Tang}, {and}
  \bibinfo{person}{Guohong Cao}.} \bibinfo{year}{2017}\natexlab{}.
\newblock \showarticletitle{Sleepmonitor: Monitoring respiratory rate and body
  position during sleep using smartwatch}.
\newblock \bibinfo{journal}{\emph{Proceedings of the ACM on interactive,
  mobile, wearable and ubiquitous technologies}} \bibinfo{volume}{1},
  \bibinfo{number}{3} (\bibinfo{year}{2017}), \bibinfo{pages}{1--22}.
\newblock


\bibitem[Sutskever et~al\mbox{.}(2014)]%
        {sutskever2014sequence}
\bibfield{author}{\bibinfo{person}{Ilya Sutskever}, \bibinfo{person}{Oriol
  Vinyals}, {and} \bibinfo{person}{Quoc~V Le}.}
  \bibinfo{year}{2014}\natexlab{}.
\newblock \showarticletitle{Sequence to sequence learning with neural
  networks}.
\newblock \bibinfo{journal}{\emph{arXiv preprint arXiv:1409.3215}}
  (\bibinfo{year}{2014}).
\newblock


\bibitem[Tak et~al\mbox{.}(2017)]%
        {tak2017novel}
\bibfield{author}{\bibinfo{person}{Rishabh~N Tak}, \bibinfo{person}{Dharmesh~M
  Agrawal}, {and} \bibinfo{person}{Hemant~A Patil}.}
  \bibinfo{year}{2017}\natexlab{}.
\newblock \showarticletitle{Novel phase encoded mel filterbank energies for
  environmental sound classification}. In
  \bibinfo{booktitle}{\emph{International Conference on Pattern Recognition and
  Machine Intelligence}}. Springer, \bibinfo{pages}{317--325}.
\newblock


\bibitem[Tan and Ng(2008)]%
        {tan2008copd}
\bibfield{author}{\bibinfo{person}{Wan~C Tan} {and} \bibinfo{person}{Tze~P
  Ng}.} \bibinfo{year}{2008}\natexlab{}.
\newblock \showarticletitle{COPD in Asia: where East meets West}.
\newblock \bibinfo{journal}{\emph{Chest}} \bibinfo{volume}{133},
  \bibinfo{number}{2} (\bibinfo{year}{2008}), \bibinfo{pages}{517--527}.
\newblock


\bibitem[Taylor et~al\mbox{.}(2018)]%
        {taylor2018estimation}
\bibfield{author}{\bibinfo{person}{Terence~E Taylor}, \bibinfo{person}{Helena
  Lacalle~Muls}, \bibinfo{person}{Richard~W Costello}, {and}
  \bibinfo{person}{Richard~B Reilly}.} \bibinfo{year}{2018}\natexlab{}.
\newblock \showarticletitle{Estimation of inhalation flow profile using
  audio-based methods to assess inhaler medication adherence}.
\newblock \bibinfo{journal}{\emph{PloS one}} \bibinfo{volume}{13},
  \bibinfo{number}{1} (\bibinfo{year}{2018}), \bibinfo{pages}{e0191330}.
\newblock


\bibitem[Teferra(2017)]%
        {teferra2017iso}
\bibfield{author}{\bibinfo{person}{Meseret~N Teferra}.}
  \bibinfo{year}{2017}\natexlab{}.
\newblock \showarticletitle{ISO 14971-Medical Device Risk Management Standard}.
\newblock \bibinfo{journal}{\emph{International Journal of Latest Research in
  Engineering and Technology (IJLRET)}} \bibinfo{volume}{3},
  \bibinfo{number}{3} (\bibinfo{year}{2017}), \bibinfo{pages}{83--87}.
\newblock


\bibitem[Thap et~al\mbox{.}(2016)]%
        {thap2016high}
\bibfield{author}{\bibinfo{person}{Tharoeun Thap}, \bibinfo{person}{Heewon
  Chung}, \bibinfo{person}{Changwon Jeong}, \bibinfo{person}{Ki-Eun Hwang},
  \bibinfo{person}{Hak-Ryul Kim}, \bibinfo{person}{Kwon-Ha Yoon}, {and}
  \bibinfo{person}{Jinseok Lee}.} \bibinfo{year}{2016}\natexlab{}.
\newblock \showarticletitle{High-resolution time-frequency spectrum-based lung
  function test from a smartphone microphone}.
\newblock \bibinfo{journal}{\emph{Sensors}} \bibinfo{volume}{16},
  \bibinfo{number}{8} (\bibinfo{year}{2016}), \bibinfo{pages}{1305}.
\newblock


\bibitem[Tomasic et~al\mbox{.}(2018)]%
        {tomasic2018continuous}
\bibfield{author}{\bibinfo{person}{Ivan Tomasic}, \bibinfo{person}{Nikica
  Tomasic}, \bibinfo{person}{Roman Trobec}, \bibinfo{person}{Miroslav Krpan},
  {and} \bibinfo{person}{Tomislav Kelava}.} \bibinfo{year}{2018}\natexlab{}.
\newblock \showarticletitle{Continuous remote monitoring of COPD
  patients—justification and explanation of the requirements and a survey of
  the available technologies}.
\newblock \bibinfo{journal}{\emph{Medical \& biological engineering \&
  computing}} \bibinfo{volume}{56}, \bibinfo{number}{4} (\bibinfo{year}{2018}),
  \bibinfo{pages}{547--569}.
\newblock


\bibitem[Townsend et~al\mbox{.}(2011)]%
        {townsend2011spirometry}
\bibfield{author}{\bibinfo{person}{Mary~C Townsend},
  \bibinfo{person}{Occupational}, \bibinfo{person}{Environmental Lung~Disorders
  Committee}, {et~al\mbox{.}}} \bibinfo{year}{2011}\natexlab{}.
\newblock \showarticletitle{Spirometry in the occupational health setting--2011
  update}.
\newblock \bibinfo{journal}{\emph{Journal of occupational and environmental
  medicine}} \bibinfo{volume}{53}, \bibinfo{number}{5} (\bibinfo{year}{2011}),
  \bibinfo{pages}{569--584}.
\newblock


\bibitem[Van~der Ent et~al\mbox{.}(1996)]%
        {van1996tidal}
\bibfield{author}{\bibinfo{person}{CK Van~der Ent}, \bibinfo{person}{HJ
  Brackel}, \bibinfo{person}{J Van~der Laag}, {and} \bibinfo{person}{Jan~M
  Bogaard}.} \bibinfo{year}{1996}\natexlab{}.
\newblock \showarticletitle{Tidal breathing analysis as a measure of airway
  obstruction in children three years of age and older.}
\newblock \bibinfo{journal}{\emph{American journal of respiratory and critical
  care medicine}} \bibinfo{volume}{153}, \bibinfo{number}{4}
  (\bibinfo{year}{1996}), \bibinfo{pages}{1253--1258}.
\newblock


\bibitem[van Schayck et~al\mbox{.}(2003)]%
        {van2003early}
\bibfield{author}{\bibinfo{person}{Onno~CP van Schayck},
  \bibinfo{person}{Anthony D'Urzo}, \bibinfo{person}{Giovanni Invernizzi},
  \bibinfo{person}{Miguel Rom{\'a}n}, \bibinfo{person}{Bj{\"o}rn
  St{\"a}llberg}, {and} \bibinfo{person}{Christopher Urbina}.}
  \bibinfo{year}{2003}\natexlab{}.
\newblock \showarticletitle{Early detection of chronic obstructive pulmonary
  disease (COPD): the role of spirometry as a diagnostic tool in primary care}.
\newblock \bibinfo{journal}{\emph{Primary Care Respiratory Journal}}
  \bibinfo{volume}{12}, \bibinfo{number}{3} (\bibinfo{year}{2003}),
  \bibinfo{pages}{90--93}.
\newblock


\bibitem[Virtanen et~al\mbox{.}(2020)]%
        {virtanen2020scipy}
\bibfield{author}{\bibinfo{person}{Pauli Virtanen}, \bibinfo{person}{Ralf
  Gommers}, \bibinfo{person}{Travis~E Oliphant}, \bibinfo{person}{Matt
  Haberland}, \bibinfo{person}{Tyler Reddy}, \bibinfo{person}{David
  Cournapeau}, \bibinfo{person}{Evgeni Burovski}, \bibinfo{person}{Pearu
  Peterson}, \bibinfo{person}{Warren Weckesser}, \bibinfo{person}{Jonathan
  Bright}, {et~al\mbox{.}}} \bibinfo{year}{2020}\natexlab{}.
\newblock \showarticletitle{SciPy 1.0: fundamental algorithms for scientific
  computing in Python}.
\newblock \bibinfo{journal}{\emph{Nature methods}} \bibinfo{volume}{17},
  \bibinfo{number}{3} (\bibinfo{year}{2020}), \bibinfo{pages}{261--272}.
\newblock


\bibitem[Viswanath et~al\mbox{.}(2018)]%
        {viswanath2018spiroconfidence}
\bibfield{author}{\bibinfo{person}{Varun Viswanath}, \bibinfo{person}{Jake
  Garrison}, {and} \bibinfo{person}{Shwetak Patel}.}
  \bibinfo{year}{2018}\natexlab{}.
\newblock \showarticletitle{SpiroConfidence: determining the validity of
  smartphone based spirometry using machine learning}. In
  \bibinfo{booktitle}{\emph{2018 40th Annual International Conference of the
  IEEE Engineering in Medicine and Biology Society (EMBC)}}. IEEE,
  \bibinfo{pages}{5499--5502}.
\newblock


\bibitem[Walker et~al\mbox{.}(1990)]%
        {walker1990clinical}
\bibfield{author}{\bibinfo{person}{H~Kenneth Walker}, \bibinfo{person}{W~Dallas
  Hall}, {and} \bibinfo{person}{J~Willis Hurst}.}
  \bibinfo{year}{1990}\natexlab{}.
\newblock \showarticletitle{Clinical methods: the history, physical, and
  laboratory examinations}.
\newblock  (\bibinfo{year}{1990}).
\newblock


\bibitem[Wang et~al\mbox{.}(2021)]%
        {wang2021smartphone}
\bibfield{author}{\bibinfo{person}{Xuyu Wang}, \bibinfo{person}{Runze Huang},
  \bibinfo{person}{Chao Yang}, {and} \bibinfo{person}{Shiwen Mao}.}
  \bibinfo{year}{2021}\natexlab{}.
\newblock \showarticletitle{Smartphone Sonar-Based Contact-Free Respiration
  Rate Monitoring}.
\newblock \bibinfo{journal}{\emph{ACM Transactions on Computing for
  Healthcare}} \bibinfo{volume}{2}, \bibinfo{number}{2} (\bibinfo{year}{2021}),
  \bibinfo{pages}{1--26}.
\newblock


\bibitem[Warne(2014)]%
        {warne2014primer}
\bibfield{author}{\bibinfo{person}{Russell~T Warne}.}
  \bibinfo{year}{2014}\natexlab{}.
\newblock \showarticletitle{A primer on multivariate analysis of variance
  (MANOVA) for behavioral scientists.}
\newblock \bibinfo{journal}{\emph{Practical Assessment, Research \&
  Evaluation}}  \bibinfo{volume}{19} (\bibinfo{year}{2014}).
\newblock


\bibitem[Whitlock et~al\mbox{.}(2020)]%
        {whitlock2020spiro}
\bibfield{author}{\bibinfo{person}{Justin Whitlock}, \bibinfo{person}{Joshua
  Sill}, {and} \bibinfo{person}{Shubham Jain}.}
  \bibinfo{year}{2020}\natexlab{}.
\newblock \showarticletitle{A-spiro: Towards continuous respiration
  monitoring}.
\newblock \bibinfo{journal}{\emph{Smart Health}}  \bibinfo{volume}{15}
  (\bibinfo{year}{2020}), \bibinfo{pages}{100105}.
\newblock


\bibitem[Winder(2002)]%
        {winder2002analog}
\bibfield{author}{\bibinfo{person}{Steve Winder}.}
  \bibinfo{year}{2002}\natexlab{}.
\newblock \bibinfo{booktitle}{\emph{Analog and digital filter design}}.
\newblock \bibinfo{publisher}{Elsevier}.
\newblock


\bibitem[Xu et~al\mbox{.}(2021)]%
        {xu2021listen2cough}
\bibfield{author}{\bibinfo{person}{Xuhai Xu}, \bibinfo{person}{Ebrahim Nemati},
  \bibinfo{person}{Korosh Vatanparvar}, \bibinfo{person}{Viswam Nathan},
  \bibinfo{person}{Tousif Ahmed}, \bibinfo{person}{Md~Mahbubur Rahman},
  \bibinfo{person}{Daniel McCaffrey}, \bibinfo{person}{Jilong Kuang}, {and}
  \bibinfo{person}{Jun~Alex Gao}.} \bibinfo{year}{2021}\natexlab{}.
\newblock \showarticletitle{Listen2Cough: Leveraging End-to-End Deep Learning
  Cough Detection Model to Enhance Lung Health Assessment Using Passively
  Sensed Audio}.
\newblock \bibinfo{journal}{\emph{Proceedings of the ACM on Interactive,
  Mobile, Wearable and Ubiquitous Technologies}} \bibinfo{volume}{5},
  \bibinfo{number}{1} (\bibinfo{year}{2021}), \bibinfo{pages}{1--22}.
\newblock


\bibitem[Yadavar(2019)]%
        {indiaspend}
\bibfield{author}{\bibinfo{person}{Swagata Yadavar}.}
  \bibinfo{year}{2019}\natexlab{}.
\newblock \bibinfo{title}{{Why India Is Struggling To Tackle Its Lung Disease
  Crisis}}.
\newblock
  \bibinfo{howpublished}{\url{https://www.indiaspend.com/why-india-is-struggling-to-tackle-its-pulmonary-disease-crisis/}}.
\newblock
\newblock
\shownote{[Online; accessed 10-March-2022]}.


\bibitem[Yap and Moussavi(2002)]%
        {yap2002acoustic}
\bibfield{author}{\bibinfo{person}{Yee~Leng Yap} {and} \bibinfo{person}{Zahra
  Moussavi}.} \bibinfo{year}{2002}\natexlab{}.
\newblock \showarticletitle{Acoustic airflow estimation from tracheal sound
  power}. In \bibinfo{booktitle}{\emph{IEEE CCECE2002. Canadian Conference on
  Electrical and Computer Engineering. Conference Proceedings (Cat. No.
  02CH37373)}}, Vol.~\bibinfo{volume}{2}. IEEE, \bibinfo{pages}{1073--1076}.
\newblock


\bibitem[Yue et~al\mbox{.}(2018)]%
        {yue2018extracting}
\bibfield{author}{\bibinfo{person}{Shichao Yue}, \bibinfo{person}{Hao He},
  \bibinfo{person}{Hao Wang}, \bibinfo{person}{Hariharan Rahul}, {and}
  \bibinfo{person}{Dina Katabi}.} \bibinfo{year}{2018}\natexlab{}.
\newblock \showarticletitle{Extracting multi-person respiration from entangled
  rf signals}.
\newblock \bibinfo{journal}{\emph{Proceedings of the ACM on Interactive,
  Mobile, Wearable and Ubiquitous Technologies}} \bibinfo{volume}{2},
  \bibinfo{number}{2} (\bibinfo{year}{2018}), \bibinfo{pages}{1--22}.
\newblock


\bibitem[Zeng et~al\mbox{.}(2018)]%
        {zeng2018understanding}
\bibfield{author}{\bibinfo{person}{Ming Zeng}, \bibinfo{person}{Haoxiang Gao},
  \bibinfo{person}{Tong Yu}, \bibinfo{person}{Ole~J Mengshoel},
  \bibinfo{person}{Helge Langseth}, \bibinfo{person}{Ian Lane}, {and}
  \bibinfo{person}{Xiaobing Liu}.} \bibinfo{year}{2018}\natexlab{}.
\newblock \showarticletitle{Understanding and improving recurrent networks for
  human activity recognition by continuous attention}. In
  \bibinfo{booktitle}{\emph{Proceedings of the 2018 ACM international symposium
  on wearable computers}}. \bibinfo{pages}{56--63}.
\newblock


\bibitem[Zhang et~al\mbox{.}(2018)]%
        {zhang2018sequence}
\bibfield{author}{\bibinfo{person}{Chaoyun Zhang}, \bibinfo{person}{Mingjun
  Zhong}, \bibinfo{person}{Zongzuo Wang}, \bibinfo{person}{Nigel Goddard},
  {and} \bibinfo{person}{Charles Sutton}.} \bibinfo{year}{2018}\natexlab{}.
\newblock \showarticletitle{Sequence-to-point learning with neural networks for
  non-intrusive load monitoring}. In \bibinfo{booktitle}{\emph{Proceedings of
  the AAAI Conference on Artificial Intelligence}}, Vol.~\bibinfo{volume}{32}.
\newblock


\bibitem[Zhou et~al\mbox{.}(2020)]%
        {zhou2020accurate}
\bibfield{author}{\bibinfo{person}{Bo Zhou}, \bibinfo{person}{Alejandro
  Baucells~Costa}, {and} \bibinfo{person}{Paul Lukowicz}.}
  \bibinfo{year}{2020}\natexlab{}.
\newblock \showarticletitle{Accurate Spirometry with Integrated Barometric
  Sensors in Face-Worn Garments}.
\newblock \bibinfo{journal}{\emph{Sensors}} \bibinfo{volume}{20},
  \bibinfo{number}{15} (\bibinfo{year}{2020}), \bibinfo{pages}{4234}.
\newblock


\bibitem[Zubaydi et~al\mbox{.}(2017)]%
        {zubaydi2017mobspiro}
\bibfield{author}{\bibinfo{person}{Fatma Zubaydi}, \bibinfo{person}{Assim
  Sagahyroon}, \bibinfo{person}{Fadi Aloul}, {and} \bibinfo{person}{Hasan
  Mir}.} \bibinfo{year}{2017}\natexlab{}.
\newblock \showarticletitle{MobSpiro: Mobile based spirometry for detecting
  COPD}. In \bibinfo{booktitle}{\emph{2017 IEEE 7th Annual Computing and
  Communication Workshop and Conference (CCWC)}}. IEEE, \bibinfo{pages}{1--4}.
\newblock


\end{thebibliography}

\clearpage
\appendix

\section{Appendix}
\label{appendix-1}

In this supplementary document, we provide additional details for our paper.

\subsection{Validating Hilbert Transformation}
\begin{figure}[ht]
    \centering
    \includegraphics[scale=.65]{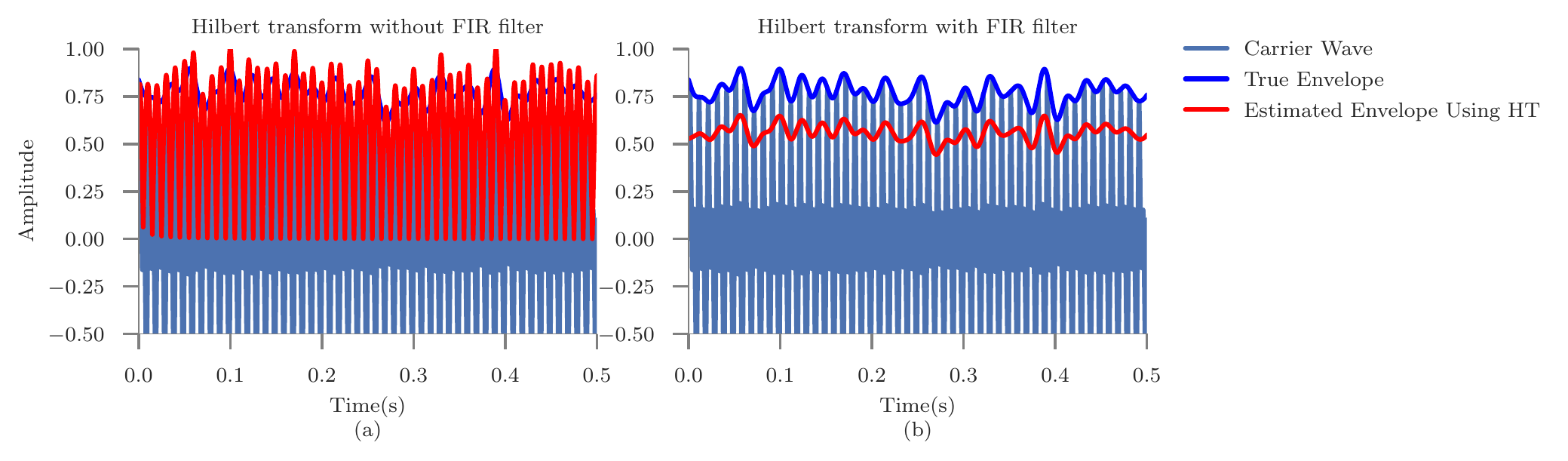}
    \caption{(a) A poor and (b) a proper fit of the signal envelope using Hilbert transform without and with FIR filter respectively}
    \label{fig:am-carrier}
\end{figure}

\noindent Figure~\ref{fig:am-carrier} below validates the Hilbert-Transform envelope detection algorithm for a synthetic signal with multiple harmonics.

\subsection{Comparison With Smartphone Spirometry}
\label{sec:comparison-with-smartphone}

\begin{figure}[ht]
\centering
\begin{subfigure}{.5\textwidth}
  \centering
  \includegraphics[width=.75\linewidth]{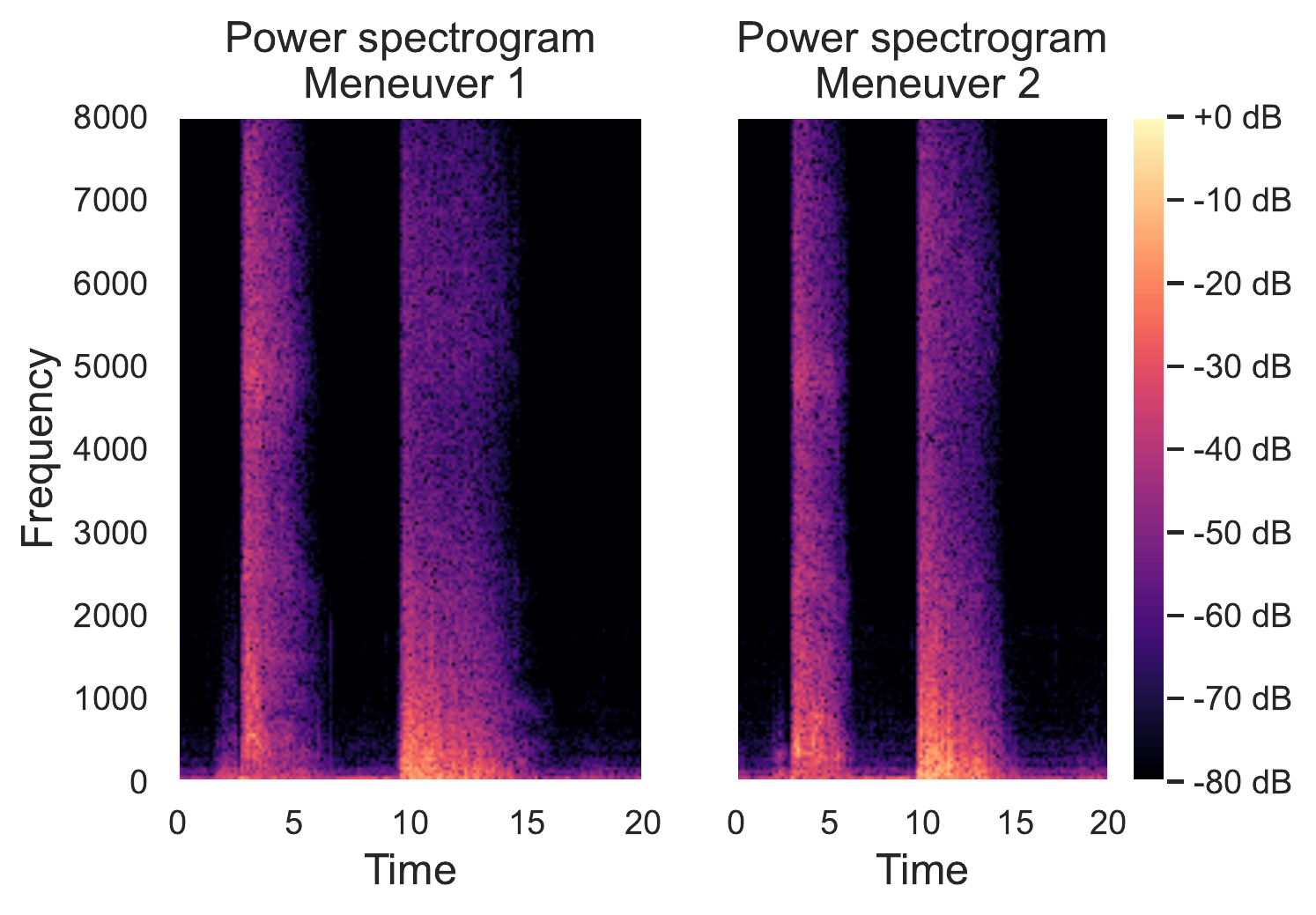}
  \caption{The information content in the spectrogram is similar\\ for both the maneuver performed in SpiroMask. Here, both\\ the inhalation followed by exhalation can be seen}
  \label{fig:multi-maneuver-spiromask}
\end{subfigure}%
\begin{subfigure}{.5\textwidth}
  \centering
  \includegraphics[width=.75\linewidth]{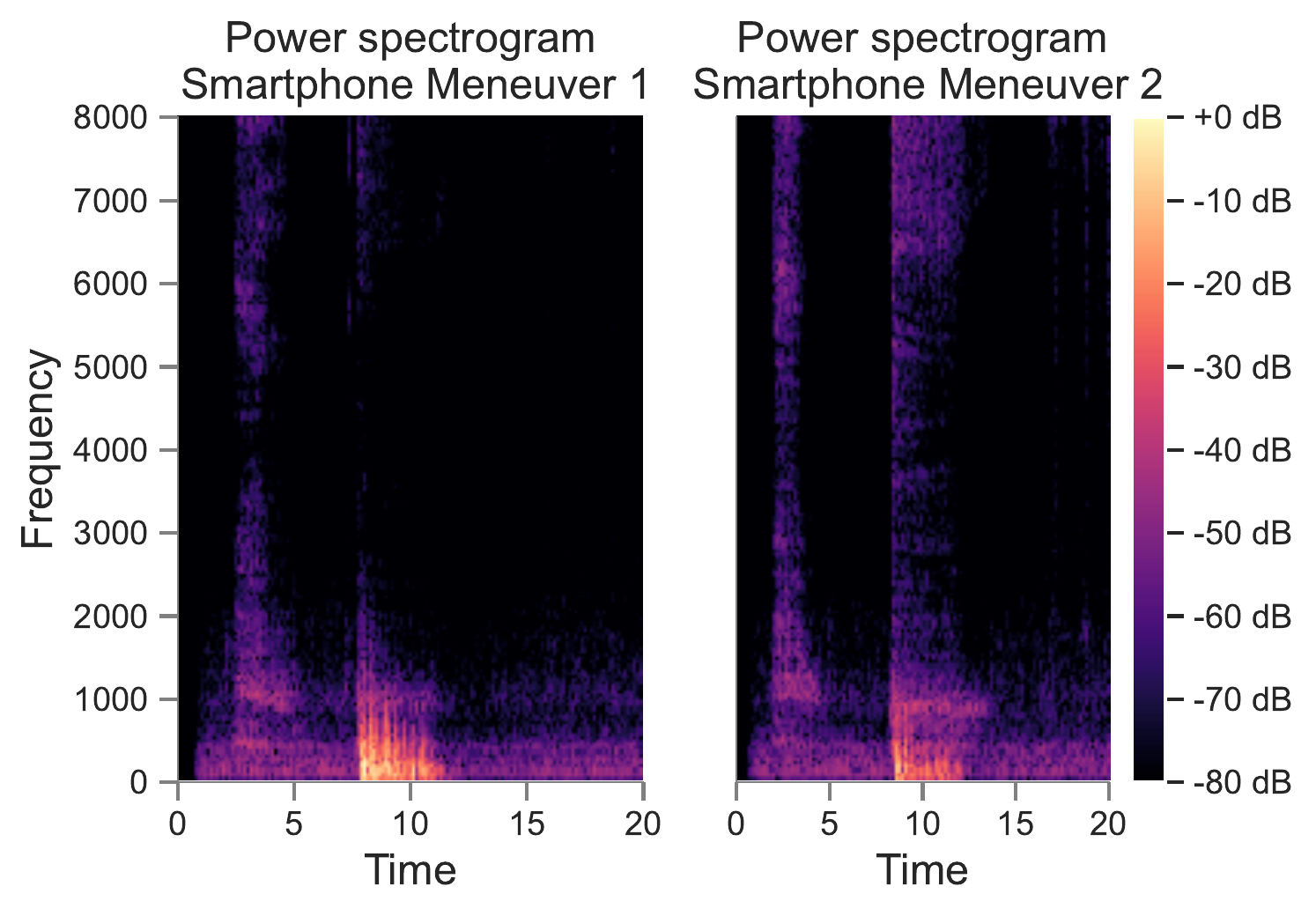}
  \caption{The information between 2-6Khz is missing in the first maneuver performed via smartphone. }
  \label{fig:multi-maneuver-smartphone}
\end{subfigure}
\caption{SpiroMask provides a more controlled environment to perform spirometry as compared to smartphone.}
\label{fig:multi-maneuver}
\end{figure}

To compare SpiroMask and smartphone spirometry,  we asked three participants to perform four forceful breathing maneuvers. Participants performed the first two maneuvers using a smartphone. They used SpiroMask for the following two maneuvers. Our objective was to quantify the difference between the two maneuvers recorded via both systems. Figure~\ref{fig:multi-maneuver} shows the comparison for one such user. The spectrogram of two maneuvers performed over SpiroMask are more similar (Figure~\ref{fig:multi-maneuver-spiromask}) than those conducted over the phone (Figure~\ref{fig:multi-maneuver-smartphone}). Particularly, the information between 2-6 kHz in one of the maneuver performed via smartphone is lost which leads to incorrect flow-volume curve~\cite{goel2016spirocall}. The variability arises in the phone because a user is unable to repeat the maneuver with the same angle and the distance between the mouth and the phone. A dissimilar flow-volume (FV) curve for the same user over a short period (< 5 Minutes) leads to a wrong lung health estimate~\cite{rubini2010daily}.  This experiment shows that SpiroMask provides a more controlled environment for spirometry. Figure~\ref{fig:multi-maneuver-spiromask} also shows that the inhalation phase of breathing 

%According to previous studies~\cite{rubini2010daily}, the Forced Vital Capacity (FVC) of two FV curves of the same person recorded over a short duration should not differ by more than 7\%. 

\subsection{Feature Set}
The list of features are described in Table~\ref{tab:all-features}
\begin{table}[ht]
\begin{tabular}{@{}lll@{}}
\toprule
  & \multicolumn{1}{c}{\textbf{Group}}                                                                            & \multicolumn{1}{c}{\textbf{Frame Level Descriptors}}                                                                                                                                                                                                                                                                                                   \\ \midrule
1 & Temporal Domain                                                                                               & \begin{tabular}[c]{@{}l@{}}Autocorrelation, Centroid, Mean absolute difference, Mean difference\\ Median absolute difference, Median difference, Distance, Sum of absolute difference,\\ Total energy,  Entropy, Peak to peak distance, Area under the curve, Absolute energy,\\ Maximum peaks, Minimum peaks, Slope, Zero crossing rate,\end{tabular} \\ \midrule
                                                       
2 & MFCC                                                                                                          & 13 Mel-Frequency Cepstral Coefficients                                                                                                        \\ \midrule
2 & MFCC (Normalised)                                                                                                          & 13  Variance Normalised Mel-Frequency Cepstral Coefficients                                                                                                        \\ \midrule
3 & Mel Spectogram                                                                                                         & 64 Mel Bands of Spectrogram                                                                      \\ \midrule
4 & Power Spectrum                                                                                                        & Power spectrum for each frame for each audio waveform                                                                      \\ \midrule
5 & Mel Filter Bank Energy                                                                                                  & 40 Mel Bands of spectrogram  
                                    \\ \midrule
4 & Log Mel Filter Bank Energy                                                                                                        & 40 Mel Bands of spectrogram    
                                                               \\ \bottomrule
\end{tabular}
\caption{List of features computed for each audio waveform. Note that the temporal domain features were also computed for the Flow-Volume curve.}
\label{tab:all-features}
\end{table}

\subsection{Length of audio signal for tidal breathing}
In Section~\ref{sec:datacollection_tidal}, we mentioned that we used 20 seconds of audio recording to estimate the respiration rate. We were limited to 20 seconds because of engineering challenges in our prototyping platform. To validate if 20 seconds of audio is enough to estimate respiration rate, we compared the 20 second recording with a 30 second recording collected from an alternate microphone. Both the microphones were placed inside the mask. One of these microphone was the onboard microphone of SpiroMask's prototyping platform\footnote{\url{https://docs.arduino.cc/hardware/nano-33-ble-sense/}} (Arduino's microphone) while the other microphone was embedded in a breakout board which was connected to the Raspberry Pi Pico (RPi Pico Microphone)~\footnote{https://www.raspberrypi.com/documentation/microcontrollers/raspberry-pi-pico.html}. We ensured that both the microphone have the same make and model. The RPi Pico with a interfaced microphone could be identified as a USB device in a Linux computer. A participant donned the two microphone mask and continued tidal breathing.\\

\begin{figure}[ht]
    \centering
    \includegraphics[scale=.65]{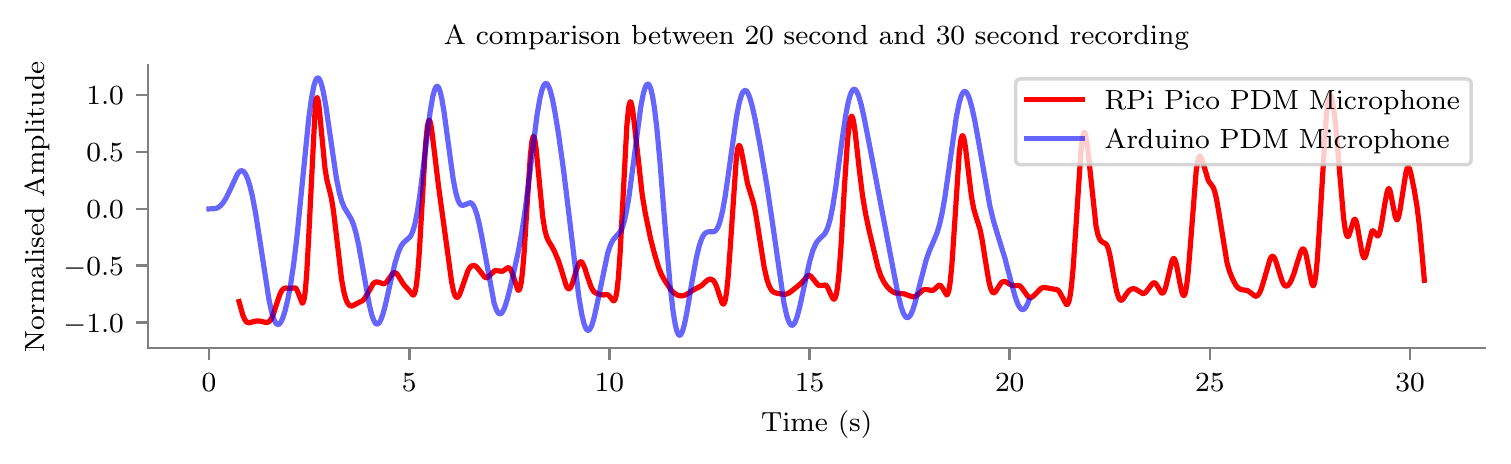}
    \caption{The average time between peak to peak between both the signal is same. A 20 second audio can thus be used to estimate respiration rate.}
    \label{fig:rpi-vs-arduino}
\end{figure}

\noindent We simultaneously started recording breathing audio from both the device. The RPi Pico microphone could record audio for a long time since it is acting as a USB microphone connected to a computer. The envelope of both the recording are shown in  Figure~\ref{fig:rpi-vs-arduino}. The Arduino microphone recorded a 20 second audio while we stopped the other recording at the end of 30 seconds. The average time between peak to peak for Arduino's 20 second recording was $2.69s$ while for the RPi Pico's 30 second recording  was $2.72s$. Both the signal had 7 peaks at the end of 20 seconds. The breathing rate estimated from the 20 second recording and the 30 second recording will be same. Thus, we conclude that 20 seconds of audio would suffice to estimate the respiration rate.  

\subsection{Ablation Studies}
Table~\ref{tab:ablation-PEF}, \ref{tab:ablation-FEV1} and \ref{tab:ablation-FVC} shows the most important features for each of the three lung parameter (PEF, FEV1 and FVC) discovered using Sequential Forward Selection (SFS) technique.

\begin{table}[ht]
\begin{tabular}{@{}llr@{}}
\toprule
  & \textbf{Category of Feature For PEF}                               & \textbf{Mean Percentage Error} \\ \midrule
1 & Temporal features of the FV curve &                         13.71\%                        \\
2 & Mel Filter Bank Energy (MFE) features                      & 13.32\%                        \\
3 & Mean and variance normalised MFCC features                 & 9.92\%                        \\ 
4 & Melspectogram features                                     & 10.12\%                         \\ 
4 & MFCC+MFE+Melspectrogram                                   & 6.89\%                         \\ 
5 & All Features (by restricting audio waveform til the peak of the signal)                                               & 6.71\%                         \\ \bottomrule
\end{tabular}
\caption{Mean Absolute Percentage Error for features selected in each category using SFS for PEF in cloth mask. The combination of thee features, i.e. MFCC, MFE and Melspectrogram gives comparable result}
\label{tab:ablation-PEF}
\end{table}

\begin{table}[ht]
\begin{tabular}{llr}
\hline
  & \textbf{Category of Feature (FEV1)}                                                                                                 & \multicolumn{1}{l}{\textbf{Mean Percentage Error}} \\ \hline

2 & Temporal features of the FV curve                                                                          & 19.25\%                                            \\
3 & Mel Filter Bank Energy (MFE) features                                                                                               & 11.58\%                                            \\
4 & Log of Mel Filter Bank Energy (MFE) features                                                                                        & 10.18\%                                            \\
5 & Mean and variance normalised MFCC features                                                                                          & 6.21\%                                            \\
                                      \\
9 & \begin{tabular}[c]{@{}l@{}}All Features (by restricting audio envelope till 1 second after the \\ start of exhalation)\end{tabular} & 5.25\%                                             \\ \hline
\end{tabular}
\caption{Mean Absolute Percentage Error for features selected in each category using SFS for FEV1 in cloth mask. MFCC alone achieves a error of 6.21\% which is acceptable as per ATS criteria.}
\label{tab:ablation-FEV1}
\end{table}

\begin{table}[ht]
\begin{tabular}{llr}
\hline
  & \textbf{Category of Feature (FVC)}                         & \multicolumn{1}{l}{\textbf{Mean Percentage Error}} \\ \hline
1 & STemporal features of the FV curve & 21.36\%                                            \\
2 & Mel Filter Bank Energy (MFE) features                      & 12.71\%                                            \\
3 & Log of Mel Filter Bank Energy (MFE) features               & 12.38\%                                            \\
4 & Mean and variance normalised MFCC features                 & 13.11\%                                            \\
5 & Power spectrum features                                    & 9.69\%                                             \\
6 & Melspectogram features+MFCC                                 & 6.16\%                                             \\

7 & All Features                                               & 5.67\%                                             \\ \hline
\end{tabular}
\caption{Mean Absolute Percantage Error for features selected in each category using SFS for FVC in cloth mask. It can be seen that melspectrogam in combination with MFCC gives comparable result}
\label{tab:ablation-FVC}
\end{table}

\subsection{Multivariate Analysis of Variance (MANOVA) Test}
\label{sec:manova}

\begin{figure}[ht]
    \centering
    \includegraphics[scale=0.52]{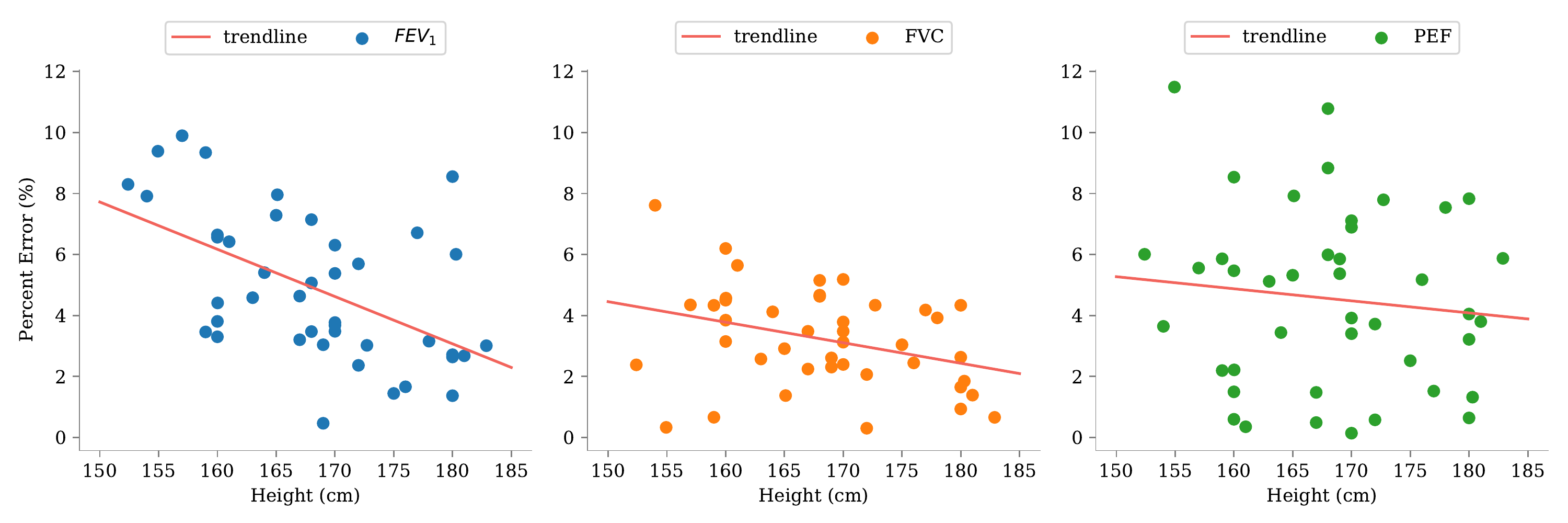}
    \caption{Variation of percent error for the forced breathing parameters with respect to height (the grouping variable with the lowest p-value in the MANOVA test) for N95 mask}
      \label{fig:scatter-height}
\end{figure}

Tables \ref{tab:manova} and \ref{tab:anova} show the results obtained in the 8-way MANOVA and 7-way ANOVA tests respectively. MANOVA was performed for forced breathing parameters and ANOVA was performed for the tidal breathing parameter (respiration rate). Both MANOVA and ANOVA were performed for cloth and N95 masks. The 8 grouping variables used in the MANOVA tests were height, weight, gender, age, whether the subject has performed spirometry tests before, whether the subject reported any lung ailments, whether the subject has a habit of smoking, and whether the subject had a meal before appearing for the experiment. The 7 grouping variables used in the ANOVA tests were height, weight, gender, age, whether the subject reported any lung ailments, whether the subject has a habit of smoking, and whether the subject had a meal before appearing for the experiment. Figure \ref{fig:scatter-height} presents the variation of percent error for the forced breathing parameters with respect to height (the grouping variable with the lowest p-value in the MANOVA test) for N95 mask. 

\begin{table}[ht]
\begin{tabular}{@{}lrr@{}}
\toprule
\multicolumn{1}{c}{\textbf{Grouping Variable}}   & \multicolumn{2}{c}{\textbf{p-value}} \\ \midrule
\multicolumn{1}{c}{} & \multicolumn{1}{c}{\textbf{Cloth Mask}} & \multicolumn{1}{c}{\textbf{N95 Mask}}\\ \midrule
Age  & 0.9886  & 0.5533  \\ 
Gender  & 0.8378 & 0.6739   \\ 
Height & 0.9365  & 0.0383         \\ 
Weight & 0.4568      & 0.5243    \\ 
Whether the subject had a meal before the experiment & 0.5802 & 0.1457  \\
Whether the subject has a habit of smoking & 0.8733  & 0.2166  \\
Whether the subject has performed spirometry before  & 0.1021 & 0.7105  \\
Whether the subject reported any lung ailment  & 0.0930  & 0.4712  \\
\bottomrule
\end{tabular}
\caption{Results of the MANOVA test for both cloth and N95 masks are shown above. The test for cloth mask shows that none of the grouping variables have a significant effect on the percent error between SpiroMask and spirometer measures (all p-values were greater than 0.05, the significance level of the test). The test for N95 mask suggests that height could be a significant variable  (p-value $\approx$ 0.04).
%\nb{combine the tables for N95 and cloth} 
%\nb{revise caption based on my earlier comment on this section in the text -- what mask is this result for?}
}
\label{tab:manova}
\end{table}

\begin{table}[ht]
\begin{tabular}{@{}lrr@{}}
\toprule
\multicolumn{1}{c}{\textbf{Grouping Variable}}   & \multicolumn{2}{c}{\textbf{p-value}} \\ \midrule
\multicolumn{1}{c}{} & \multicolumn{1}{c}{\textbf{Cloth Mask}} & \multicolumn{1}{c}{\textbf{N95 Mask}}\\ \midrule
Age  & 0.9436  & 0.1311  \\ 
Gender  & 0.6492 & 0.7258   \\ 
Height & 0.0796  & 0.7958         \\ 
Weight  & 0.4298      & 0.5990    \\ 
Whether the subject had a meal before the experiment & 0.2749 & 0.0881  \\
Whether the subject has a habit of smoking & 0.3807  & 0.8628  \\
Whether the subject reported any lung ailment  & 0.4152  & 0.8152  \\
\bottomrule
\end{tabular}
\caption{Results of the ANOVA test for both cloth and N95 masks are shown above. None of the grouping variables have a significant effect on the percent error between SpiroMask and smartphone measures (all p-values are greater than 0.05, the significance level of the test). 
%\nb{combine the tables for N95 and cloth} 
%\nb{revise caption based on my earlier comment on this section in the text -- what mask is this result for?}
}
\label{tab:anova}
\end{table}

\subsection{Sensitivity Analysis on Position of Sensor}

\begin{figure}[ht]
    \centering
    \includegraphics[scale=0.3]{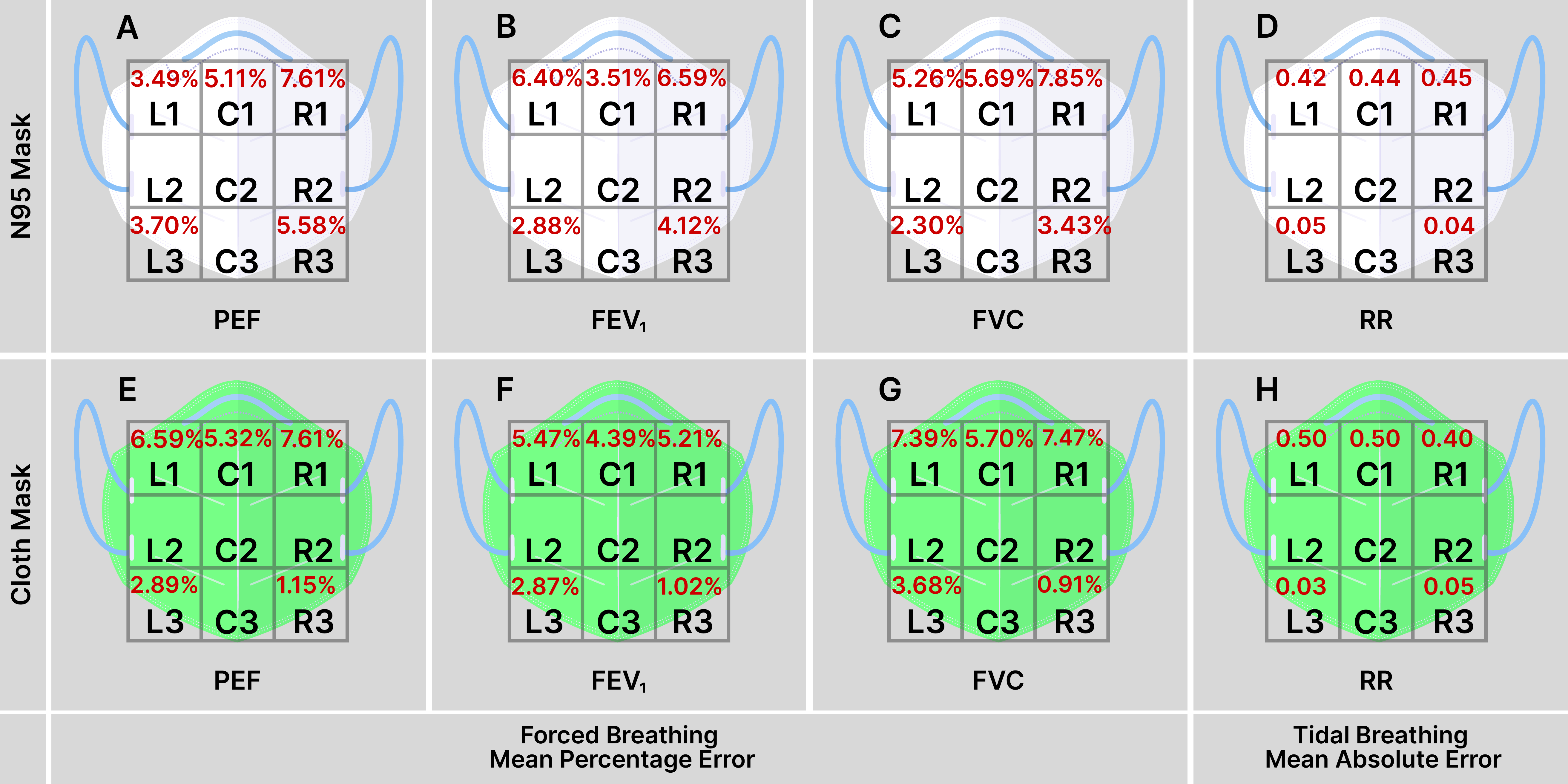}
    \caption{Errors in forced breathing parameters are presented for 4 healthy subjects (for both types of masks) and errors in tidal breathing parameter (respiration rate) are presented for 10 healthy subjects (N95 mask) and 7 healthy subjects (cloth mask)
    %\nb{do we not have sensor position sensitivity for tidal?}%
    }
      \label{fig:position-appendix_1}
\end{figure}

\begin{figure}[ht]
    \centering
    \includegraphics[scale=0.3]{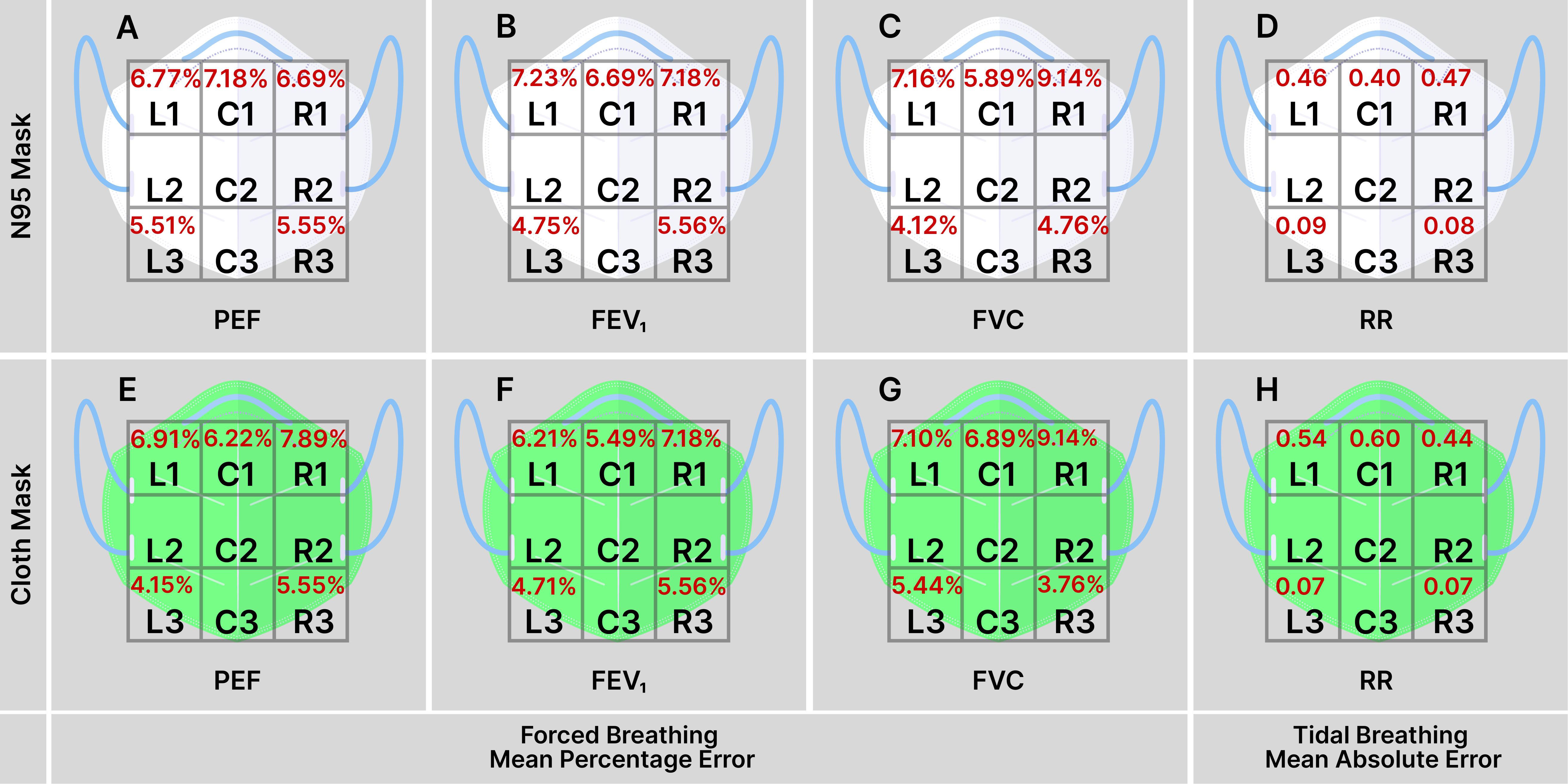}
    \caption{Errors in forced breathing parameters are presented for 8 unhealthy subjects (for both types of masks) and errors in tidal breathing parameter (respiration rate) are presented for 8 unhealthy subjects (for both types of masks)
    %\nb{do we not have sensor position sensitivity for tidal?}
    }
      \label{fig:position-appendix}
\end{figure}

Figures \ref{fig:position-appendix_1} and \ref{fig:position-appendix} present a detailed break up of MPE and MAE for forced and tidal breathing parameters among healthy and unhealthy participants respectively.

\end{document}